\documentclass[
reprint,
superscriptaddress,
 amsmath,amssymb,showpacs,showkeys,
 aps,
pra,twocolumn,
floatfix,
]{revtex4}
\makeatletter

\newcommand{\Rmnum}[1]{\expandafter\@slowromancap\romannumeral #1@}
\makeatother
\usepackage{graphicx}% Include figure files
\usepackage{dcolumn}% Align table columns on decimal point
\usepackage{bm}% bold math
\usepackage[colorlinks,citecolor=blue,linkcolor=blue]{hyperref}
\usepackage{amssymb}
\usepackage{subfigure}
\begin{document}

\title{Engineering the phase-robust topological router in a chiral-symmetric dimerized superconducting circuit lattice with long-range hopping}% Force line breaks with \\

\author{Li-Na Zheng}
\affiliation{Center for Quantum Sciences and School of Physics, Northeast Normal University, Changchun 130024, China}
\author{Hong-Fu Wang}
\email{hfwang@ybu.edu.cn}
\affiliation{Center for Quantum Sciences and School of Physics, Northeast Normal University, Changchun 130024, China}
\affiliation{Department of Physics, College of Science, Yanbian University, Yanji, Jilin 133002, China}
\author{Xuexi Yi}
\email{yixx@nenu.edu.cn}
\affiliation{Center for Quantum Sciences and School of Physics, Northeast Normal University, Changchun 130024, China}


\date{\today}% It is always \today, today,

\begin{abstract}
We propose a scheme to implement the phase-robust topological router based on a one-dimensional dimerized superconducting circuit lattice with long-range hopping. We show that the proposed dimerized superconducting circuit lattice can be mapped into an extended chiral-symmetric Su-Schrieffer-Heeger (SSH) model with long-range hopping, in which the existence of long-range hopping induces a special zero-energy mode. The peculiar distribution of the zero-energy mode enables us to engineer a phase-robust topological router, which can achieve quantum state transfer (QST) from one site (input port) to multiple sites (output ports). Benefiting from the topological protection of chiral symmetry, we demonstrate that the presence of the mild disorder in nearest-neighbor and long-range hopping has no appreciable effects on QST in the lattice. Especially, after introducing another new long-range hopping into the extended SSH lattice, we propose an optimized protocol of the phase-robust topological router, in which the number of the output ports can be efficiently increased. Resorting to the Bose statistical properties of the superconducting circuit lattice, the input port and output ports assisted by the zero-energy mode can be detected via the mean distribution of the photons. Our work breaks the traditional QST form with only one outport by the zero-energy mode and opens a pathway to construct large-scale quantum information processing in the SSH chains with long-range hopping.
\end{abstract}
\maketitle

\section{Introduction}
The quantum Hall (QH) effect in two-dimensional (2D) electron gas is a great discovery~\cite{Klitzing1980New, Laughlin1981Quantized, Thouless1982Quantized}, which breaks the iron rule of classifying different phases according to Landau's approach in condensed matter physics. And following the profound study in QH effect, a brand-new classification paradigm named topological order~\cite{Wen1995Topological} is introduced. Generally, QH effect needs to meet two rigorous conditions: strong magnetic field and low temperature, which limits its popularization and application. Therefore, a looming problem of condensed matter physics appears to explore QH effect in real materials~\cite{Bernevig2004Quantum, Hsieh2008Dirac, Zhang2009Topological}. By combining the spin orbit interaction and time reversal symmetry~\cite {Kane2005Quantum, Kane2005Topological,Fu2007Topological,Roy2009Topological}, topological insulators~\cite{Hasan2010Topological, Qi2011Topological} are one class of derivatives and achieve an electronic state with a similar physical phenomenon of QH effect. Moreover, several interesting characters of topological insulators have been recognized. For example, the electrical edge states on the surface are protected by the energy gap and they propagate in a single direction only along the edge~\cite{Hasan2010Topological, Cao2020Band, Schnyder2008Classification}. These edge states present robust transfer properties which are immune to the mild disorder and perturbation added into the whole system~\cite{Wu2013Robust, Chen2013Robustness, Alvarez2018Non-Hermitian}. Accordingly, it has been shown that the topological edge states are the promising candidates for implementing quantum state transfer (QST)~\cite{Dlaska2017Robust, Longhi2017Robust, Lemonde2019Quantum, Tan2020High, Qi2020Controllable, Zheng2020Defect, Verbin2015Topological, Mei2018Robust}.

The QST is a crucial process, which allows the quantum state of encoded information to be transferred between remote nodes, and directly or indirectly determines the reliability of quantum information processing (QIP)~\cite{Duan2010Quantum, Galindo2002Information, Zheng2000Efficient, Stannigel2012Optomechanical, Monroe2002Quantum}. Recently, given the structural simplicity and the abundant physical phenomenon concurrently, the Su-Schrieffer-Heeger (SSH) model~\cite{Schrieffer1979Solitons, Takayama1980Continuum, Kivelson1982Hubbard}, as one of the simplest topological insulator models, has been widely investigated~\cite{Zheng2020Defect, Mei2018Robust, Qi2021Topological, Han2021Large, Palaiodimopoulos2021Fast, Nie2020Bandgap, Qi2020Engineering} towards utilizing the edge state to realize QST. For instance, based on an odd-sized SSH model, a single-qubit state has been accurately transferred from the left- to the right-edge in one-dimensional (1D) superconducting Xmon qubits chain~\cite{Mei2018Robust}. Notably, the QST with only one output port seems to keep solidified, which is not enough to build large-scale QIP. Motivated by multiple output ports in QST, Qi {\it et al.}~\cite{Qi2020Engineering} proposed a scheme to implement the topological beam splitter in an even-sized SSH chain, in which the particle at the right-edge can be transferred to the first two sites with equal weight. After that, they also designed a topological router with multiple output ports~\cite{Qi2021Topological}, which further expands and deepens the research of QST by adding the specific long-range hopping between odd sites in the SSH model. We note that the above scheme for constructing large-scale QST with multiple output ports focuses on only the probability distribution of the gap state and ignores the phase information. This may limit the application of tasks related to phase information in QIP, such as quantum interference~\cite {Harris1998Photon, Yan2001Observation} and quantum logical gate~\cite{Chen2015Fast, Torosov2014High, Palmero2017Fast}. Besides that, the existence of the long-range hopping also destroys the chiral symmetry, which greatly weakens the topological protection originating from chiral symmetry~\cite{Han2020Valleylike, Mochizuki2020Topological, Martinez Alvarez2019Edge, Obuse2011Topological}.

To overcome these limitations, in this paper, we propose a scheme to realize the phase-robust topological router in a chiral-symmetric dimerized superconducting circuit lattice with long-range hopping. We demonstrate that the introduction of the long-range hopping with the same strength as the intercell hopping added on the first site and all of the other even sites~(except the second site) cannot break the chiral symmetry of the system, which further supports a special topological zero-energy gap state in the extended SSH lattice. We derive the wave function of the gap state theoretically and analytically, which reveals that topological gap state has the equal probability distribution at odd sites~(except the third site) accompanied with different phase distributions. Via utilizing the probability distribution and phase distribution of zero-energy gap state, we show that the topologically protected quantum channel can be constructed, which realizes the function of phase topological router. To verify the robustness of the above function, we examine the effects of different types of disorder on the phase topological router. We find that the present scheme is immune to the mild disorder added in nearest-neighbor and long-range hopping due to the topological protection of chiral symmetry.

Significantly, we also demonstrate that the number of output ports can be increased via increasing the long-range hopping with the same strength as the intercell hopping between the second site and arbitrary odd site~(except the first and the third site) in the previous extended SSH model. Furthermore, we make a comparison of the minimum energy gap versus the new long-range hopping that connects different sites. We find that, when the new long-range hopping is added on the second site and the third site from the right edge, the phase-robust topological router with $N+1$ output ports can be realized due to the existence of  gap state with a wide energy gap. In addition, we show that the input and output ports assisted by the gap state can be detected via the mean distribution of the photons.

The phase-robust topological router we proposed has the following advantages. First, as the core resource of implementing the phase-robust topological router, the gap state possesses a unique zero-energy mode and is strictly protected by chiral symmetry, which provides a natural barrier against the mild disorders and perturbations. Second, we investigate simultaneously the probability distribution and phase distribution of the gap state. These may promote the applications that depend on quantum phase, such as quantum logic gate and quantum interference. Finally, the gap state is localized at more sites with the same probability and the present scheme may contribute to development of quantum communication technology~\cite{Gisin2007Quantum, Vaziri2002Experimental}.

The paper is organized as follows: In Sec.~\ref{sec.2}, we present the extended SSH model of the 1D superconducting circuit lattice with the long-range hopping and analyse the distribution of the special gap state induced by the long-range hopping. In Sec.~\ref{sec.3}, we demonstrate that the phase-robust topological router can be engineered via the gap state. In Sec.~\ref{sec.4}, we show that the detection and evolution of the gap state. Finally, a conclusion is given in Sec.~\ref{sec.5}.

\section{\label{sec.2}System and Hamiltonian}
\subsection{\label{sec.2A} The extended SSH-type superconducting circuit lattice with long-range hopping}
\begin{figure}
	\centering
	\includegraphics[width=1\linewidth]{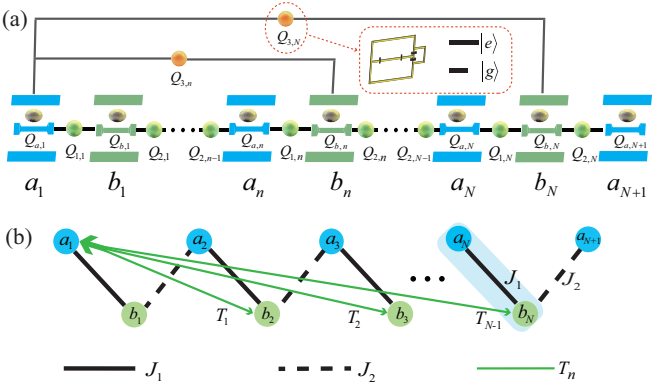}\\
	\caption{\label{f1} (a) Schematic of the phase-robust topological router in a chiral-symmetric dimerized superconducting circuit lattice with long-range hopping. Each subsite contains a resonator and an embedded flux qubit, labeled as $a_{n}$~($b_{n}$) and $Q_{a,n}$~($Q_{b,n}$). Two adjacent resonators, $a_{n}$ and $b_{n}$, constitute a unit cell, which connected by the flux qubit $Q_{1,n}$. While the another two resonators, $a_{n+1}$ and $b_{n}$ belonging to different units can be connected by the flux qubit $Q_{2,n}$. In addtion, the two resonators $a_{1}$ and $b_{n}~(n=2,3,4,\cdots,N)$ are both coupled with each other via the flux qubit $Q_{3,n}$. Their coupling strengths are set as $g_{a,n}$($g_{b,n}$), $g_{1,n}$,  $g_{2,n}$, and $g_{3,n}$, respectively. (b) The extended SSH-type superconducting circuit lattice with long-range hopping. Each color represents a class of sites.  The solid line in black describes the intracell hopping amplitude $J_{1}$ between $a_{n}$ and $b_{n}$. The dotted line in black describes the intercell hopping amplitude $J_{2}$ between $a_{n+1}$ and $b_{n}$. The appointed long-range hopping in green line connecting the site $a_{1}$ and site $b_{n}~(n = 2, 3, ..., N)$ with hopping amplitude $T_{n}~(n = 1, 2, ..., N-1)$.}\label{fig1}
\end{figure}

The setup of the superconducting circuit lattice for engineering the phase-robust topological router is illustrated in Fig.~\ref{fig1}(a). This lattice is composed of $2N+1$ resonators and $5N$ flux qubits (the energy levels $|e\rangle$ and $|g\rangle$), in which the $N$ is the total number of unit cells and we stipulate it as even in the following.  Each unit cell contains two resonators, labeled $a_{n}$ and $b_{n}$, which are both coupled with the flux qubit $Q_{1,n}$. The resonators $a_{n+1}$ and $b_{n}$ belonging to the two nearest-neighbor unit cells are both coupled with the flux qubit $Q_{2,n}$. Here, the intracell and intercell coupling strengths are $g_{1,n}$ and $g_{2,n}$, respectively. The flux qubit $Q_{a,n}$ ($Q_{b,n}$) is embedded inside $a_{n}$ ($b_{n}$) with the coupling strength $g_{a,n}$ ($g_{b,n}$). Especially, two resonators $a_{1}$ and $b_{n}~(n=2,3,4,\cdots,N)$ 
are connected by the flux qubit $Q_{3,n}$ with the coupling strength $g_{3,n}$. Then, the system can be dominated via the following Hamiltonian,

 \begin{eqnarray}\label{e01}
H_{total}&=&\sum_{n}\sum_{i=a,b,1,2,3} \left ( \omega_{a,n}a_{n}^{\dag}a_{n}+\omega_{b,n}b_{n}^{\dag}b_{n}+\frac{\omega_{qi,n}}{2}\sigma_{qi,n}^{z}\right )\cr\cr
&&+\sum_{n} \left (\Omega_{a,n}e^{-i\omega_{d,n}t}a_{n}^{\dag}+\Omega_{b,n}e^{-i\omega_{d,n}t}b_{n}^{\dag}+\mathrm{H.c.} \right )\cr\cr
&&+\sum_{n} \left (g_{a,n}\sigma_{qa,n}^{\dag}a_{n}
+g_{b,n}\sigma_{qb,n}^{\dag}b_{n}+\mathrm{H.c.} \right)\cr\cr
&&+\sum_{n} \left [ g_{1,n}\sigma_{q1,n}^{\dag}(a_{n}+b_{n})\right.\cr\cr
&&+\left.g_{2,n}\sigma_{q2,n}^{\dag}(b_{n}+a_{n+1})+\mathrm{H.c.} \right] \cr\cr
&&+\sum_{n} \left [ g_{3,n}\sigma_{q3,n}^{\dag}(a_{1}+b_{n})+\mathrm{H.c.} \right]
\end{eqnarray}
where the first summation represents the free energy of the superconducting resonators and qubits, the second summation is the external driving of the resonators with frequency $\omega_{d,n}$ and amplitude $\Omega_{k,n}$ $(k=a,b)$, the third summation represents the coupling between the resonators and the embedded qubits, the fourth summation indicates the intracell (intercell) hopping between two adjacent resonators assisted via the intermediate qubits, and the last summation represents the long-range hopping via the flux qubits.

In the rotating frame with respect to the externel driving frequency $\omega_{d,n}$ and qubit frequency $\omega_{qi,n}$, when all of the qubits are prepared in their ground states~\cite{Mei2016Witnessingt}, the Hamiltonian in Eq.~(\ref{e01}) can be written as [see Appendix A]
\begin{eqnarray}\label{e02}
H_{eff}^{\prime}&=& \left( \Delta_{a,1}-\frac{g_{a,1}^{2}}{\Delta_{qa,1}} -\frac{g_{1,1}^{2}}{\Delta_{q1,1}}-\frac{g_{2,1}^{2}}{\Delta_{q2,1}}-\sum_{n}\frac{g_{3,n}^{2}}{\Delta_{q3,n}} \right)a_{1}^{\dag}a_{1}\cr\cr
&&+\left(\Delta_{b,1}-\frac{g_{b}^{2}}{\Delta_{qb,1}} -\frac{g_{1}^{2}}{\Delta_{q1,1}}-\frac{g_{2}^{2}}{\Delta_{q2,1}}\right)b_{1}^{\dag}b_{1}\cr\cr
&&+\sum_{n}\left [ \left (\Delta_{a,n}-\frac{g_{a,n}^{2}}{\Delta_{qa,n}} -\frac{g_{1,n}^{2}}{\Delta_{q1,n}}-\frac{g_{2,n}^{2}}{\Delta_{q2,n}} \right)a_{n}^{\dag}a_{n}\right.\cr\cr
&&\left.+\left(\Delta_{b,n}-\frac{g_{b,n}^{2}}{\Delta_{qb,n}} -\frac{g_{1,n}^{2}}{\Delta_{q1,n}}-\frac{g_{2,n}^{2}}{\Delta_{q2,n}}-\frac{g_{3,n}^{2}}{\Delta_{q3,n}}\right) b_{n}^{\dag}b_{n} \right]\cr\cr
&&+\sum_{n}\left (-\frac{g_{1,n}^{2}}{\Delta_{q1,n}}b_{n}^{\dag}a_{n}-\frac{g_{2,n}^{2}}{\Delta_{q2,n}}a_{n+1}^{\dag}b_{n}+\mathrm{H.c.}\right)\cr\cr
&&+\sum_{n}\left (-\frac{g_{3,n}^{2}}{\Delta_{q3,n}}b_{n}^{\dag}a_{1}-\frac{g_{3,n}^{2}}{\Delta_{q3,n}}a_{1}^{\dag}b_{n}\right)\cr\cr
&&+\sum_{n} \left (\Omega_{a,n}a_{n}^{\dag}+\Omega_{b,n}b_{n}^{\dag}+\mathrm{H.c.}\right).
\end{eqnarray}
where $\Delta_{k,n}=\omega_{k,n}-\omega_{d,n}$ is the detuning between the resonators and external driving, $\Delta_{i,n}=\omega_{qi,n}-\omega_{d,n}$ ($i=a, b, 1, 2, 3$) represents the detuning of the qubits. 
Benefiting from the fact that the coupling between resonators and qubits or the detuning of qubits can be tuned individually and adiabatically via the external control, e.g., the flux-bias line or the external coupling circuit~\cite{Neeley2008Neeley, You2005Superconducting, Schmidt2013Circuit}, we modulate the 
coupling strengths and detuning as $-\frac{g_{1,n}^{2}}{\Delta_{q1,n}}=J_{1}=J-\cos\theta$, $-\frac{g_{2,n}^{2}}{\Delta_{q2,n}}=-\frac{g_{3,n}^{2}}{\Delta_{q3,n}}=J_{2}=J+\cos\theta$, $\frac{g_{a,1}^{2}}{\Delta_{qa,1}}=J_{1}+NJ_{2}$, $\frac{g_{b,1}^{2}}{\Delta_{qb,1}}=\frac{g_{a,n}^{2}}{\Delta_{qa,n}}=J_{1}+J_{2}$, and $\frac{g_{b,n}^{2}}{\Delta_{qb,n}}=J_{1}+2J_{2}$. After resetting the detuning of resonators $\Delta_{a,n}=\Delta_{b,n}$ as zero-energy point and further implementing the standard linearization process [see Appendix A],  the final Hamiltonian becomes 
\begin{eqnarray}\label{e03}
H&=&\sum_{n=1}^{N}\left(J_{1}a_{n}^{\dag}b_{n}+J_{2}a_{n+1}^{\dag}b_{n}+\mathrm{H.c.}\right)\cr\cr
&+&\sum_{n=1}^{N-1}\left(T_{n}a_{1}^{\dag}b_{n+1}+\mathrm{H.c.}\right),
\end{eqnarray}
The above Hamiltonian implies that the coupling terms between the resonator and the flux qubit can be removed, leading to the superconducting circuit lattice is equivalent to an effective coupled resonator lattice, as shown in Fig.~\ref{fig1}(b). The coupled resonator lattice is composed of $N$ unit cells ($2N+1$ sites), each of the unit cell contains two sublattice sites $a_{n}$ and $b_{n}$, in which the intracell (intercell) hopping amplitude is defined as $J_{1}=J+\cos\theta$ ($J_{2}=J-\cos\theta$). And, the long-range hopping connecting the first $a$-type site $a_{1}$ and $b$-type sites in the $n$th ($n = 2, 3, \cdots, N$) unit cell accompanied with the hopping amplitude $T_{n}=J_{2}=J-\cos\theta~(n = 1, 2, \cdots, N-1)$. Apparently, the superconducting circuit lattice can be completely mapped into the extended SSH model with long-range hopping.

\subsection{\label{sec.2B} Chiral symmetry and the wave function of zero-energy mode}
The system with Hamiltonian $H$ has chiral symmetry \cite{JK2016Short}, i.e., $\hat{\Gamma}\hat{H}\hat{\Gamma}^{+}=-\hat{H}$. Here, chiral operator $\hat{\Gamma}$ can be written in a diagonal matrix form with
\begin{eqnarray}\label{e04}
	\Gamma =\mathrm{Diag}\left[(-1)^{0}, (-1)^{1}, (-1)^{2}, \cdots, (-1)^{L-1}\right].
\end{eqnarray}
The chiral symmetry results in the energy spectrum of the extended SSH model is symmetric, each eigenvalue $E$ accompanied by a chiral symmetric partner with eigenvalue $-E$. Such symmetry indicates the existence of a zero-energy eigenvalue that is paired with itself in the present odd-sized extended SSH lattice, namely, $0_{+}=-0_{-}=0$. Resorting to the topological protection of energy gap and chiral symmetry,  the zero-energy mode is insensitive to mild disorder and perturbation added into the whole system. 

In the basis of lattice sites, the wave function of the zero-energy mode can be described by the following form
\begin{eqnarray}\label{e05}
	|\Psi_{E=0}\rangle &=&| \psi\rangle_{a_{1}}\otimes| \psi\rangle_{b_{1}}\otimes| \psi\rangle_{a_{2}}\otimes| \psi\rangle_{b_{2}} \otimes\cr\cr
	&&\cdots\otimes| \psi\rangle_{a_{N}}\otimes|\psi\rangle_{b_{N}}\otimes| \psi\rangle_{a_{N+1}}\cr\cr
	&=&|\psi_{a_{1}}, \psi_{b_{1}}, \psi_{a_{2}}, \psi_{b_{2}}, \cdots, \psi_{a_{N}}, \psi_{b_{N}}, \psi_{a_{N+1}}\rangle,
\end{eqnarray}
where $E=0$ represents the eigenvalue with zero energy, $|\psi_{a_{n}}\rangle=\rho _{a,n} e^{i\phi_{a,n} } $ ($|\psi_{b_{n}}\rangle=\rho _{b,n} e^{i\phi_{b,n} } $) shows the probability density and phase information of the eigenstate at $a_{n}$ ($b_{n}$) site. To illustrate the distribution of the zero-energy eigenstate, we now turn to perform an analytical solutions. Substituting Eq.~(\ref{e03})  and Eq.~(\ref{e05}) into the eigenvalue equation $H|\Psi_{E=0}\rangle= E|\Psi_{E=0}\rangle$, we obtain a series of coupled equations about $\psi_{a_{n}}$ and $\psi_{b_{n}}$ at $a$-type and $b$-type sites, respectively. For example, at $b$-type sites, 
\begin{eqnarray}\label{e06}
	J_{2} \psi_{b_{n}} + J_{1} \psi_{b_{n+1}}=0~~(n=1,\cdots,N-1),
\end{eqnarray}
with the boundary condition $J_{1}\psi_{b_{1}} + J_{2} \psi_{b_{2}} +J_{2} \psi_{b_{3}} + J_{2} \psi_{b_{4}} + \cdots +J_{2} \psi_{b_{N}} =  0$ and $J_{2} \psi_{b_{N}}  =  0$. For the nonvanishing hopping amplitude $J_{1}\ne0$ and $J_{2}\ne0$, the boundary condition determines $\psi_{b_{N}}=0$, which further leads $b$-type probability density to satisfy $\rho _{b,n}=0~(n=1,2,\cdots,N-1)$. The results reveal that, the zero-energy eigenstate has no probability distribution at all $b$-type sites, with $|\Psi_{E=0}\rangle=|\psi_{a_{1}}, 0, \psi_{a_{2}}, 0, \cdots, \psi_{a_{N}}, 0, \psi_{a_{N+1}}\rangle$. To further obtain the form of the zero-energy eigenstate, we need to estimate the  distribution of zero-energy eigenstate at $a$-type sites. For $a$-type sites, they also satisfy a set of coupled equations
\begin{eqnarray}\label{e07}
	J_{2} \psi_{a_{1}} + & J_{1} \psi_{a_{n}} + J_{2} \psi_{a_{n+1}}  =  0~~(n=2,\cdots,N),
\end{eqnarray}
with the boundary condition $J_{1}\psi_{a_{1}}  + J_{2} \psi_{a_{2}}  =  0$. Similarly, for the nonvanishing hopping amplitude $J_{1}\ne0$ and $J_{2}\ne0$, the boundary condition $J_{1}\psi_{a_{1}}  + J_{2} \psi_{a_{2}}  =  0$ ensures that
\begin{eqnarray}\label{e08}
	\psi_{a_{2}}&=&\lambda\psi_{a_{1}},\cr\cr
	\psi_{a_{n}}&=&-\psi_{a_{1}}+\lambda^{n-1}\psi_{a_{1}}+\left(\frac{\lambda^{n-2}-\lambda}{1-\lambda} \right)\psi_{a_{1}}\cr\cr
	&&(n=3,\cdots,N+1).
\end{eqnarray}
Here, $\lambda=-J_{1}/J_{2}$ represents the localization index. Obviously, the eigenstate of zero-energy mode at site $a_{n}$~($N=3,\cdots,N+1$) mainly exhibits an exponential behavior. Specifically, for $J_{1}\gg J_{2}$~($|\lambda|\sim \infty$), the eigenstate at site $a_{n}$~($N=3,\cdots,N+1$) satisfies $\psi_{a_{n}}\approx(\lambda^{n-1}-\lambda^{n-3})\psi_{a_{1}}\sim\lambda^{n-1}\psi_{a_{1}}$. Thus, when $J_{1}\gg J_{2}$, the zero-energy eigenstate can be approximately expressed as
\begin{eqnarray}\label{e09}
	|\Psi_{E=0}^{(1)}\rangle=|\psi_{a_{1}}, 0, \lambda\psi_{a_{1}}, 0, \cdots, \lambda^{N-1}\psi_{a_{1}}, 0, \lambda^{N}\psi_{a_{1}}\rangle.\cr&
\end{eqnarray}
The special distribution indicates that, when $J_{1}\gg J_{2}$, the zero-energy eigenstate (after normalization) is mainly localized at the rightmost site $a_{N+1}$ in an exponential way, which just corresponds the distribution characteristics of the topological right edge state. However, when $J_{1}\ll J_{2}$ ($|\lambda|\sim 0$), $\psi_{a_{2}}$ and $\psi_{a_{n}}$ ($N=3,\cdots,N+1$) satisfy $\psi_{a_{2}}\sim 0$ and $\psi_{a_{n}}\sim-\psi_{a_{1}}$ due to the existence of exponential decay factor.
Thus, when $J_{1}\ll J_{2}$, the zero-energy eigenstate can be expressed as
\begin{eqnarray}\label{e10}
	|\Psi_{E=0}^{(2)}\rangle=|\psi_{a_{1}}, 0, 0, 0, -\psi_{a_{1}}, 0, \cdots, -\psi_{a_{1}}, 0, -\psi_{a_{1}}\rangle.\cr&
\end{eqnarray}
The result clearly reveals that, when $J_{1}\ll J_{2}$, the zero-energy eigenstate (after normalization) is mainly localized at the sites $a_{1}$, $a_{3}$, $\cdots$, $a_{N}$, and $a_{N+1}$ with the equal probability $1/N$. Further, note that the zero-energy eigenstate also has a $\pi$ phase difference between the site $a_{1}$ and other sites $a_{n}$ ($n=3,\dots,N+1$).

According to the above analytical results, we can easily infer that, when $J_{1}>J_{2}=T_{n}$~($\theta\in[0,\pi/2]\cup[3\pi/2,2\pi]$), the zero-energy eigenstate is mainly localized at the right edge site, while when $J_{1}<J_{2}=T_{n}$~($\theta\in[\pi/2,3\pi/2]$), the zero-energy eigenstate is mainly localized at the sites $a_{1}$, $a_{3}$, $\cdots$, $a_{N}$, and $a_{N+1}$ uniformly. 

\section{\label{sec.3}The phase-robust topological router assisted by the zero-energy  mode}
\begin{figure}
	\centering
	\subfigure{\includegraphics[width=0.48\linewidth]{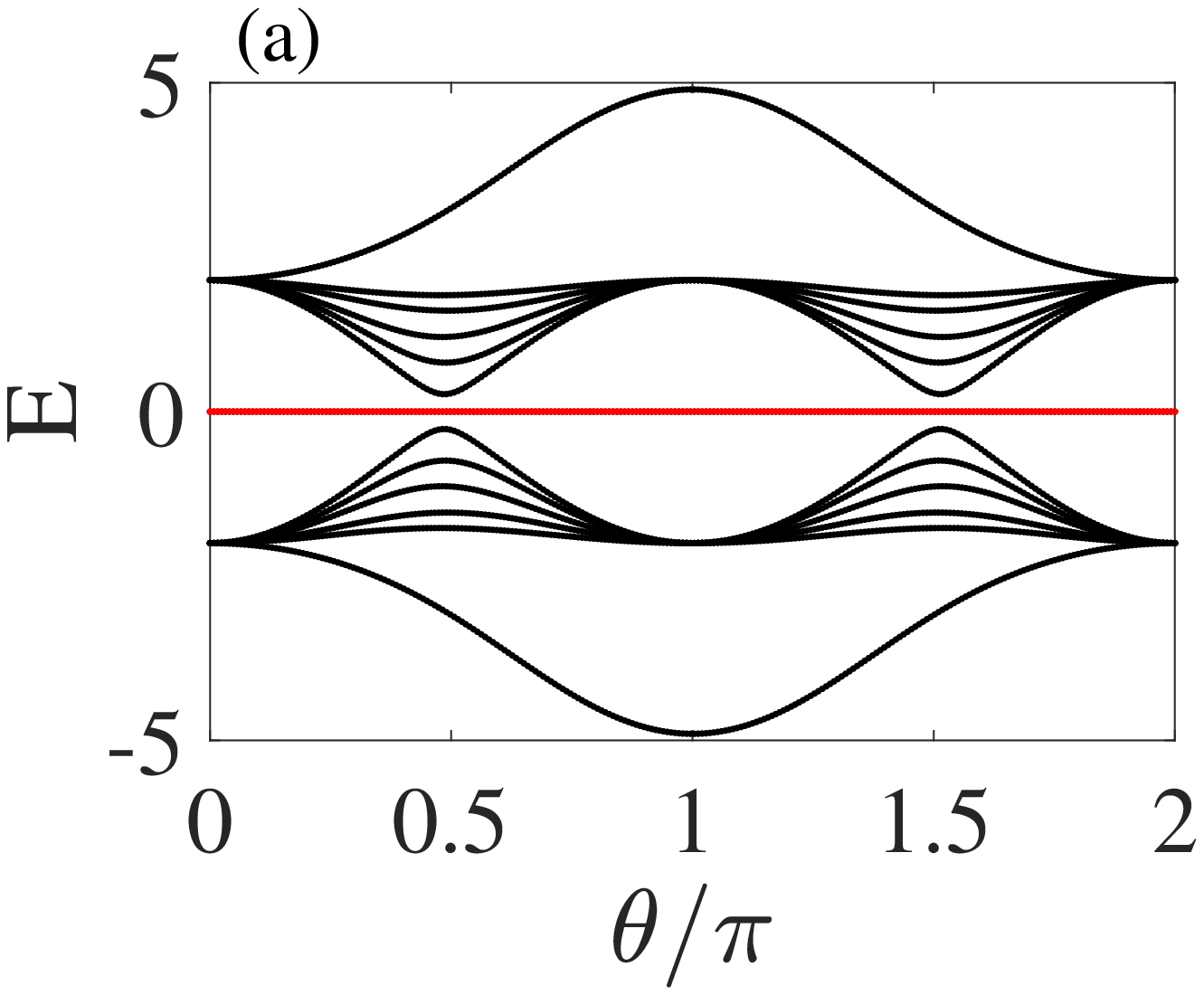}}
	\subfigure{\includegraphics[width=0.48\linewidth]{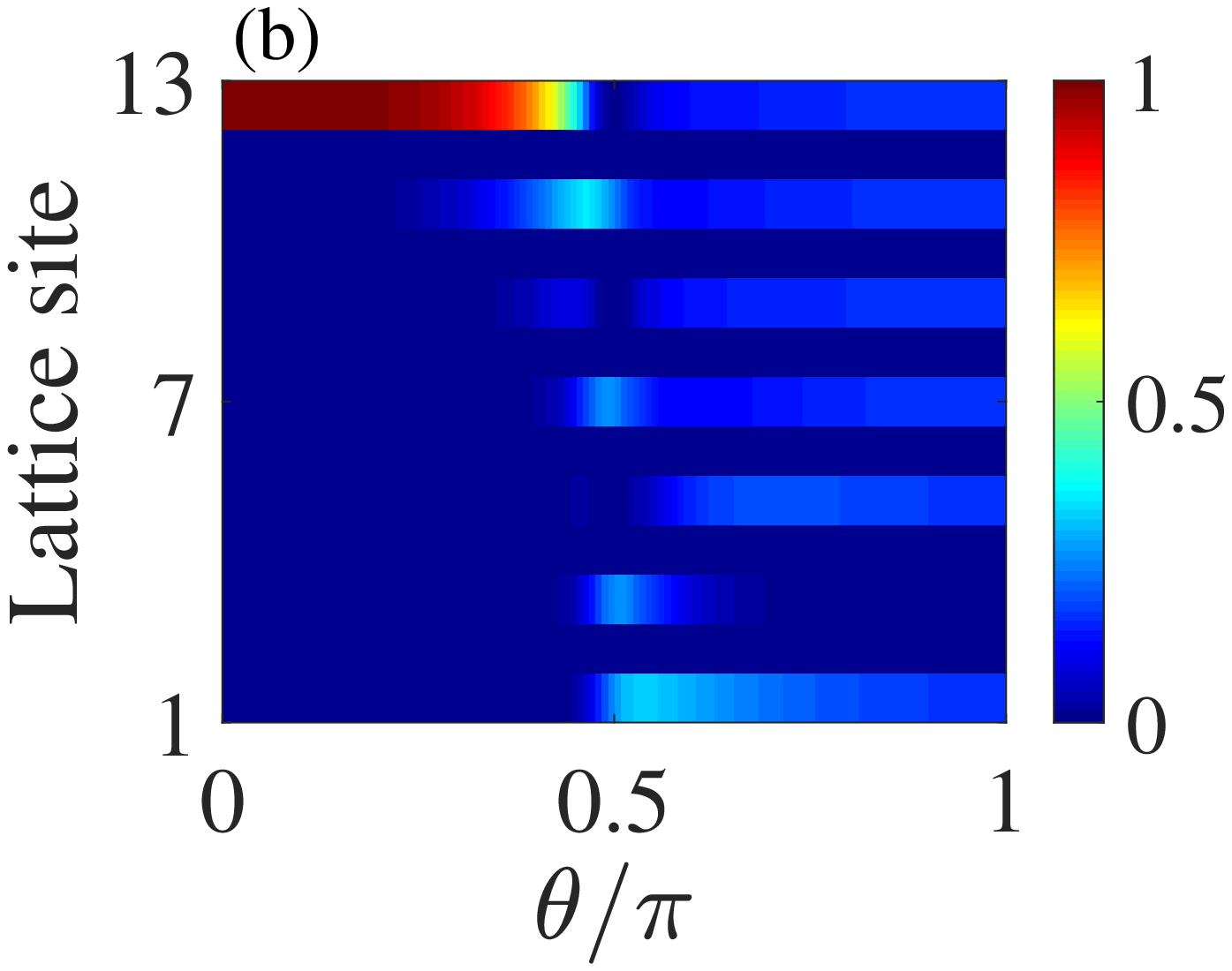}}
	
	\subfigure{\includegraphics[width=0.48\linewidth]{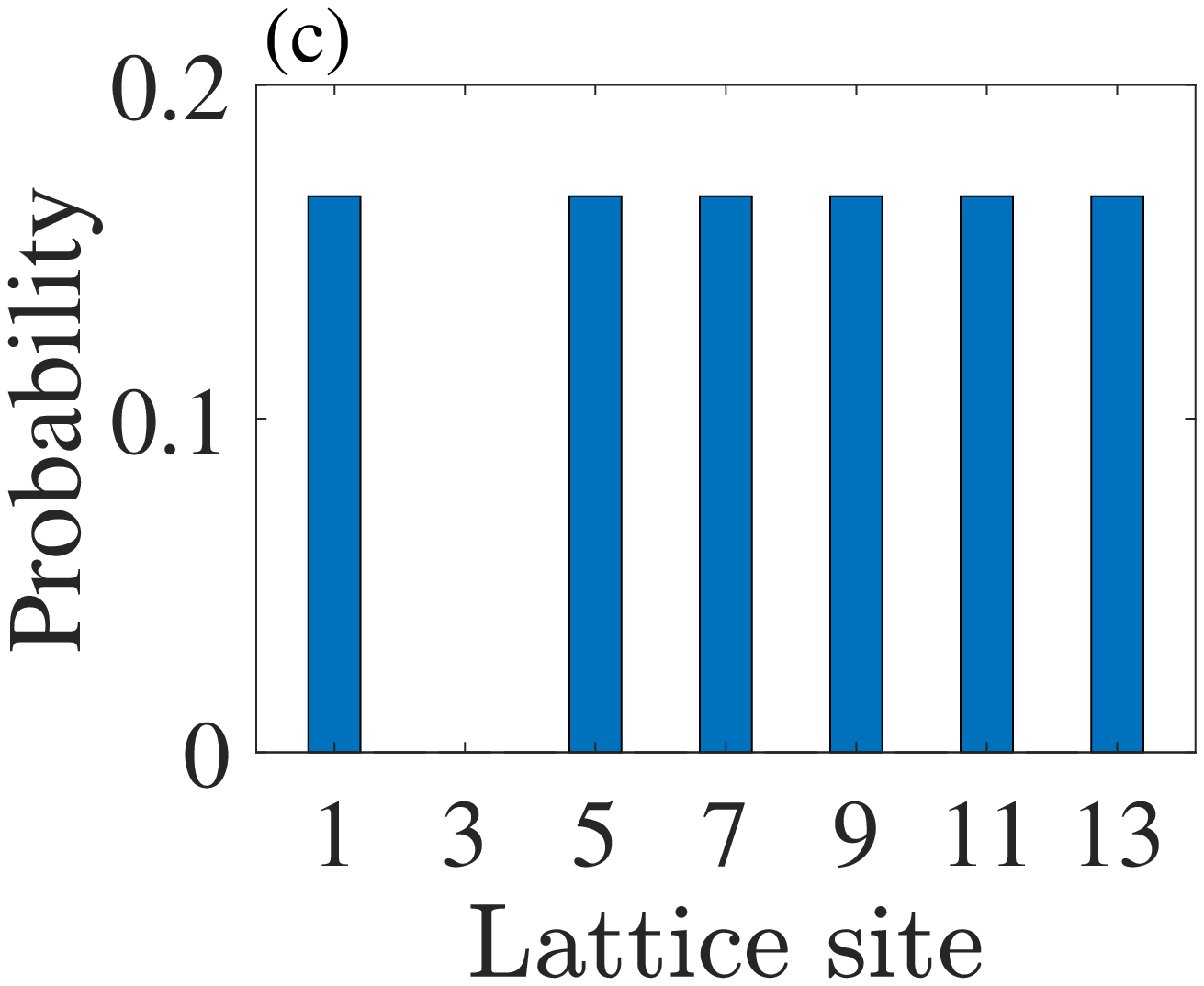}}
	\subfigure{\includegraphics[width=0.48\linewidth]{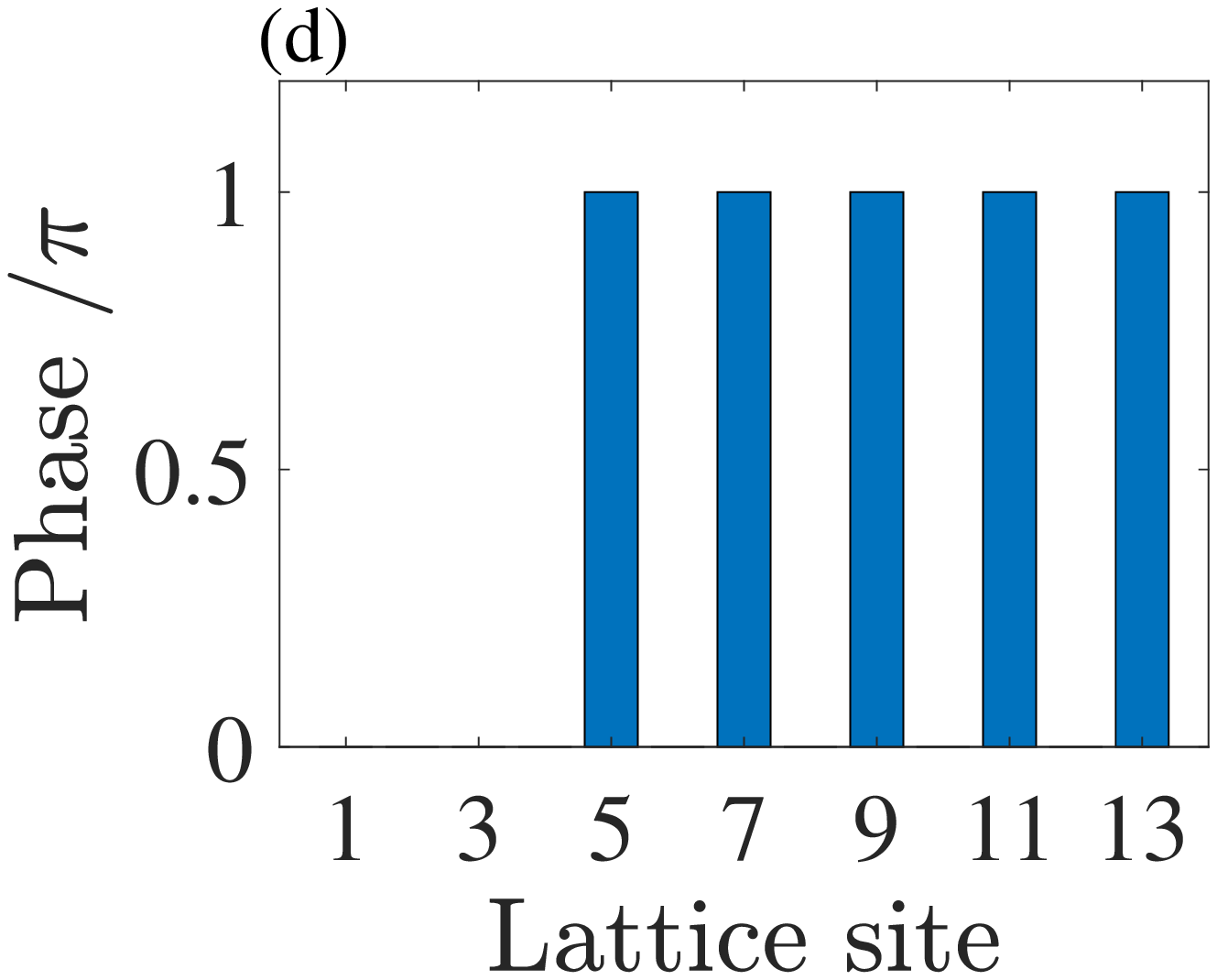}}
	\caption{(a) Energy spectrum of the 1D extended SSH lattice versus the parameter $\theta$. The energy spectrum has a zero-energy mode remaining unchanged when $\theta\in [0, 2\pi]$. (b) Distribution of the zero-energy mode versus the parameter $\theta$. ($\theta\in [0, \pi]$) (c) Probability distribution of the zero-energy mode when $\theta=\pi$. (d) Phase distribution of the zero-energy mode when $\theta=\pi$. The size of the lattice is $L=2N+1=13$. The unit is $J=1$.}\label{fig2}
\end{figure}

\subsection{\label{sec.3A} A phase-robust topological router with $N$ output ports}
In Sec.~\ref{sec.2B}, we have analytically worked out the wave function of its topological zero-energy mode. Here, we will further verify the above inference and engineer the phase-robust topological router assisted by the zero-energy mode. The energy spectrum of the system and the distribution of the zero-energy mode are shown in Fig.~\ref{fig2}. For simplicity, we choose $L=2N+1=13$ as example in all of the paper. As shown in Fig.~\ref{fig2}(a), the extended SSH model possesses symmetric modes and exhibits a topological zero-energy mode. The smallest space between the zero-energy mode and the up band appears at the point $\theta=\pi/2$ and $\theta=3\pi/2$, where the topological zero-energy mode may touch the bulk energy bands. Due to the symmetry structure of energy spectrum on both sides of $\theta=\pi$, one can focus on the distribution of the zero-energy gap state within $\theta\in[0,\pi]$. In Fig.~\ref{fig2}(b), we can conclude that the gap state is localized at the last site $a_{N+1}$ within $\theta\in[0,\pi / 2]$ while it is distributed at odd sites $a_{1}$, $a_{3}$, $...$, $a_{N}$, and $a_{N+1}$ within $\theta\in[\pi / 2, \pi]$. To make the distribution of gap state more accurately, we plot probability distribution and phase distribution with $\theta=\pi$ in Figs.~\ref{fig2}(c) and ~\ref{fig2}(d). As shown in Fig.~\ref{fig2}(c), the zero-energy gap state has equal probability $1/N$ at sites $a_{1}$, $a_{3}$, $...$, $a_{N}$, and $a_{N+1}$. Besides, the zero-energy gap state at sites $a_{3}$, $a_{4}$, $...$, $a_{N}$, and $a_{N+1}$ possess the same phase with $\pi$ while at site $a_{1}$ possesses $0$, as shown in Fig.~\ref{fig2}(d). Obviously, these results are consistent with the conclusion obtained from Eq.~(\ref{e09}) and Eq.~(\ref{e10}), meaning that the analytical results of the topological zero-energy mode are exact.

\begin{figure}
	   \centering
        \subfigure{\includegraphics[width=0.99\linewidth]{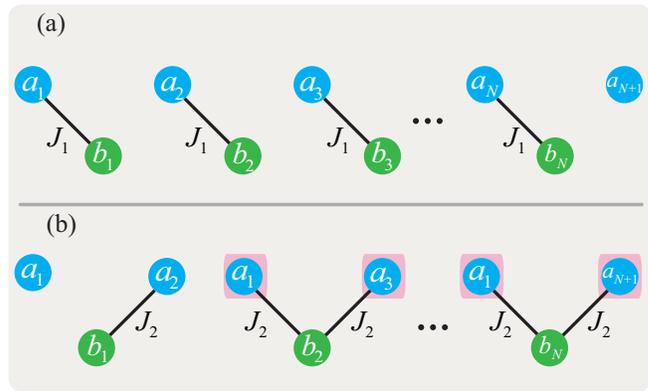}}
        \caption{The physical mechanism of the distribution for the gap state when $\theta =0$ and $\theta = \pi$, where the gap state occurs at the end of the extended SSH lattice and domain walls. The top represents the NN and long-range hopping amplitudes satisfy $J_{1}>J_{2}=T_{n}$ when $\theta =0$ and the gap state can be localized on a single site $a_{N+1}$. The bottom represents the NN and long-range hopping amplitudes satisfy $J_{1}<J_{2}=T_{n}$ when $\theta = \pi$ and the gap state can be localized on a superposition of sites $a_{1}$ and  $a_{3} ~(a_{4},a_{5},...,a_{N+1})$.  (The domain walls host zero-energy eigenstates as shown in pink shading.)}\label{fig3}
\end{figure}

Note that, the localized distribution of zero-energy mode at multiple sites is totally different from the traditional distribution in the previous SSH model. The peculiar distribution originates from the introduction of the long-range hopping. Specifically, when $\theta \in [0,\pi / 2]$, the NN and long-range hopping amplitudes satisfy $J_{1}>J_{2}=T_{n}$, leading that the two sites ($a_{n}$ and $b_{n}$) in one unit cell to be paired and the last unpaired site $a_{N+1}$ to be isolated. Then, the zero-energy mode is localized at the site $a_{N+1}$, which just corresponds to the right edge state, as shown in Fig.~\ref{fig3}(a). On the contrary, when $\theta \in [\pi / 2, \pi]$, the NN and long-range hopping amplitudes meet $J_{1}<J_{2}=T_{n}$. The weak hopping amplitude $J_{1}$ and the strong hopping amplitudes $J_{2}=T_{n}$ lead that the lattice is divided into several trimers composed by three sites $a_{1}$, $b_{n}$, and $a_{n+1}$ ($n=2,\cdots,N$) as shown in Fig.~\ref{fig3}(b). In addition, there are two sites $b_{1}$ and $a_{2}$ form a dimer with the strong hopping amplitude $J_{2}$. For the trimer, the strong hopping amplitudes $J_{2}=T_{n}$ indicate that the three sites, such as $a_{1}$, $b_{n}$, and $a_{n+1}$ ($n=2,\cdots,N$) generate a domain wall~\cite{JK2016Short}, which further induces the sites $a_{1}$ and $a_{n+1}$ ($n=2,\cdots,N$) to be isolated. As a result, the zero-energy mode is mainly localized at sites $a_{1}$ and $a_{n+1}$ ($n=2,\cdots,N$) when $\theta \in [\pi / 2, \pi]$, namely, the sites $a_{1}$ and $a_{n+1}$ ($n=2,\cdots,N$) become the special edge state. The peculiar distribution of the gap state indicates that, the particle or quantum state initially prepared at the right edge may be transferred to sites $a_{1}$ and $a_{n+1}$ ($n=2,\cdots,N$) with the equal probability $1/N$  when the periodic parameter $\theta$ varies from $0$ to $\pi$. 
\begin{figure*}
	    \centering
	    \subfigure{\includegraphics[width=0.3\linewidth]{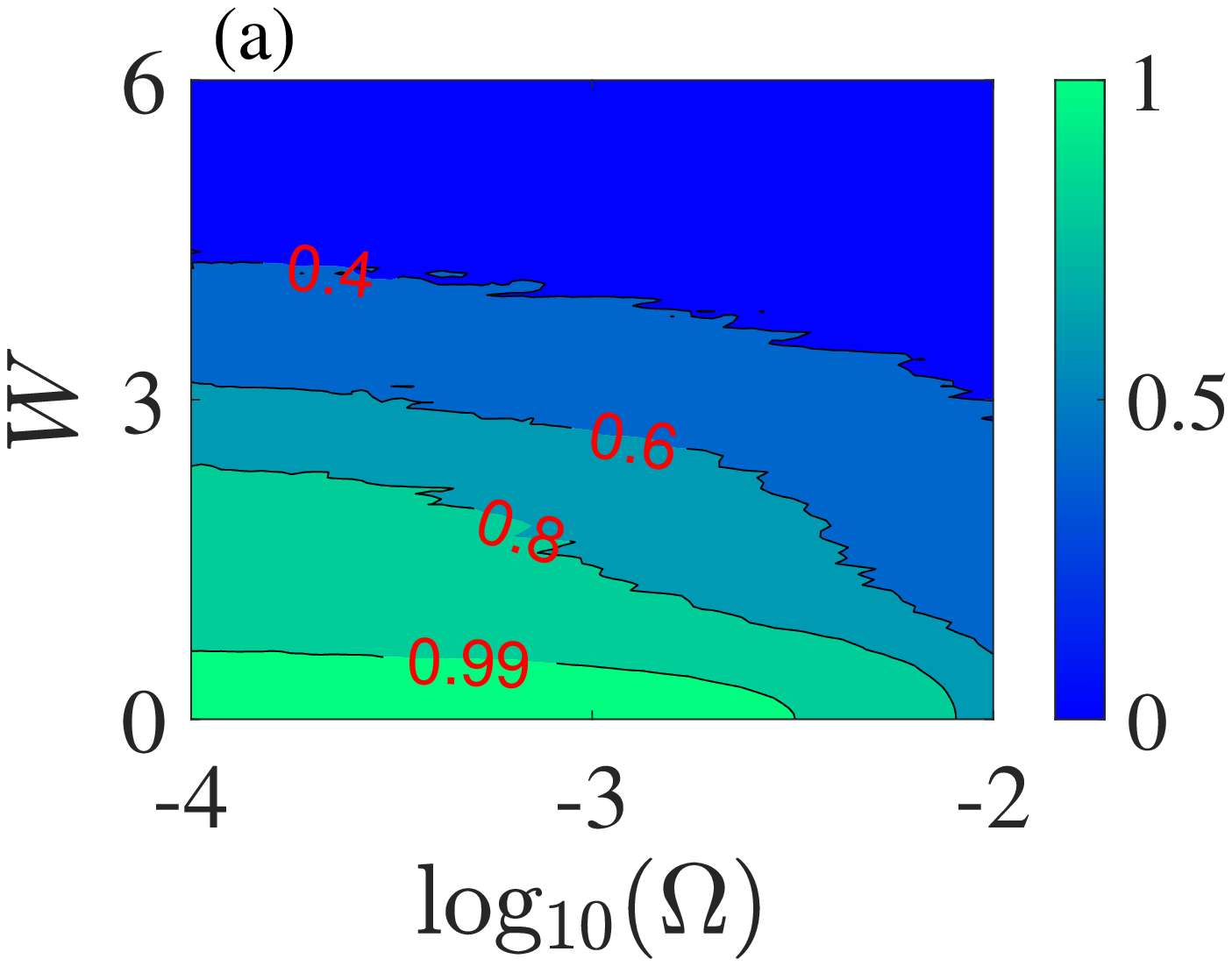}}
	    \subfigure{\includegraphics[width=0.3\linewidth]{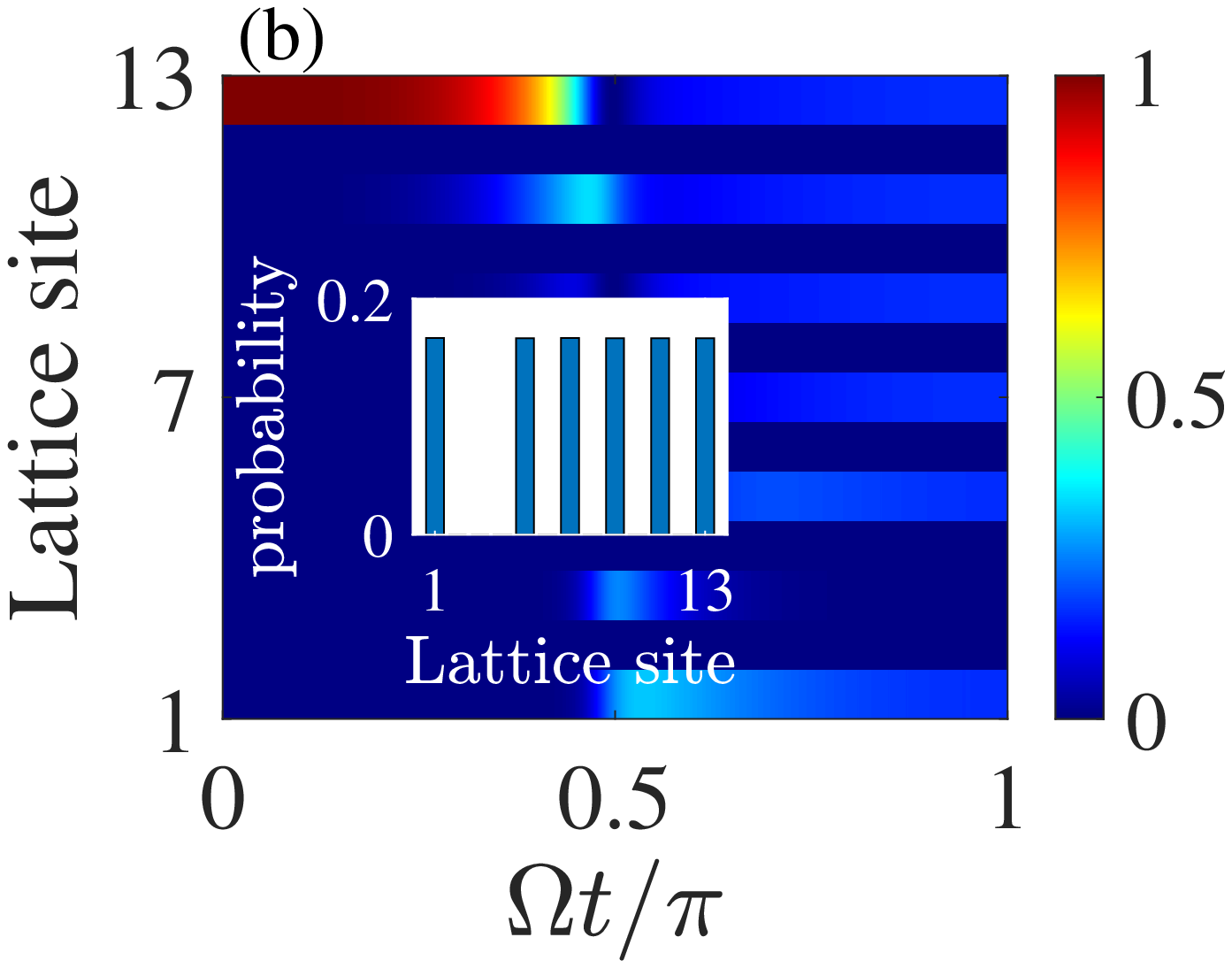}}
         \subfigure{\includegraphics[width=0.3\linewidth]{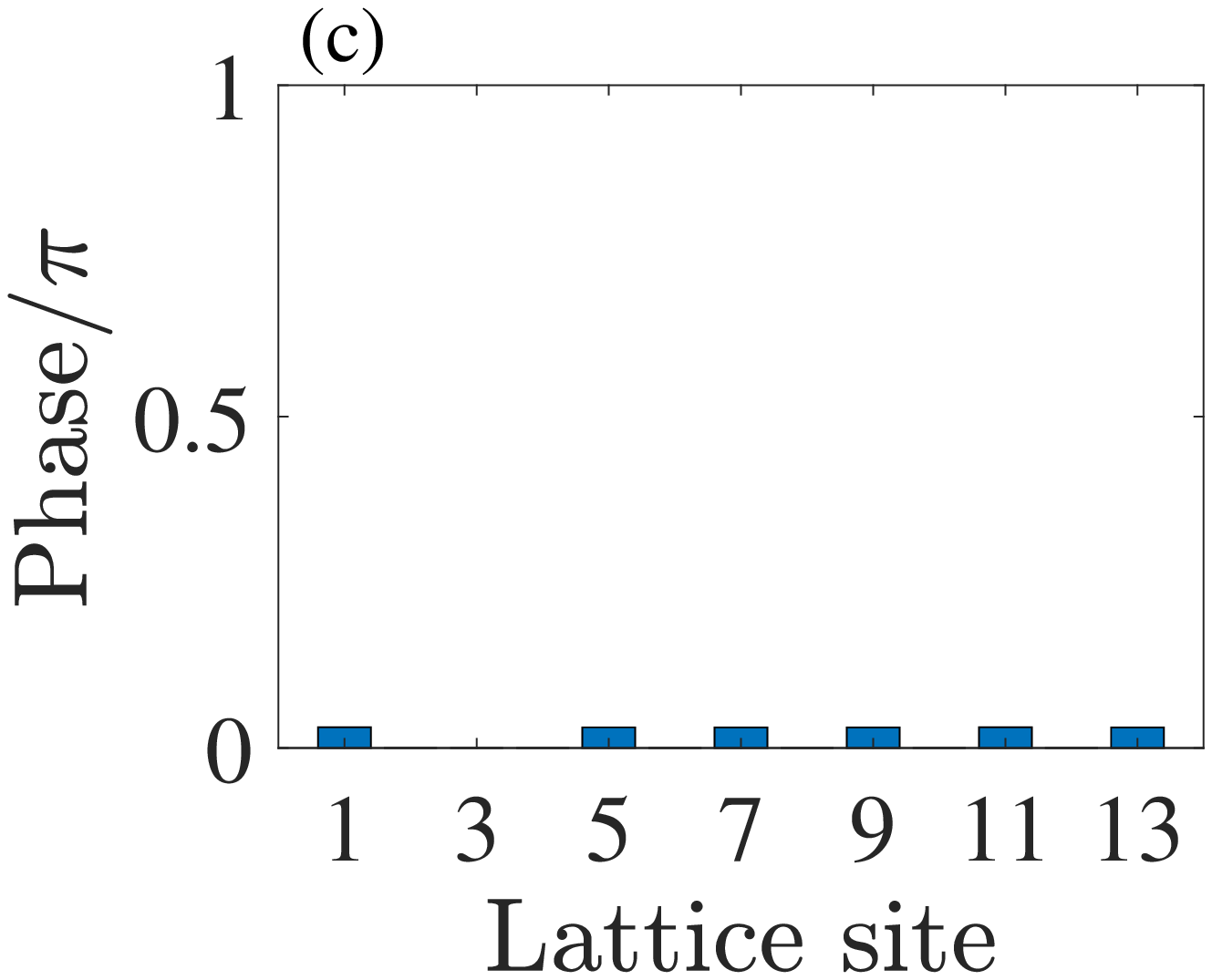}}

         \subfigure{\includegraphics[width=0.3\linewidth]{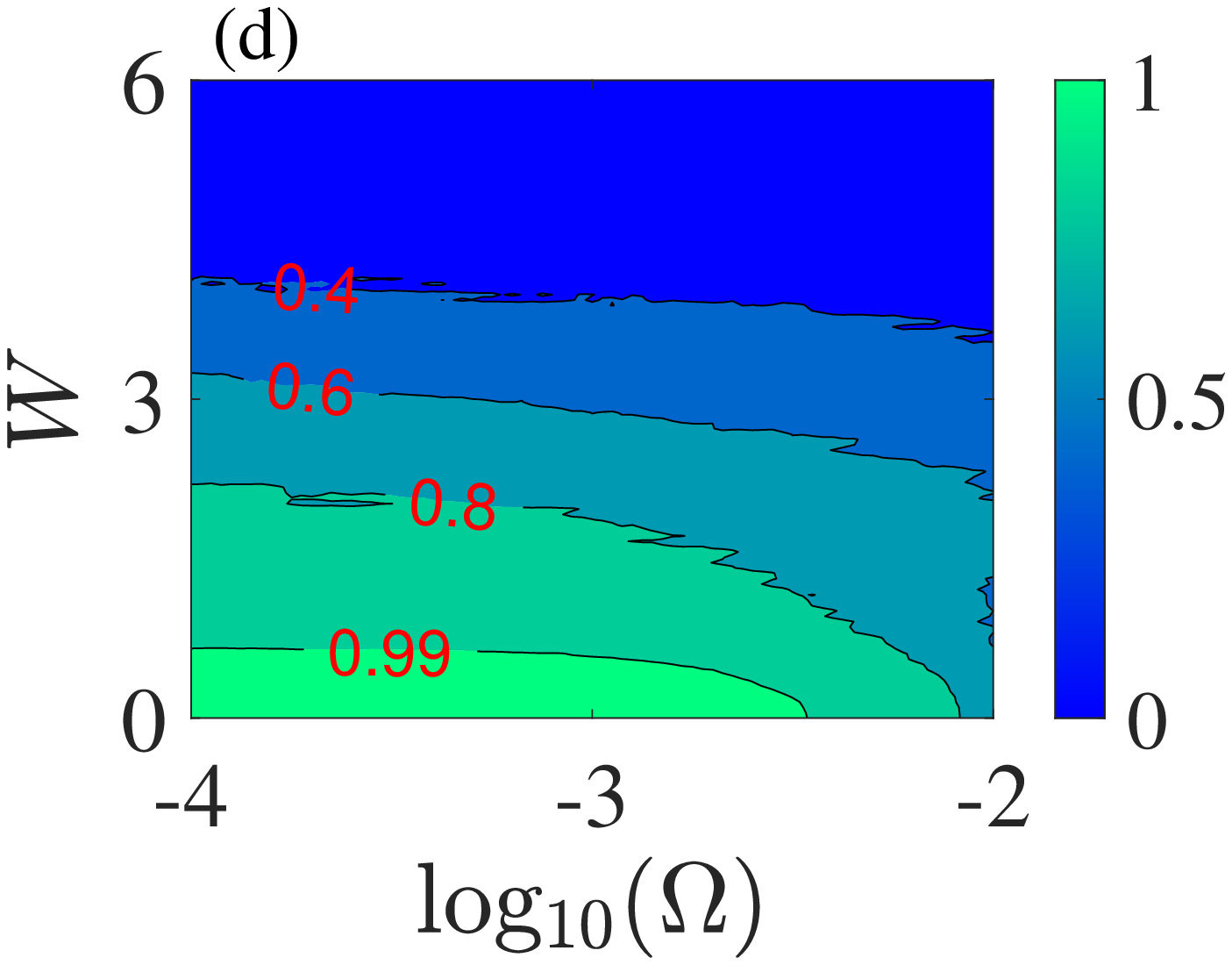}}
	    \subfigure{\includegraphics[width=0.3\linewidth]{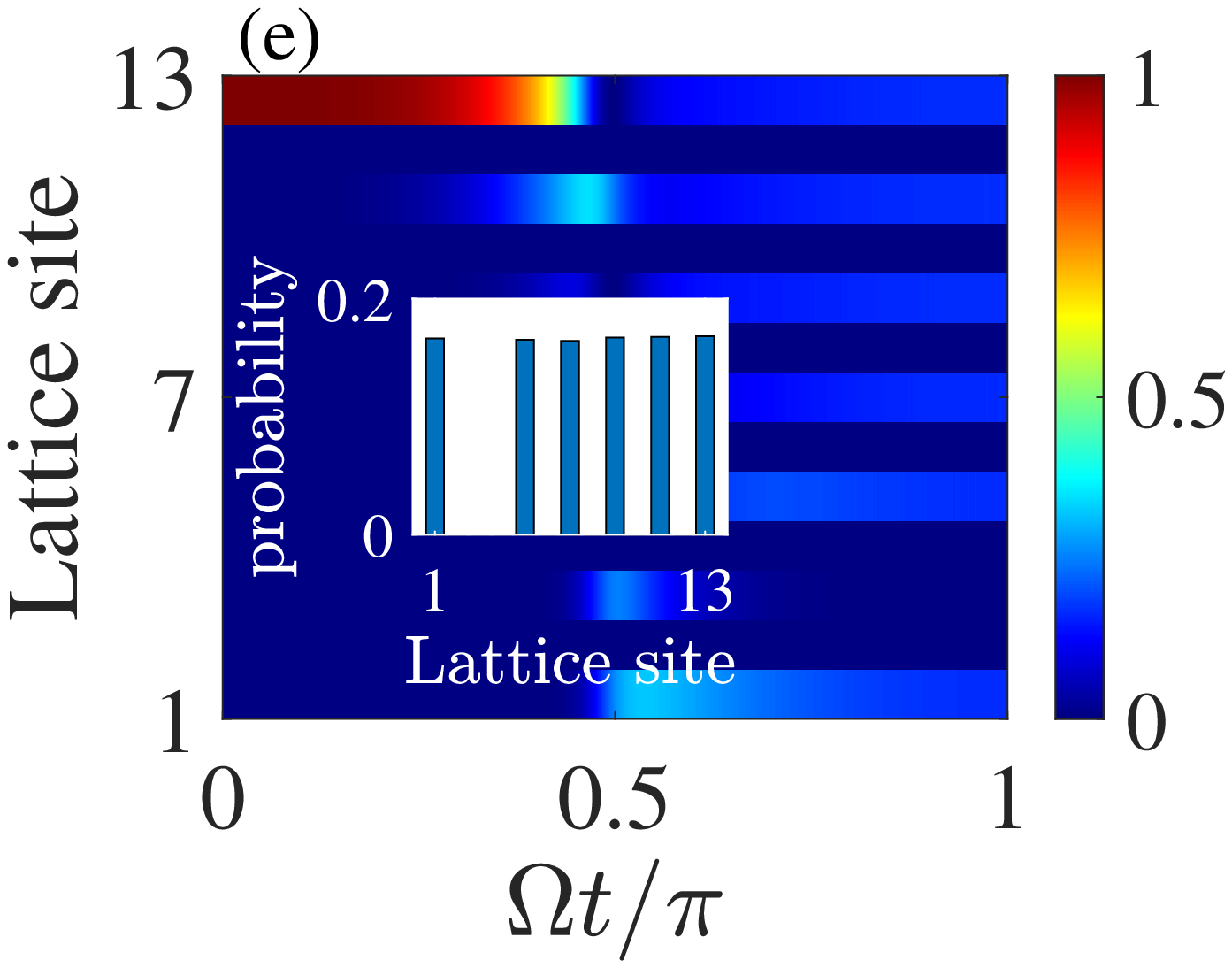}}
         \subfigure{\includegraphics[width=0.3\linewidth]{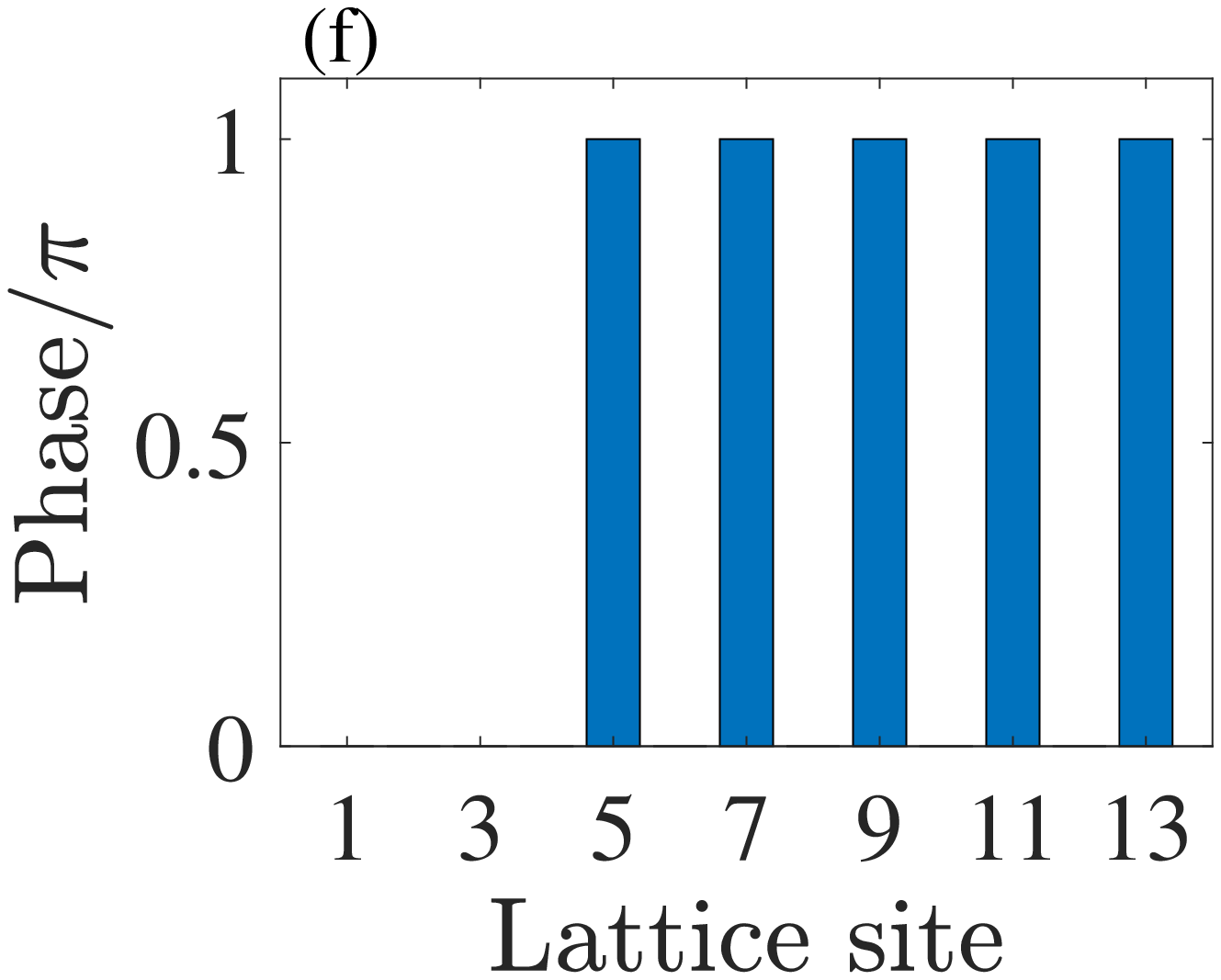}}

	    \subfigure{\includegraphics[width=0.3\linewidth]{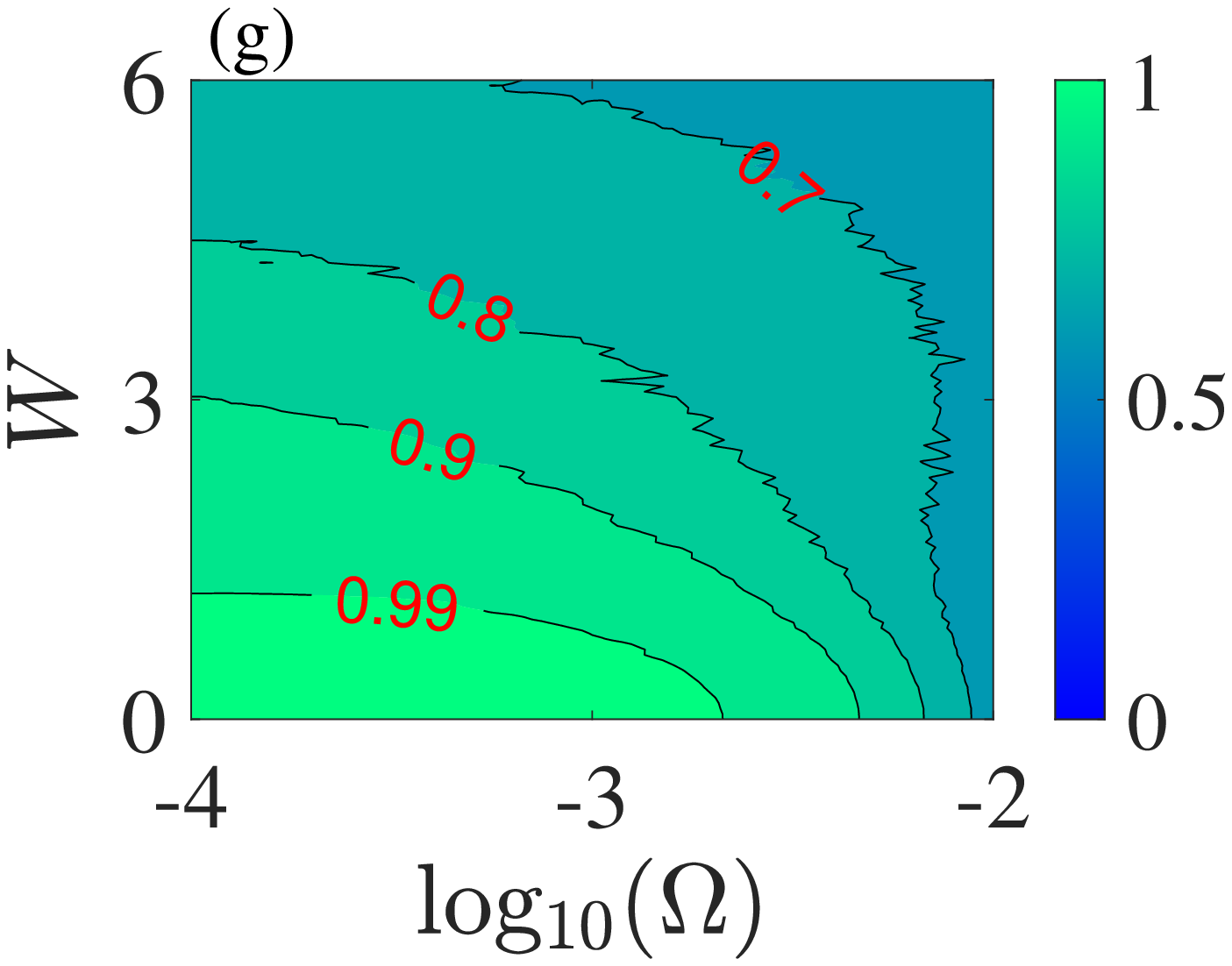}}
         \subfigure{\includegraphics[width=0.3\linewidth]{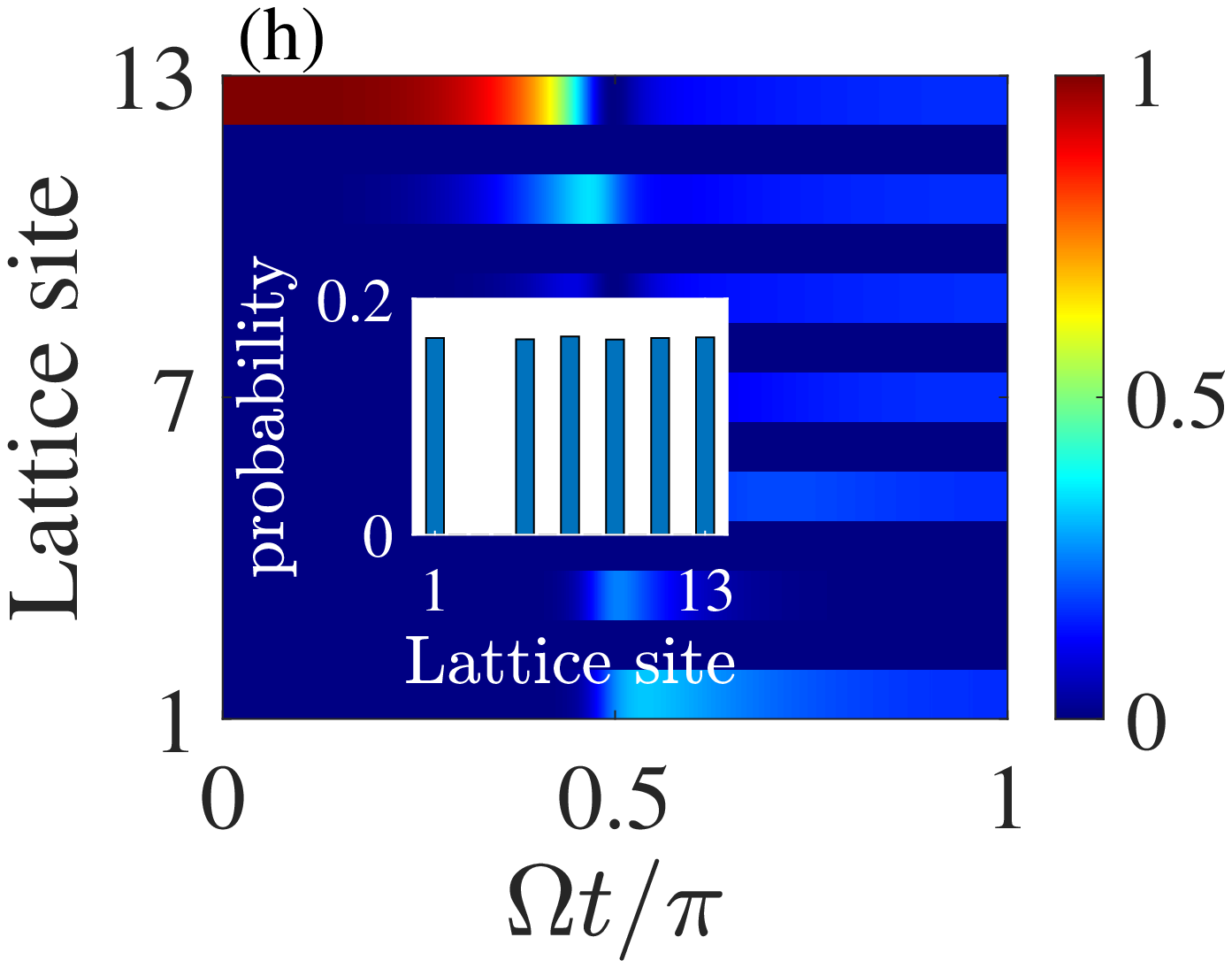}}
	    \subfigure{\includegraphics[width=0.3\linewidth]{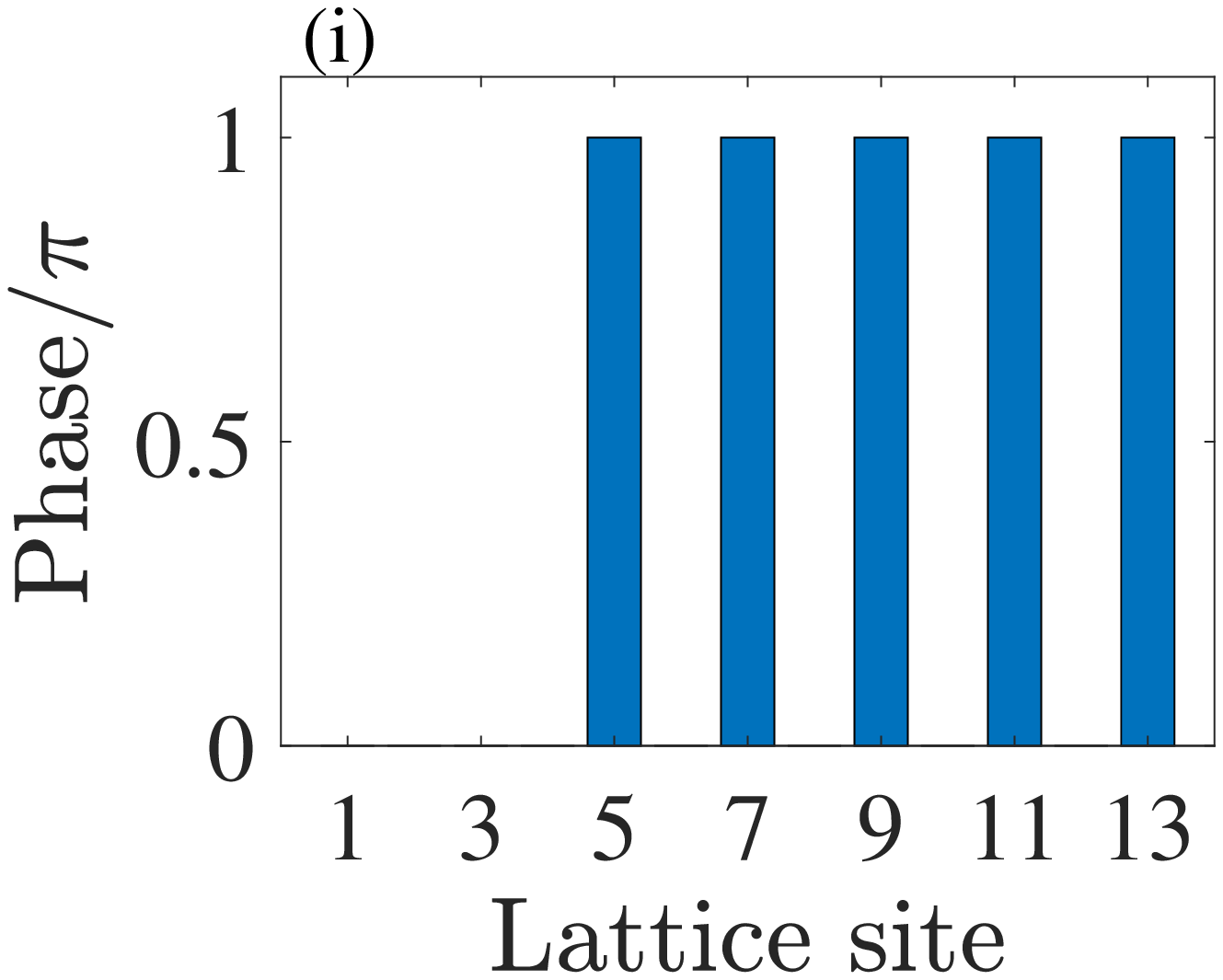}}
         \caption{\label{f4} The evolution of the initial state $|\Psi_{E=0}^{(1)}\rangle$ when the system is an imperfect lattice with random disorder. (a) The fidelity between the evolved final state $|\psi_{f}^{\prime}\rangle$ and the ideal final state $|\Psi_{E=0}^{(2)\prime}\rangle$ versus the ramping speed $\Omega$ and on-site disorder with strength $W$. (b) The evolution of the initial state $|\Psi_{E=0}^{(1)}\rangle$ with on-site disorder strength $W=0.2$, in which the ramping speed $\Omega$ satisfies $\Omega=0.0001$. The subgraph represents the probability distribution of evolved final state. (c) The phase distribution of the evolved final state in Fig.~\ref{fig4}(b). (d) The fidelity between the evolved final state $|\psi_{f}^{\prime}\rangle$ and the ideal final state $|\Psi_{E=0}^{(2)\prime}\rangle$ versus the ramping
speed $\Omega$ and NN disorder with strength $W$. (e) The evolution of the initial state $|\Psi_{E=0}^{(1)}\rangle$ with NN disorder strength $W=0.2$, in which the ramping speed $\Omega$ satisfies $\Omega=0.0001$. The subgraph represents the probability distribution of evolved final state.
(f) The phase distribution of the evolved final state in Fig.~\ref{fig4}(e). (g) The fidelity between the evolved final state $|\psi_{f}^{\prime}\rangle$ and the ideal final state $|\Psi_{E=0}^{(2)\prime}\rangle$ versus the ramping speed $\Omega$ and long-range disorder with strength $W$. (h) The evolution of the initial state $|\Psi_{E=0}^{(1)}\rangle$ with long-range disorder strength $W=0.2$, in which the ramping speed $\Omega$ satisfies $\Omega=0.0001$. The subgraph represents the probability distribution of evolved final state. (i) The phase distribution of the evolved final state in Fig.~\ref{fig4}(h). The size of the lattice is $L=2N+1=13$. The unit is $J=1$.}\label{fig4}
\end{figure*}

To implement the above process, we should rewrite the parameter $\theta$ as time-dependent version, i.e., $\theta=\Omega t$, in which $\Omega$ is the ramping speed of $\theta$ and $t$ is time. In this way, for the small enough $\Omega$, the initial state $|\Psi_{E=0}^{(1)}\rangle$ will evolve along the zero-energy gap state under the domination of $i\frac{\partial }{\partial t}|\Psi_{E=0}^{(1)}\rangle=H|\Psi_{E=0}^{(1)}\rangle$. Ideally, when the time reaches $t=\pi/\Omega$, the initial state $|\Psi_{E=0}^{(1)}\rangle$ can be evolved to the ideal final state $|\Psi_{E=0}^{(2)}\rangle$, indicating that the topological channel assisted by the gap state is effective to implement the special state transfer between $|\Psi_{E=0}^{(1)}\rangle$ and $|\Psi_{E=0}^{(2)}\rangle$. Note that, due to the topological protection of the energy gap and chiral symmetry,  the evolution process of gap state is naturally immune to the mild disorders and fluctuations in the system. To evaluate the robustness of the special state transfer, we study the probability distribution and phase distribution of the evolved final state when the disorder is added into the system. Here, the fidelity is defined as $F=|\langle \Psi_{E=0}^{(2)\prime}|\psi_{f}^{\prime}\rangle|$, in which $|\Psi_{E=0}^{(2)\prime}\rangle=||\Psi_{E=0}^{(2)}\rangle|$ represents the probability density for $|\Psi_{E=0}^{(2)}\rangle$ after ignoring the phase information and $|\psi_{f}^{\prime}\rangle=||\psi_{f}\rangle|$ is the probability density for evolved final state $|\psi_{f}\rangle$ after ignoring the phase information. In this way, we can explore the effects of the disorder on the probability distribution and phase distribution respectively.

Figure~\ref{fig4} shows the different effects when the system is an imperfect lattice possesses the different-type disorders. When the on-site disorder $H_{\mathrm{OD}}=\Sigma_{n}W(\delta_{a,n}a_{n}^{\dagger}a_{n}+\delta_{b,n}b_{n}^{\dagger}b_{n})$ ($W$ is the disorder strength while $\delta_{a,n}$ and $\delta_{b,n}$ are the random number in the range of $[-0.5,0.5]$) is added into the system, the fidelity $F$ versus the evolution speed $\Omega$ and on-site disorder strength $W$ is shown in Fig.~\ref{fig4}(a). Obviously, the probability distribution of evolved final state $|\psi_{f}\rangle$ infinitely approaches the ideal final state $|\Psi_{E=0}^{(2)}\rangle$ with $F>0.99$ when the evolution speed satisfies $\log_{10}(\Omega)< -2.5$ and on-site disorder strength satisfies $W<0.5$. The numerical results reveal that, the probability distribution of gap state is robust to the mild on-site disorder with the reasonable evolution speed $\Omega$. Further, we depict the detailed state transfer process of the initial state $|\Psi_{E=0}^{(1)}\rangle$ with $\Omega=0.0001$ and $W=0.2$. As shown in Fig.~\ref{fig4}(b), the evolved final state possesses the equal probability $1/N$ at sites $a_{1}$, $a_{3}$, $\cdots$, $a_{N}$, and $a_{N+1}$. It shows that the probability distribution of evolved final state at different sites is equivalent and the mild on-site disorder almost has no effect in probability distribution.  We also plot the phase distribution of the evolved final state in Fig.~\ref{fig4}(c). The phase information at sites $a_{3}$, $\cdots$, $a_{N}$, and $a_{N+1}$ possess 0 is different with the ideal final state $|\Psi_{E=0}^{(2)}\rangle$ when $t=\pi / \Omega$, which means the phase information is destroyed with the presence of the mild on-site disorder. The reason is that, the existence of on-site disorder (diagonal disorder) leads to the chiral symmetry break, namely, $\hat{\Gamma}[H+H_{\mathrm{OD}}]\hat{\Gamma}^{+}\ne-[H+H_{\mathrm{OD}}]$. In this way, the topological channel assisted by the gap state is frail to implement the special state transfer between $|\Psi_{E=0}^{(1)}\rangle$ and $|\Psi_{E=0}^{(2)}\rangle$.

In Fig.~\ref{fig4}(d), we consider the effects of evolution speed $\Omega$ and NN hopping disorder $H_{\mathrm{ND}}=\Sigma_{n}W \delta (a_{n}^{\dagger}b_{n}+a_{n+1}^{\dagger}b_{n})+$H.c. on fidelity $F$. A similar conclusion is obtained, the probability distribution of the gap state is robust to the mild NN hopping disorder with $W<0.5$ and the ramping speed slow enough with $\log_{10}(\Omega)< -2.5$. In Fig.~\ref{fig4}(e), we depict the state transfer process when $\Omega=0.0001$ and $W=0.2$. Obviously,  the state transfer between $|\Psi_{E=0}^{(1)\prime}\rangle$ and $|\Psi_{E=0}^{(2)\prime}\rangle$ can indeed be implemented, and the sites $a_{1}$, $a_{3}$, $\cdots$, $a_{N}$, and $a_{N+1}$ possess the equal probability $1/N$. In Fig.~\ref{fig4}(f), we investigate the phase information of the evolved final state $|\psi_{f}\rangle$ when $t=\pi / \Omega$. The phase distribution in site $a_{1}$ possesses 0 while in sites $a_{3}$, $\cdots$, $a_{N}$, and $a_{N+1}$ possess $\pi$. These results are completely consistent with the ideal phase distribution [see Fig.~\ref{fig2}(d)], indicating that the phase information of gap state is insensitive to the mild NN hopping disorder. Clearly, the existence of mild disorder in NN hopping has no appreciable effects on probability distribution and phase distribution of the gap state due to the system with chiral symmetry. Then, the special state transfer between $|\Psi_{E=0}^{(1)}\rangle$ and $|\Psi_{E=0}^{(2)}\rangle$ assisted by the gap state is resilient to mild NN hopping disorder. Further, in Figs.~\ref{fig4}(g)-\ref{fig4}(i), we investigate the effects of long-range hopping disorder $H_{\mathrm{LD}}=\Sigma_{n}W \delta (a_{1}^{\dagger}b_{n}+b_{n}^{\dagger}a_{1}) $ on the special state transfer, which reveals the similar conclusions with the NN disorder added into the system. The mild long-range hopping disorder in system keeps the chiral symmetry, implying that the probability and phase information for gap state are robust to the mild long-range hopping disorder. 

Combined with the above analysis, based on the topological protection originating from the energy gap and chiral symmetry, the state transfer between $|\Psi_{E=0}^{(1)}\rangle$ and $|\Psi_{E=0}^{(2)}\rangle$ assisted by gap state is naturally immune to the mild disorders and fluctuations of NN hopping and long-range hopping in the system. The robustness of the state transfer provides much more convenience for the experimental realization and the practical application of the scalability or the large-scale distribution of the quantum network. The special state transfer provides a certain distribution capabilities, which is analogous to a traditional router.  If we prepare the particle or quantum state at last site $a_{N+1}$ initially [regard as input port], with the varying of parameter $\theta$ from 0 to $\pi$, the initial particle or quantum state will be finally transferred into the sites $a_{1}$, $a_{3}$, $\cdots$, $a_{N}$, and $a_{N+1}$ with equal probability [regard as output ports], namely, phase-robust topological router. 

\subsection{\label{sec.3B} A phase-robust topological router with $N+1$ output ports}
In the above subsection, we have shown the phase-robust topological router with $N$ output ports. Observably, almost all $a$-type sites act as the output ports besides the site $a_{2}$. The reason can be traced back to the Sec.~\ref{sec.3A} that when $\theta \in [\pi / 2, \pi]$, the strong hopping amplitude $J_{2}$ impel the site $b_{1}$ and the site $a_{2}$ to be paired, which makes it impossible to decouple from the pair and play an isolated site. As a result, the topological zero-energy gap state does not occur at the site $a_{2}$. 

\begin{figure}
	     \centering
          \scalebox{0.4}{\includegraphics{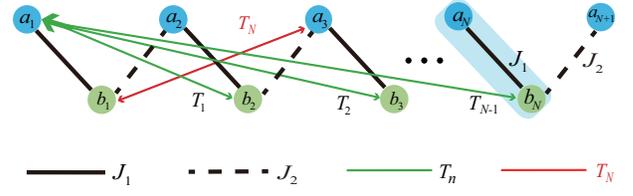}}\caption{\label{f5} Schematic of an extended SSH lattice to implement the phase-robust topological router with $N+1$ output ports. The solid line in red describes the new long-range hopping between site $a_{3}$ and $b_{1}$ with amplitude $T_{N}$. Other parameters are consistent with Fig.~\ref{fig1}(b)}\label{fig5}
\end{figure}

Next, to break the above obstacle, we put emphasis on the gap state localized at all $a$-type sites and implementing an optimized phase topological router with $N+1$ output ports. We still consider the extended SSH lattice owns the long-range hopping connecting the first $a$-type site $a_{1}$ and $b$-type sites in the $n$th ($n = 2, 3, \cdots, N$) unit cell. The difference of the optimized scheme lies in, we increase the long-range hopping $T_{N}$ between the site $a_{3}$ and the site $b_{1}$.  As shown in Fig.~\ref{fig5},  the extended SSH lattice chain can be redescribed as
\begin{eqnarray}\label{e11}
H&=&\sum_{n=1}^{N}\left(J_{1}a_{n}^{\dag}b_{n}+J_{2}a_{n+1}^{\dag}b_{n}+\mathrm{H.c.}\right)\cr\cr
&+&\sum_{n=1}^{N-1}\left(T_{n}a_{1}^{\dag}b_{n+1}+\mathrm{H.c.}\right)+T_{N}\left(a_{3}^{\dag}b_{1}+b_{1}^{\dag}a_{3}\right),
\end{eqnarray}
where the new long-range hopping amplitude $T_{N}$ satisfies $T_{N}=T_{n}=J_{2}$. The present system also possesses chiral symmetry and there is a zero-energy gap state in the energy gap. Theoretically, when $\theta \in [\pi / 2, \pi]$, the strong intercell and new long-range hoppings ensure that three sites $a_{2}$, $b_{1}$ and $a_{3}$ generate a new domain wall. Then, the site $a_{2}$ to be isolated and the gap state is localized at site $a_{2}$ with a certain probability. In this way, the present gap state with zero-energy has great potential to realize a phase topological router with $N+1$ output ports.

\begin{figure}
	 \centering
	 \subfigure{\includegraphics[width=0.48\linewidth]{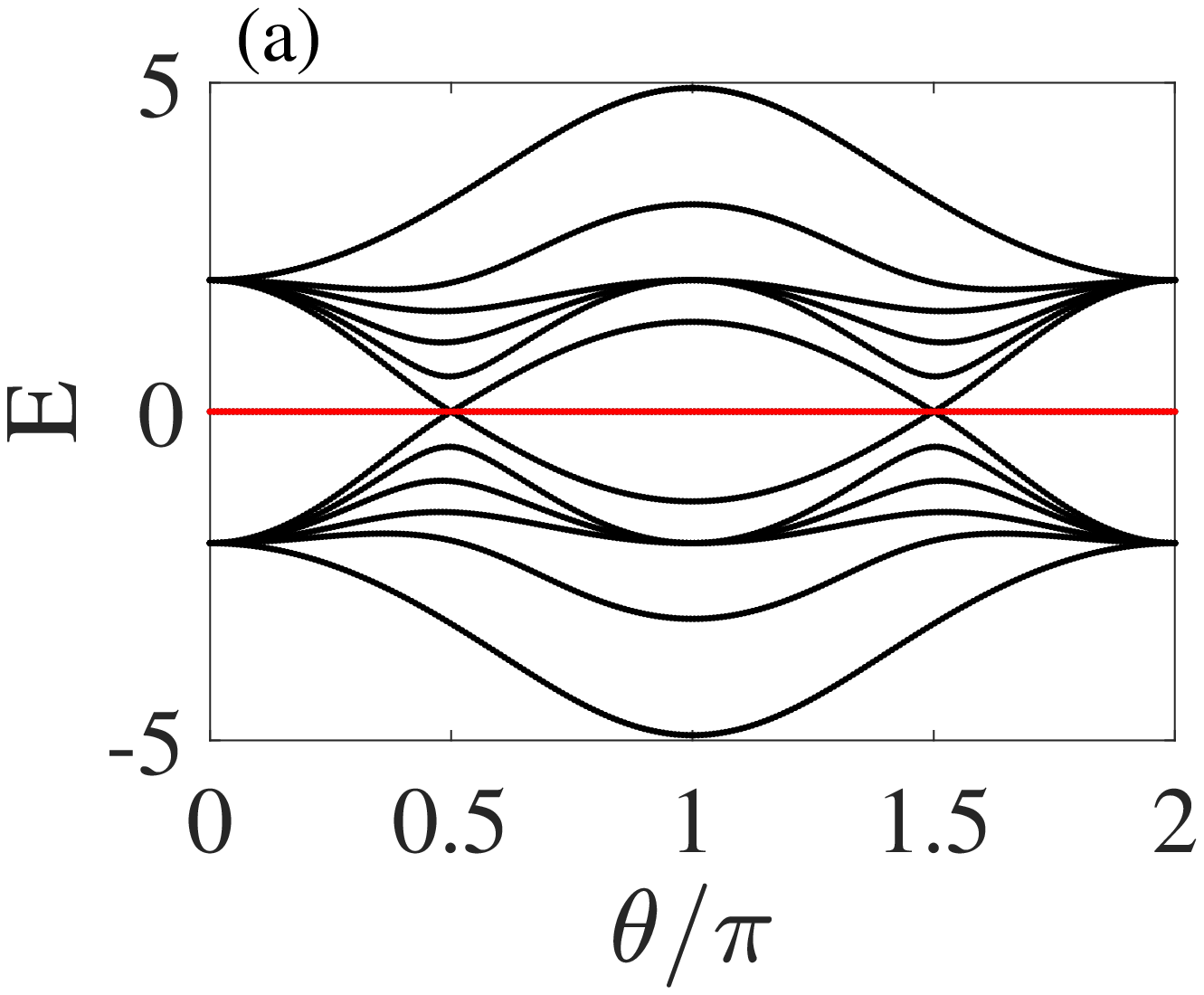}}
	 \subfigure{\includegraphics[width=0.48\linewidth]{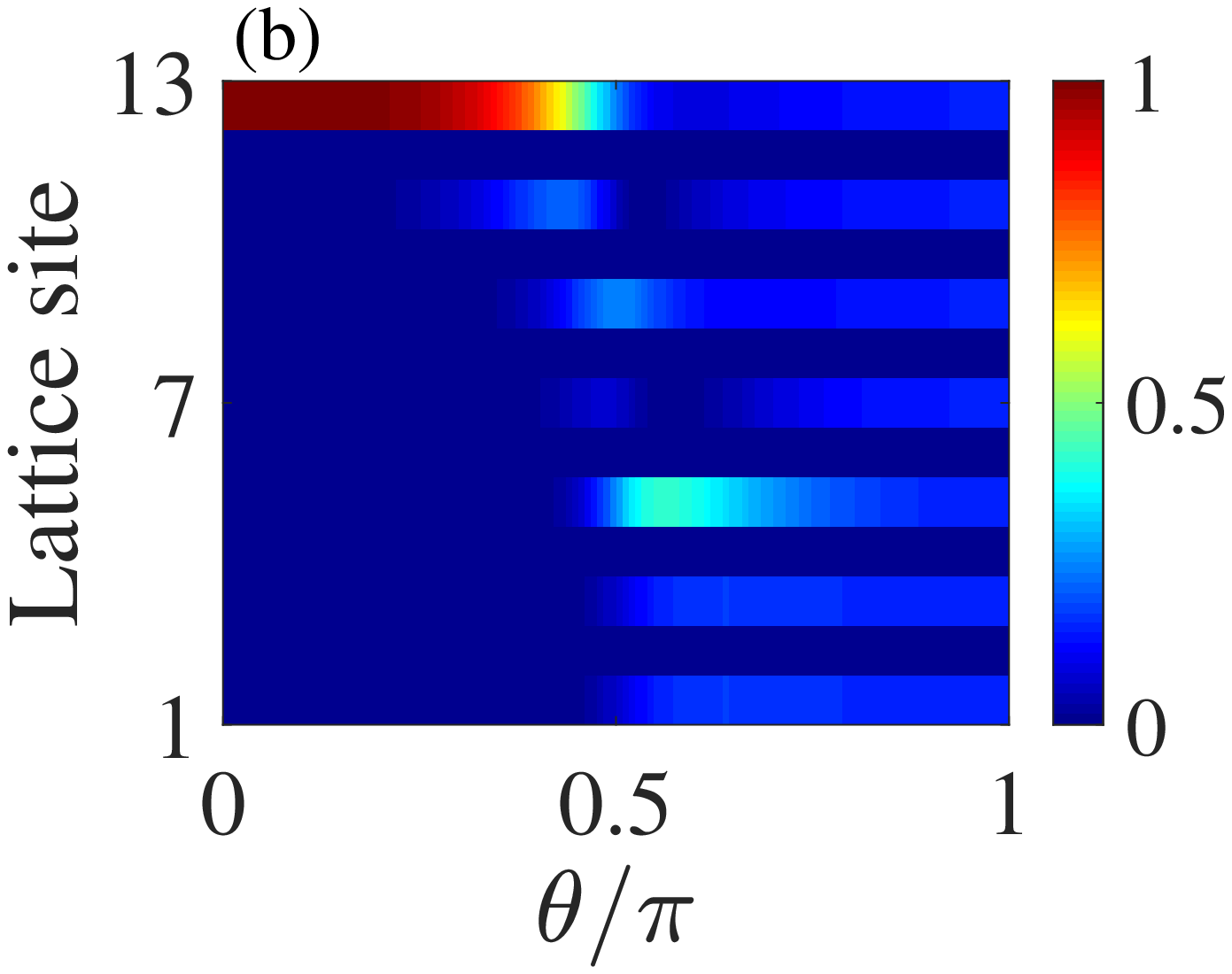}}

      \subfigure{\includegraphics[width=0.48\linewidth]{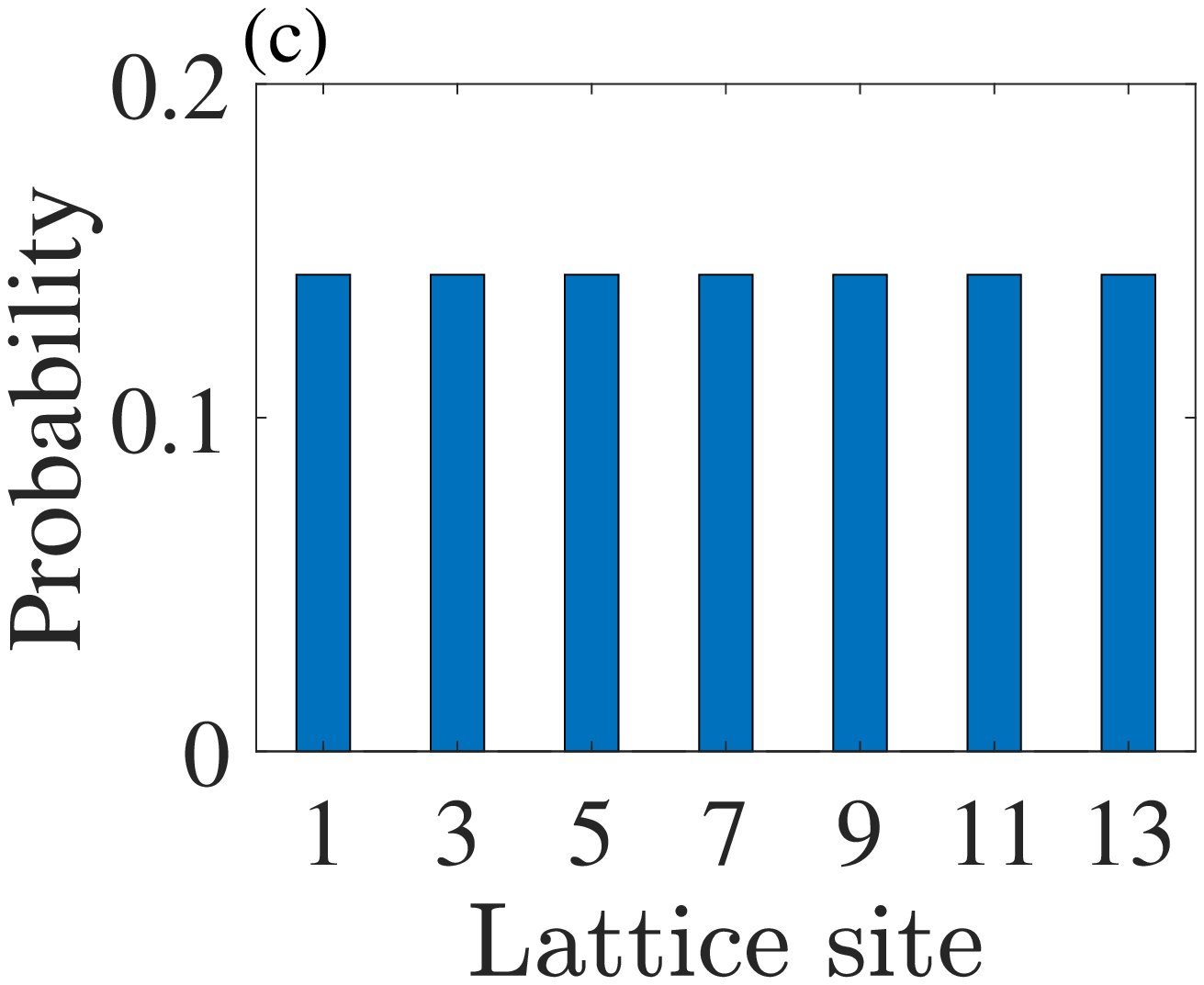}}
	  \subfigure{\includegraphics[width=0.48\linewidth]{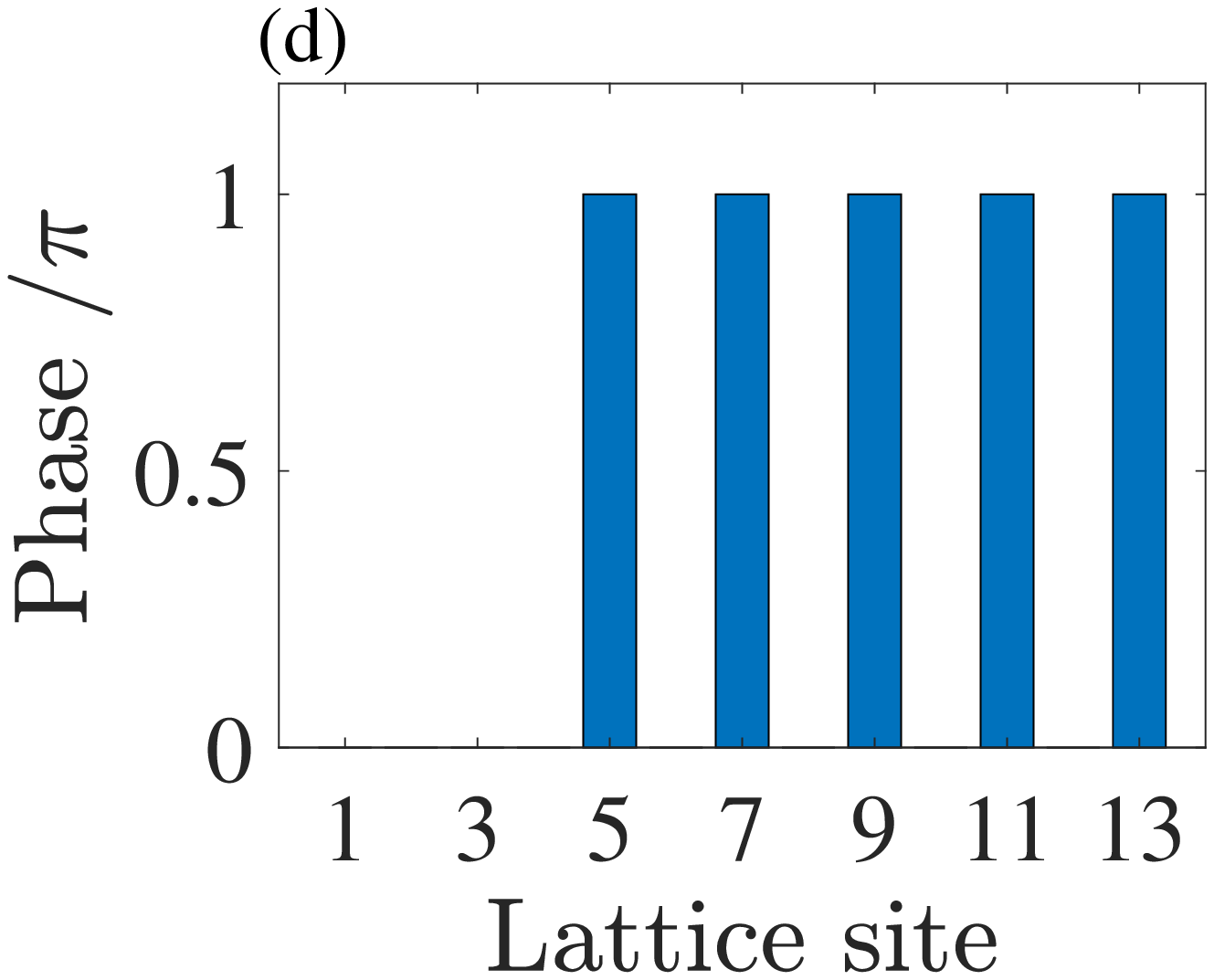}}
      \caption{\label{f6} Energy spectrum of the 1D extended SSH lattice versus the parameter $\theta$. The energy spectrum has a zero-energy mode remaining unchanged when $\theta\in [0, 2\pi]$. (b) Distribution of the zero-energy mode versus the parameter $\theta$. ($\theta\in [0, \pi]$) (c) Probability distribution of the zero-energy mode when $\theta=\pi$. (d) Phase distribution of the zero-energy mode when $\theta=\pi$. The size of the lattice is $L=2N+1=13$. The unit is $J=1$.}\label{fig6}
\end{figure}

To verify the inference mentioned above, we plot the energy spectrum of system and the gap state distribution in Fig.~\ref{fig6}(a) and Fig.~\ref{fig6}(b). For the energy spectrum, it still possesses a zero-energy mode with the appearance of the new long-range hopping $T_{N}$. As shown in Fig.~\ref{fig6}(b),  the gap state is occupied the last site $a_{N+1}$ when $\theta\in [0,\pi/2]$ and concentrated near the sites $a_{1}$, $a_{2}$, ..., $a_{N}$, $a_{N+1}$ when $\theta\in [\pi/2, \pi]$. It should be stressed here that the gap state in the site $a_{2}$ with a certain probability. To validate that the gap state is equally distributed at each $a$-type site, in Fig.~\ref{fig6}(c), we perform simulation for the probability distribution of the gap state. Clearly, the gap state is distributed at sites $a_{1}, a_{2}, ..., a_{N}$, and $a_{N+1}$ with the same probability of $1/(N+1)$.  Lastly, Fig.~\ref{fig6}(d) depicts the corresponding phase information acting in 0 (at sites $a_{1}$ and $a_{2}$) and $\pi$ (at sites $a_{3}$, $a_{4}$,...,$a_{N}$, and $a_{N+1}$). Generally, with the peculiar distribution of the gap state induced by the long-range hopping $T_{N}$ in the SSH lattice, if we regard the last site $a_{N+1}$ as the input port and regard the sites $a_{1}$, $a_{2}$, ..., $a_{N}$, and $a_{N+1}$ as multiple output ports, the state transfer process between between $|\Psi_{E=0}^{(1)}\rangle=|\psi_{a_{1}}, 0, \lambda\psi_{a_{1}}, 0, \cdots, \lambda^{N-1}\psi_{a_{1}}, 0, \lambda^{N}\psi_{a_{1}}\rangle$ and $|\Psi_{E=0}^{(3)}\rangle=|\psi_{a_{1}}, 0, \psi_{a_{1}}, 0, -\psi_{a_{1}}, 0, \cdots, -\psi_{a_{1}}, 0, -\psi_{a_{1}}\rangle$ is actually equivalent to a phase topological router with $N+1$ output ports.

\begin{figure}
	    \centering
         \subfigure{\includegraphics[width=0.45\linewidth]{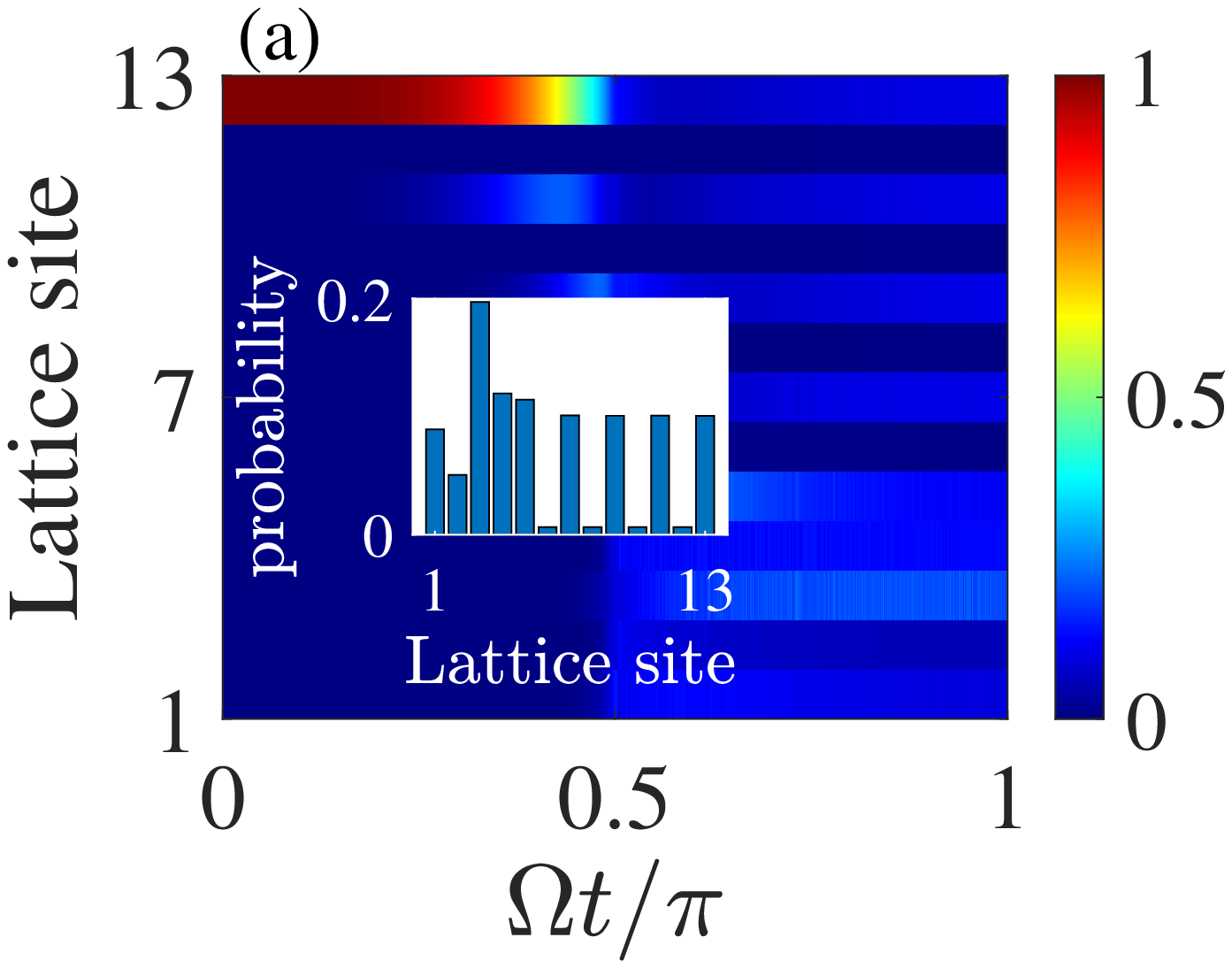}}
	    \subfigure{\includegraphics[width=0.45\linewidth]{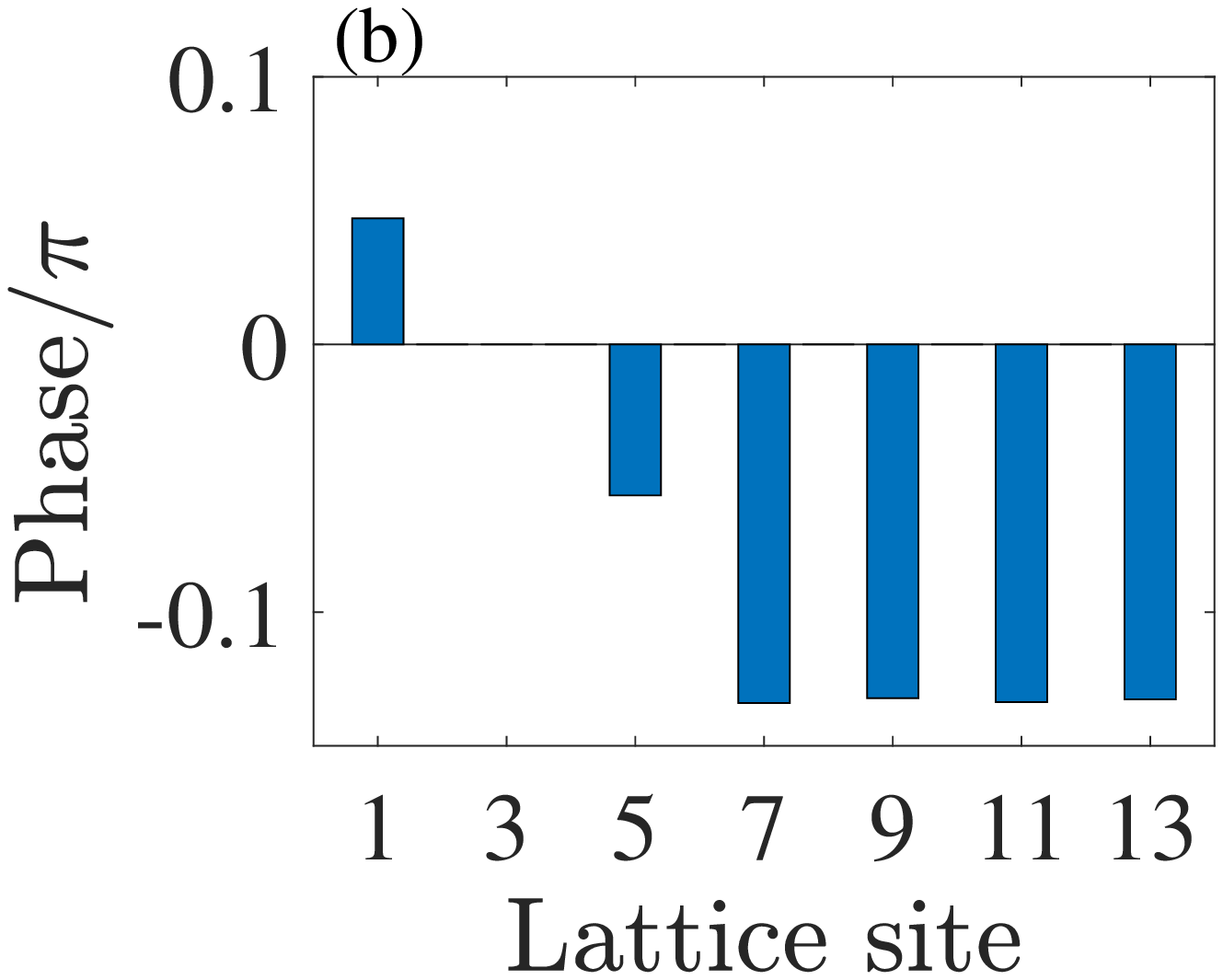}}

          \subfigure{\includegraphics[width=0.45\linewidth]{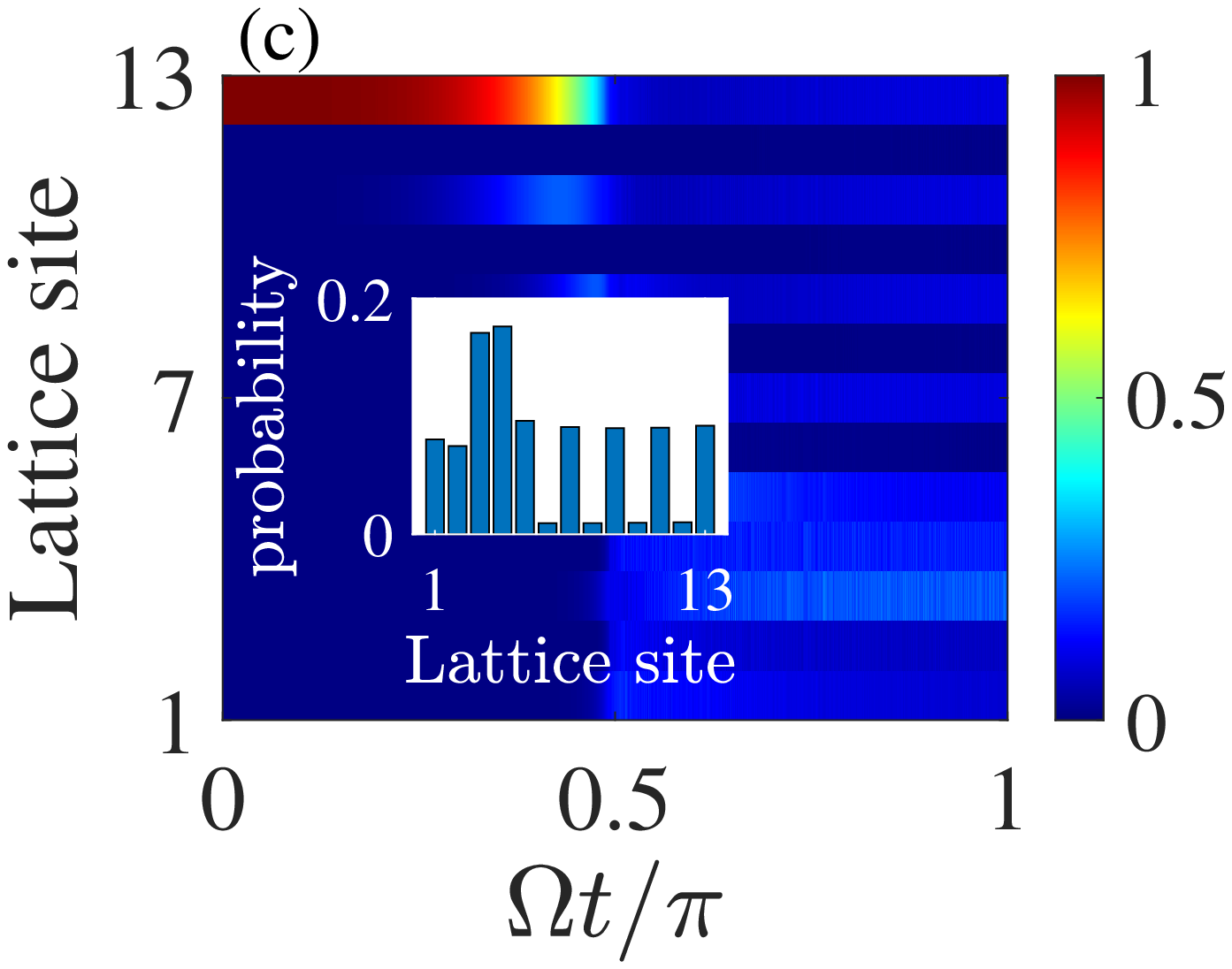}}
	    \subfigure{\includegraphics[width=0.45\linewidth]{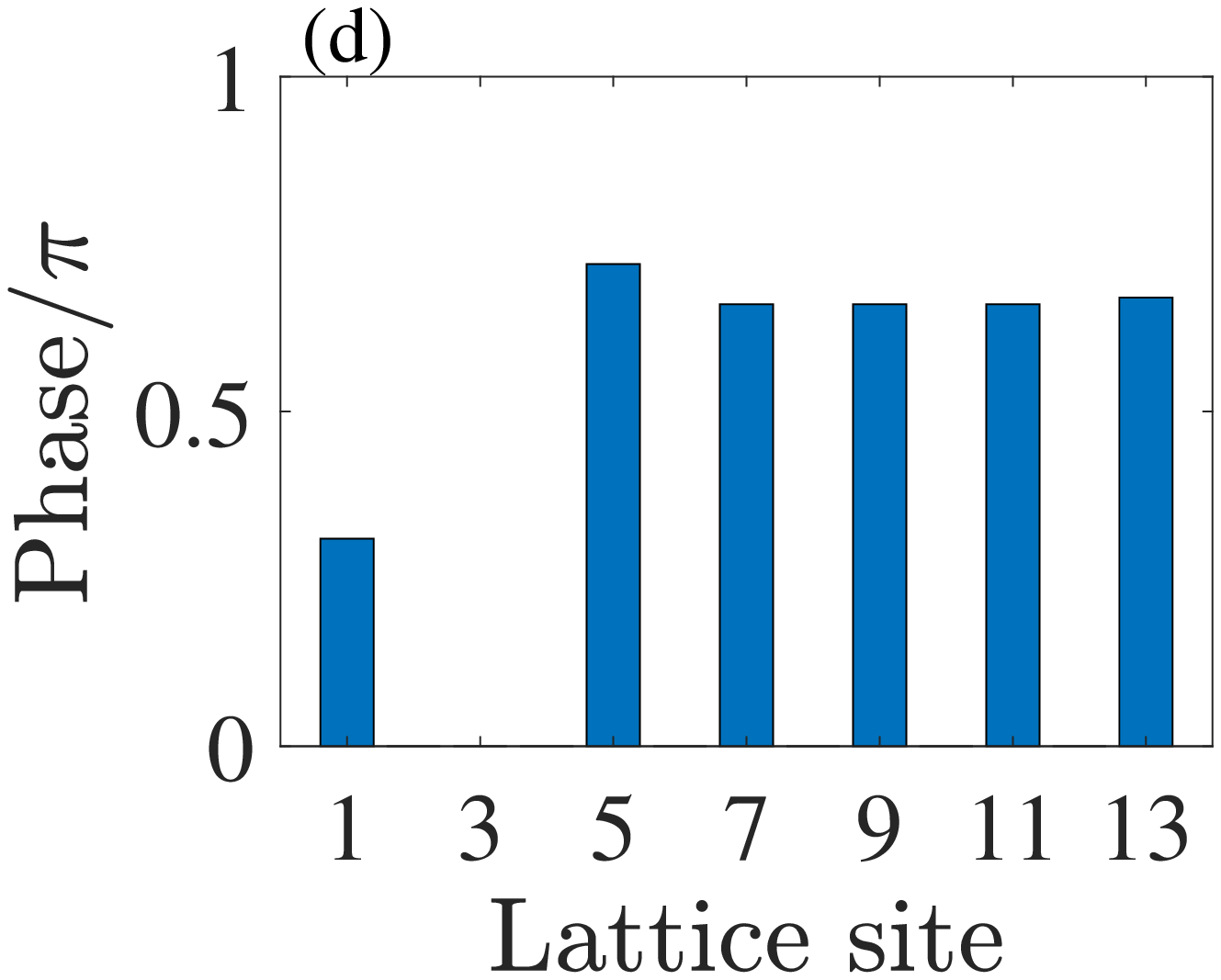}}

         \subfigure{\includegraphics[width=0.45\linewidth]{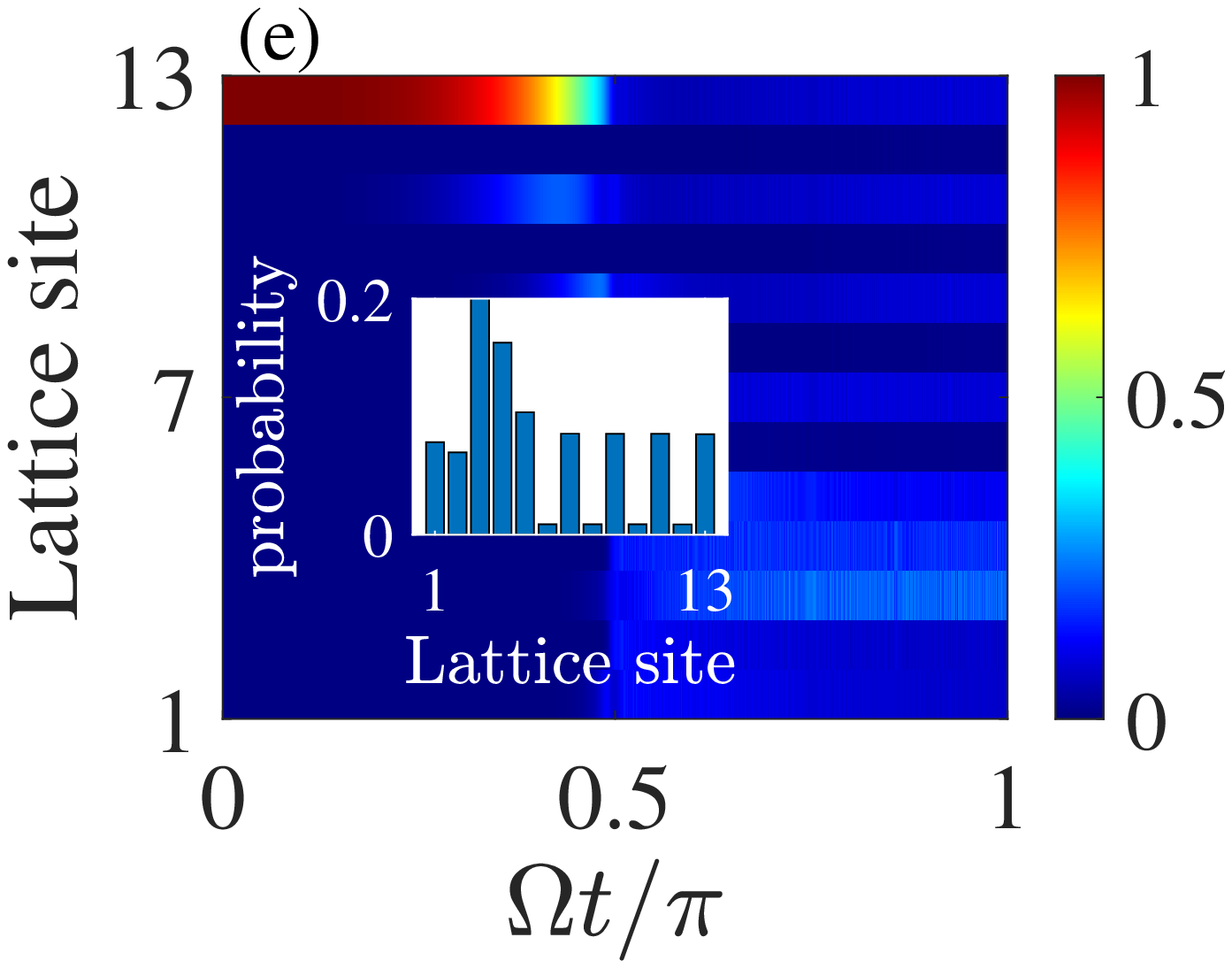}}
	    \subfigure{\includegraphics[width=0.45\linewidth]{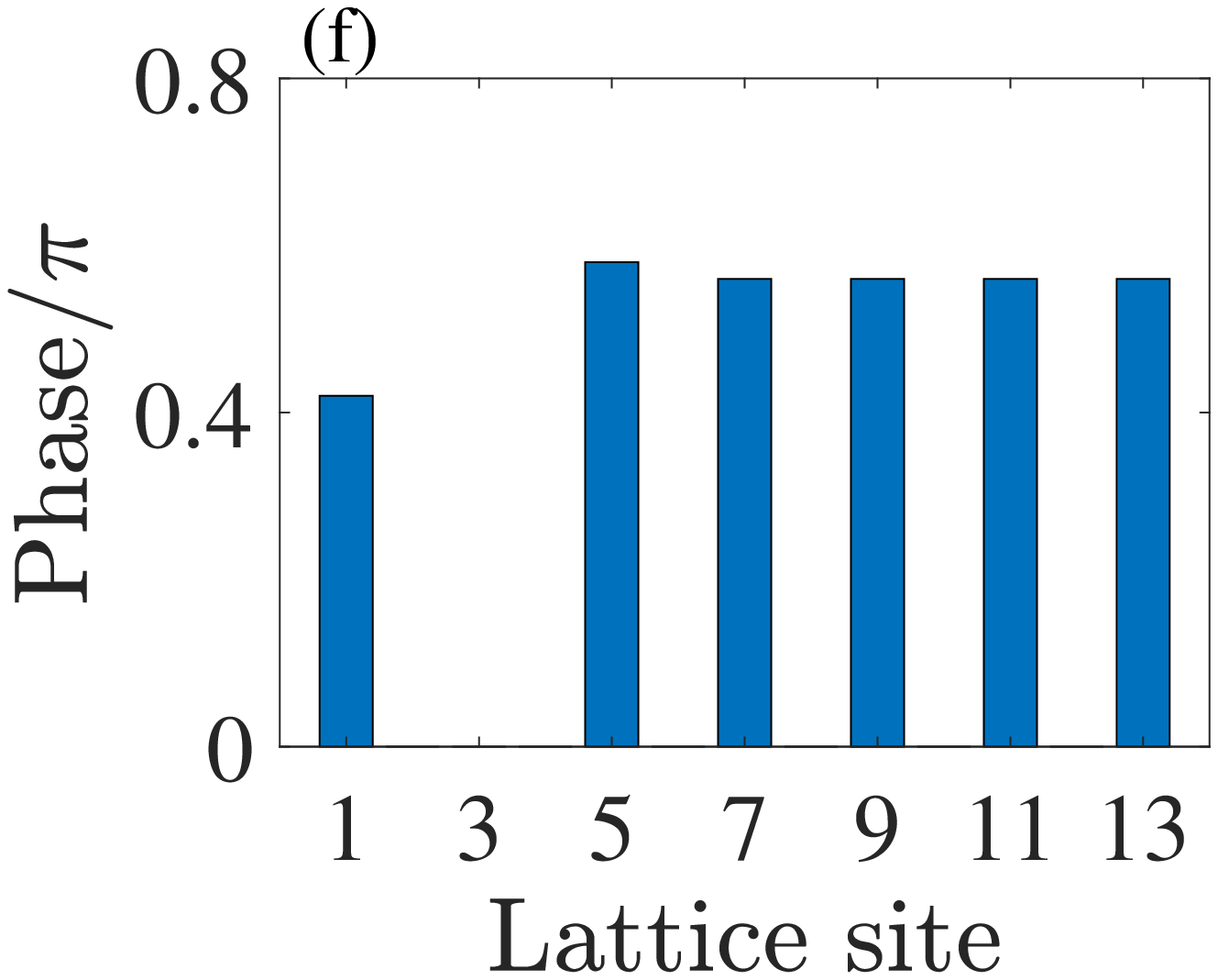}}\caption{\label{f7} The evolution of the initial state $|\Psi_{E=0}^{(1)}\rangle$ when the system is an imperfect lattice with random disorder. (a) (b) The evolution of the initial state $|\Psi_{E=0}^{(1)}\rangle$ with on-site disorder strength $W=0.2$, in which the ramping speed $\Omega$ satisfies $\Omega=0.0001$. The corresponding phase distribution of the evolved final state when $\theta=\pi$. (c) (d)The evolution of the initial state $|\Psi_{E=0}^{(1)}\rangle$ with NN disorder strength $W=0.2$, in which the ramping speed $\Omega$ satisfies $\Omega=0.0001$. The corresponding phase distribution of the evolved final state when $\theta=\pi$  (e) (f)The evolution of the initial state $|\Psi_{E=0}^{(1)}\rangle$ with long-range disorder strength $W=0.2$, in which the ramping speed $\Omega$ satisfies $\Omega=0.0001$. The corresponding phase distribution of the evolved final state when $\theta=\pi$.The size of the lattice is $L=2N+1=13$. The unit is $J=1$.}\label{fig7}
\end{figure}

\begin{figure}
        \centering
	    \subfigure{\includegraphics[width=0.45\linewidth]{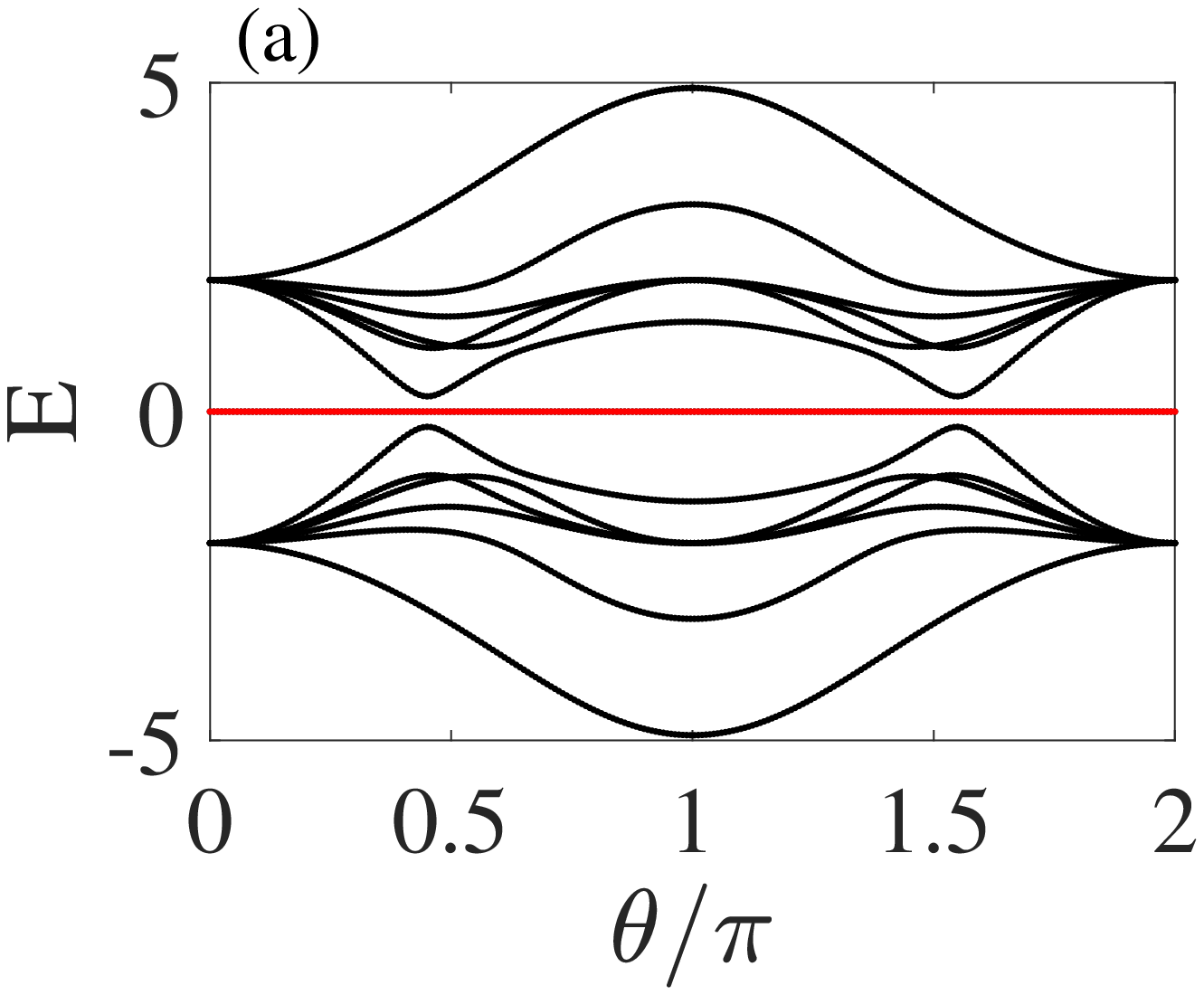}}
	    \subfigure{\includegraphics[width=0.45\linewidth]{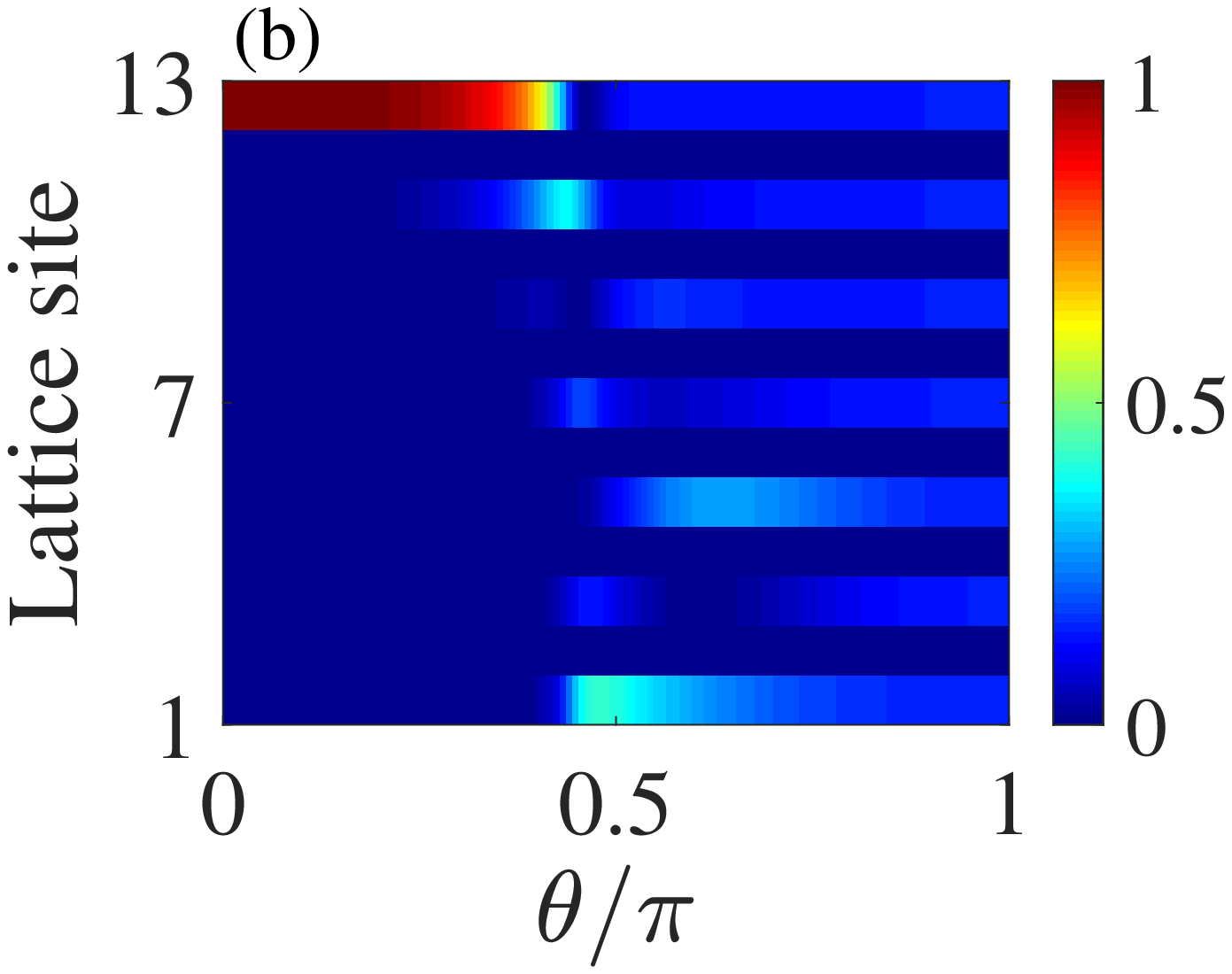}}

         \subfigure{\includegraphics[width=0.45\linewidth]{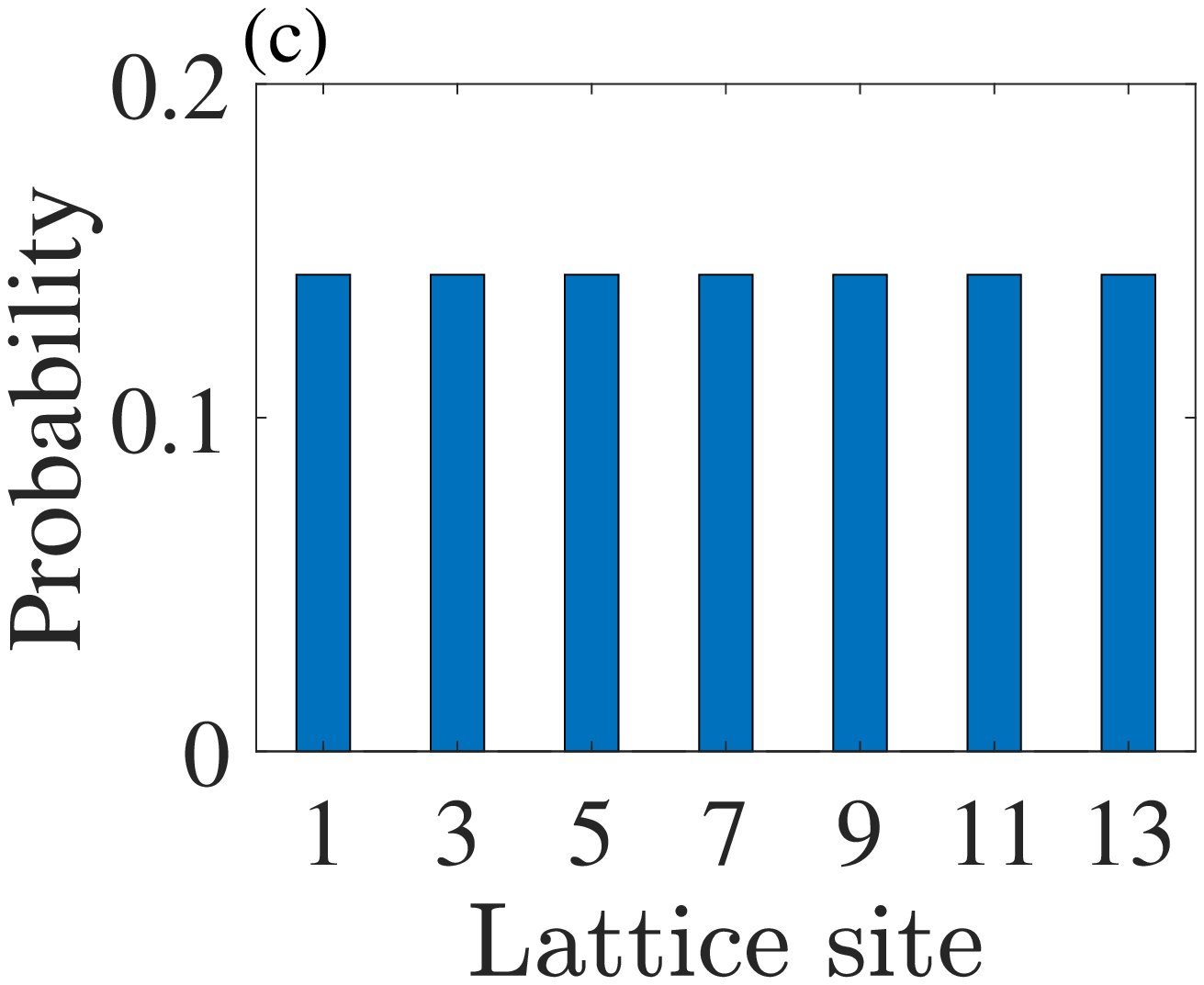}}
	    \subfigure{\includegraphics[width=0.45\linewidth]{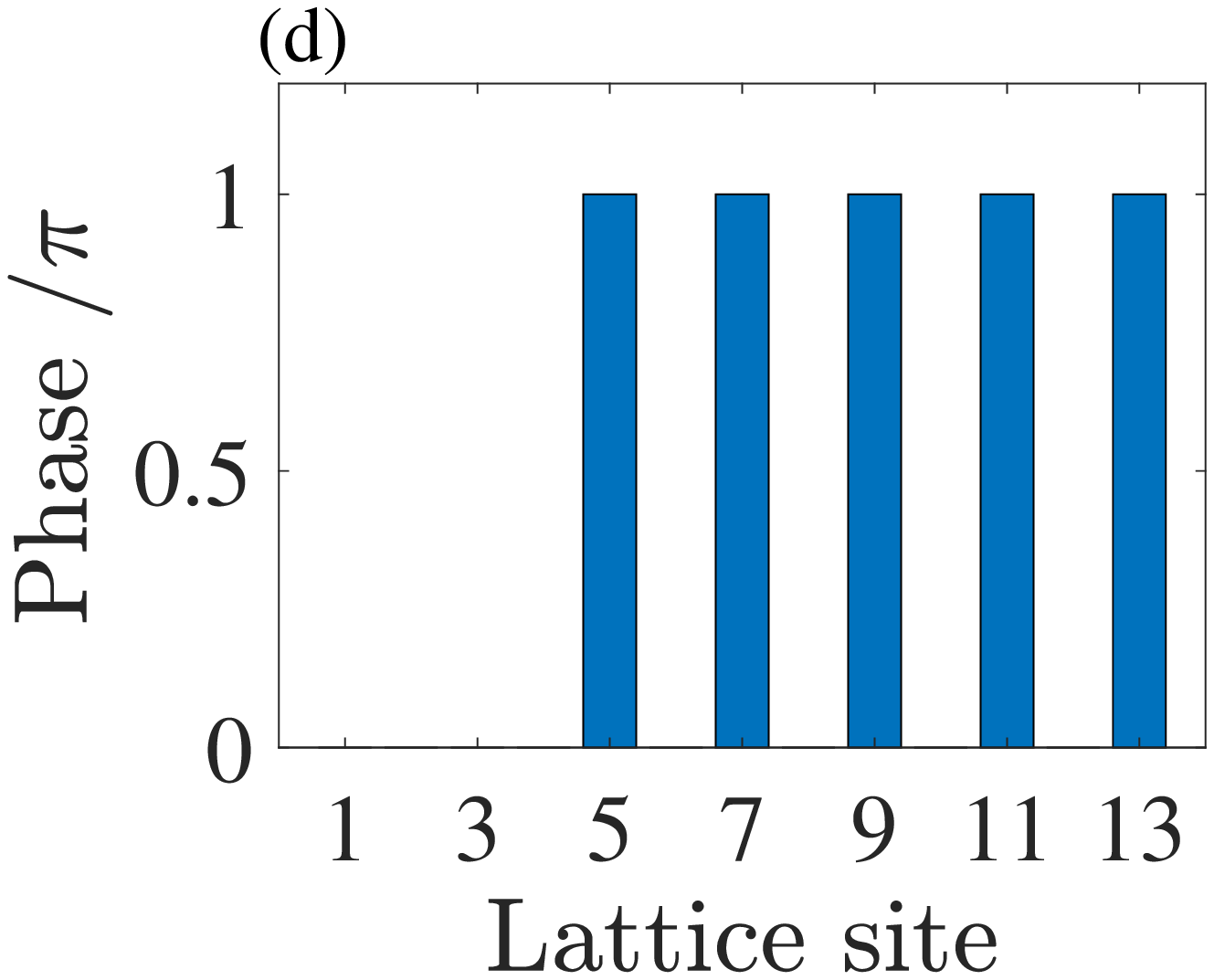}}\caption{\label{f8} (a) Energy spectrum of the 1D extended SSH lattice versus the parameter $\theta$. The energy spectrum has a zero-energy mode remaining unchanged when $\theta\in [0, 2\pi]$. (b) Distribution of the zero-energy mode versus the parameter $\theta$. ($\theta\in [0, \pi]$) (c) Probability distribution of the zero-energy mode when $\theta=\pi$. (d) Phase distribution of the zero-energy mode when $\theta=\pi$. The size of the lattice is $L=2N+1=13$. The unit is $J=1$.}\label{fig8}
\end{figure}

\begin{figure*}
       \centering
       \subfigure{\includegraphics[width=0.3\linewidth]{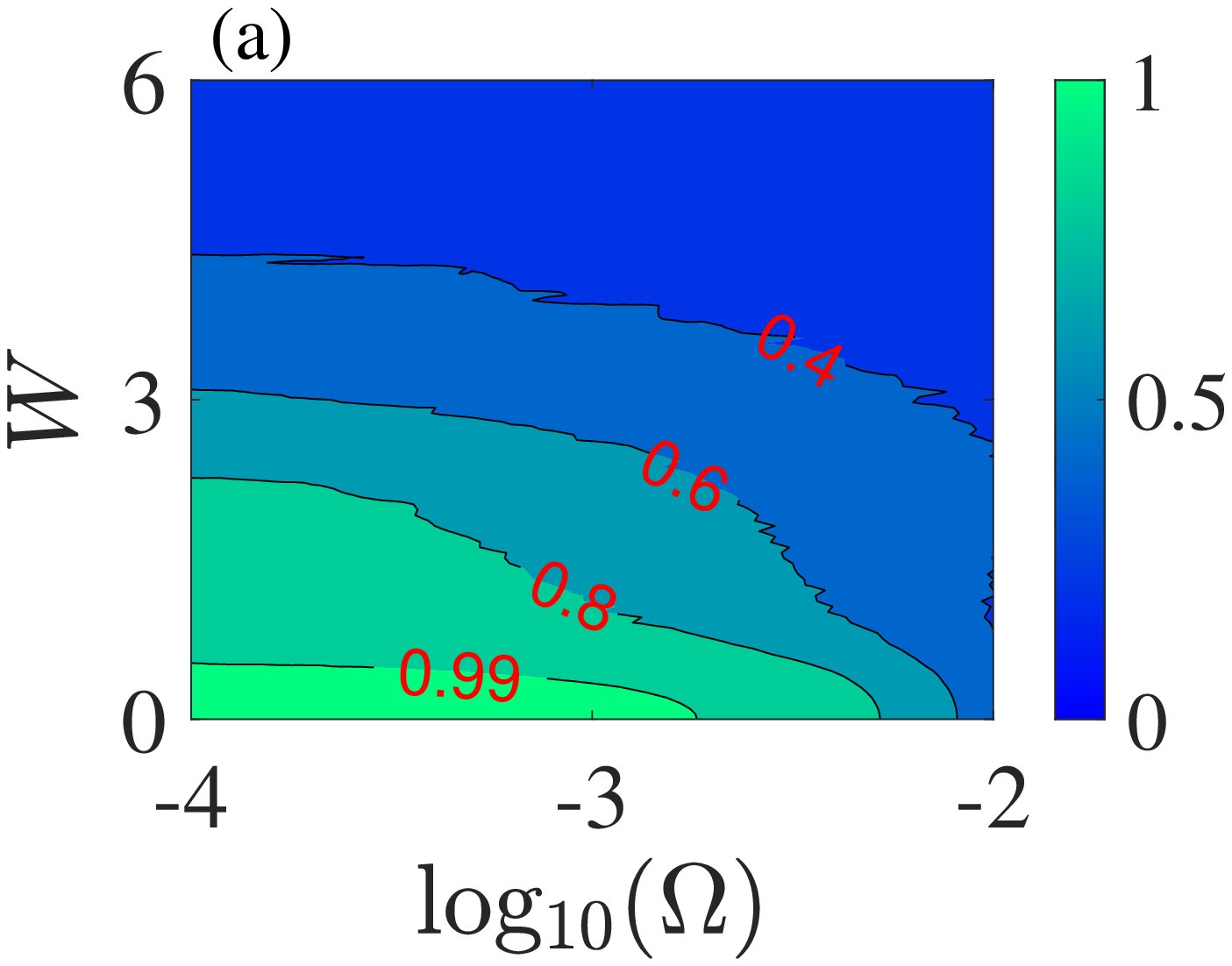}}
       \subfigure{\includegraphics[width=0.3\linewidth]{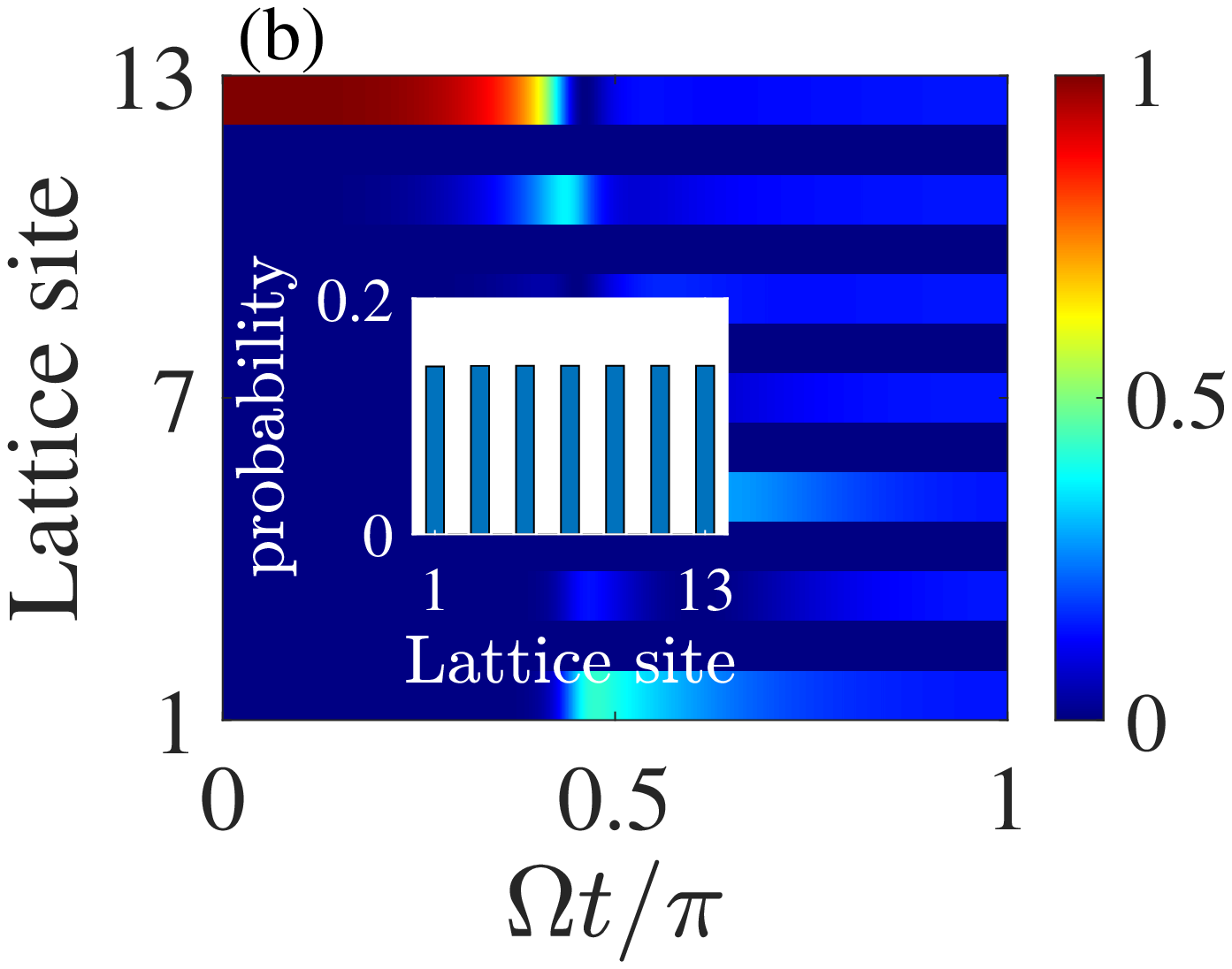}}
       \subfigure{\includegraphics[width=0.3\linewidth]{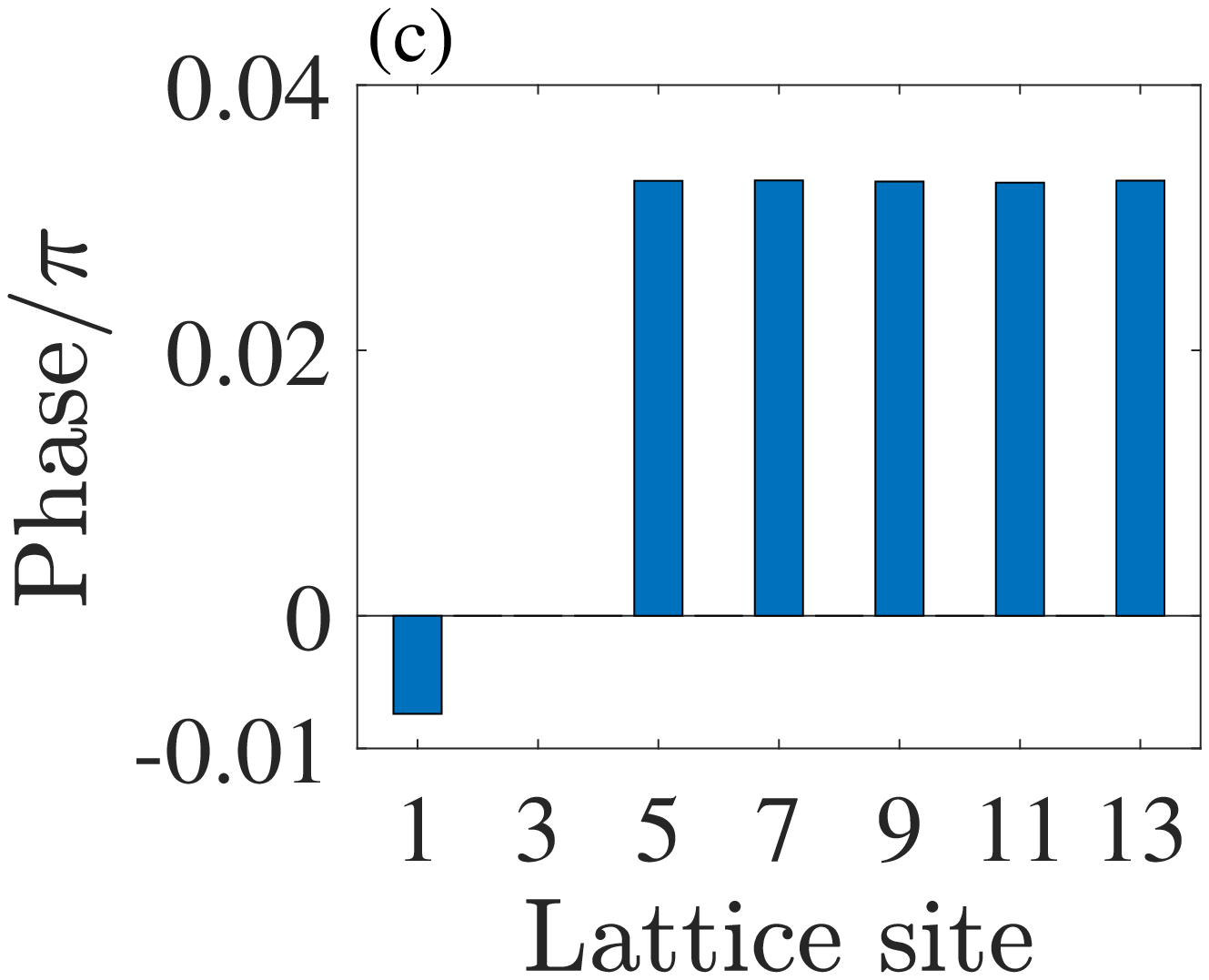}}

       \subfigure{\includegraphics[width=0.3\linewidth]{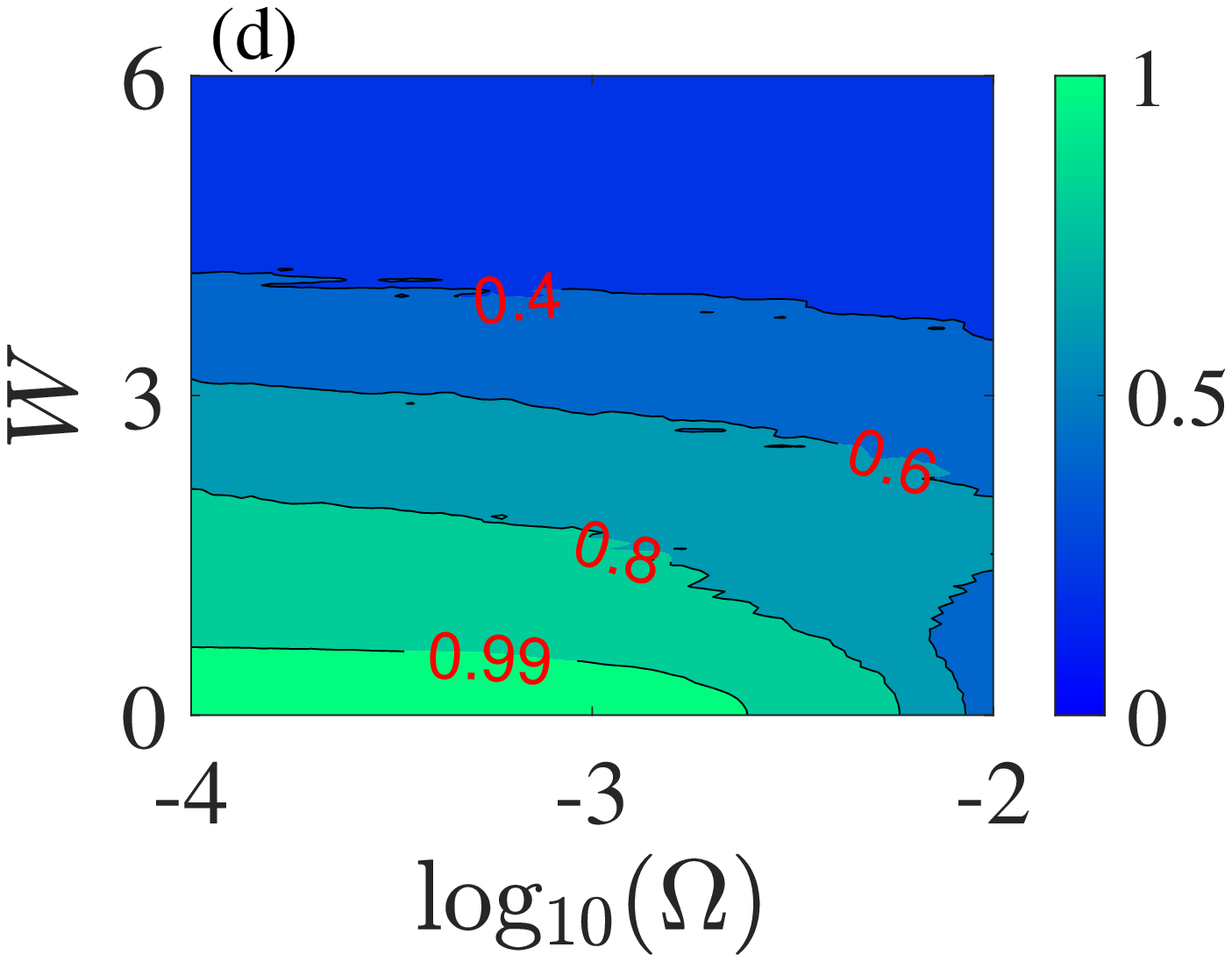}}
       \subfigure{\includegraphics[width=0.3\linewidth]{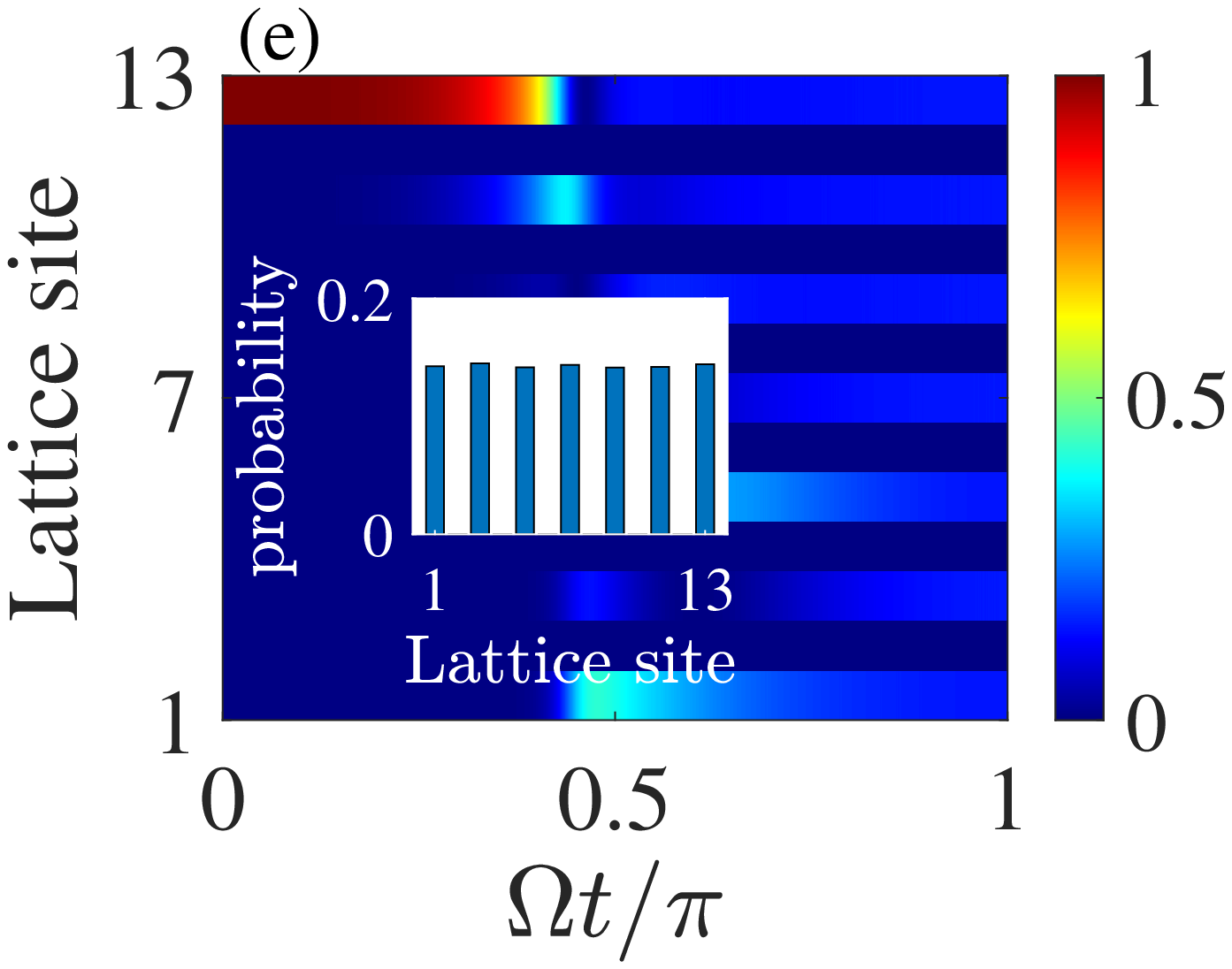}}
       \subfigure{\includegraphics[width=0.3\linewidth]{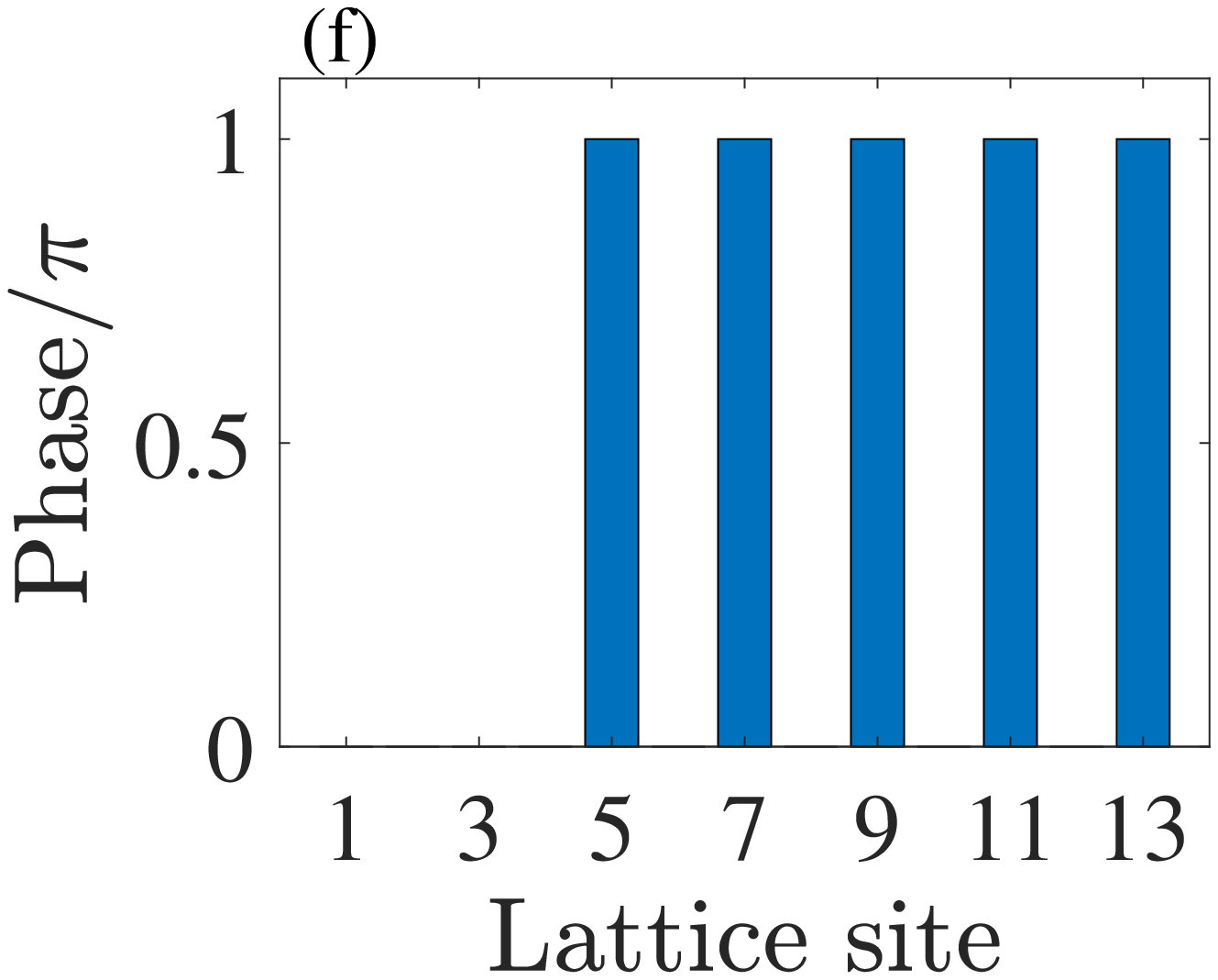}}

       \subfigure{\includegraphics[width=0.3\linewidth]{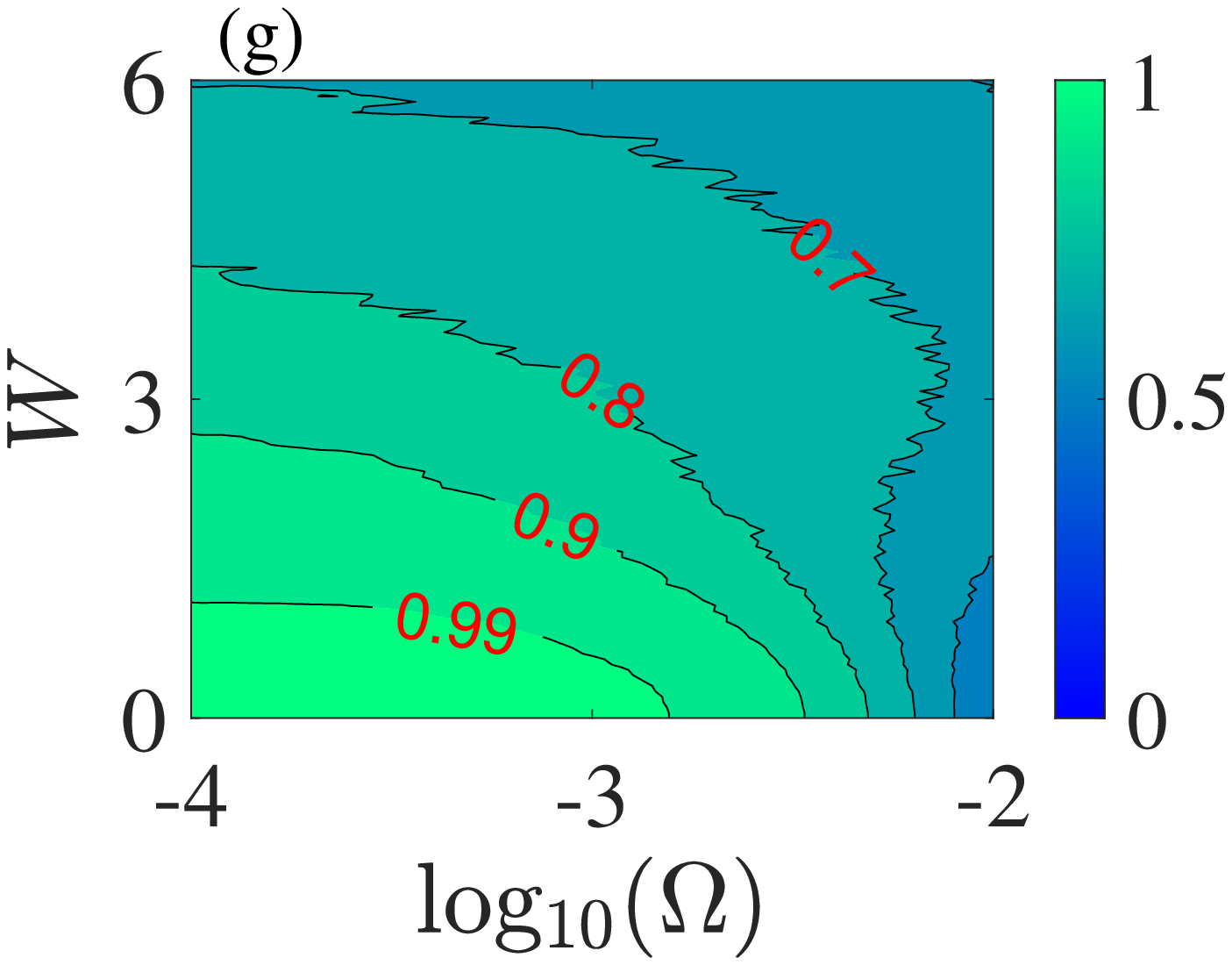}}
       \subfigure{\includegraphics[width=0.3\linewidth]{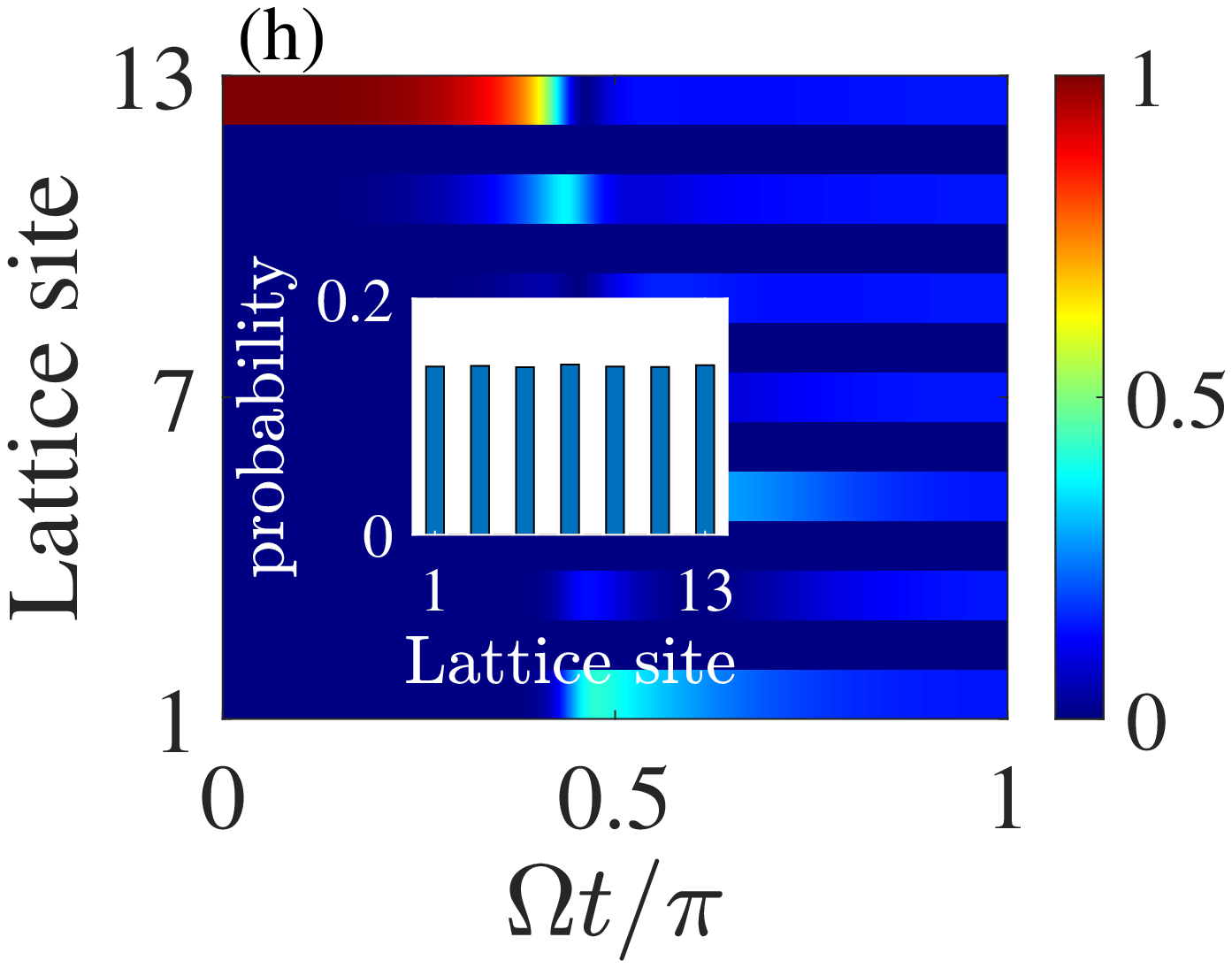}}
       \subfigure{\includegraphics[width=0.3\linewidth]{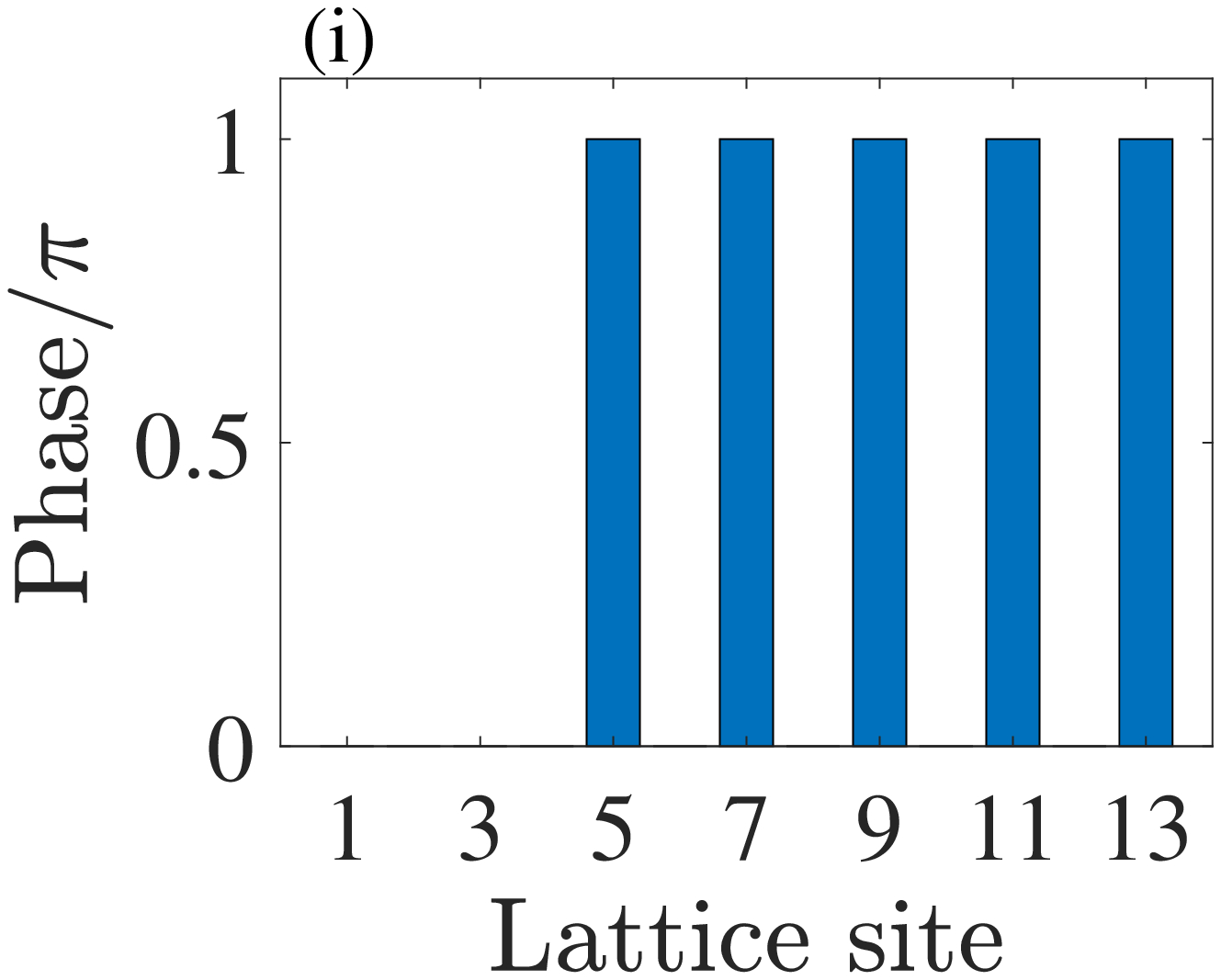}}
       \caption{\label{f9}  The evolution of the initial state $|\Psi_{E=0}^{(1)}\rangle$ when the system is an imperfect lattice with  random disorder. (a) The fidelity between the evolved final state $|\psi_{f}^{\prime}\rangle$ and the ideal final state $|\Psi_{E=0}^{(3)\prime}\rangle$ versus the ramping
speed $\Omega$ and an random on-site disorder with strength $W$. (b) The evolution of the initial state $|\Psi_{E=0}^{(1)}\rangle$ with on-site disorder strength $W=0.2$, in which the ramping speed $\Omega$ satisfies $\Omega=0.0001$. The subgraph represents the probability distribution of evolved final state.
(c) The phase distribution of the evolved final state in Fig.~\ref{fig9}(b). (d) The fidelity between the evolved final state $|\psi_{f}^{\prime}\rangle$ and the ideal final state $|\Psi_{E=0}^{(3)\prime}\rangle$ versus the ramping speed $\Omega$ and an random NN disorder strength $W$. (e) The evolution of the initial state $|\Psi_{E=0}^{(1)}\rangle$ with NN disorder strength $W=0.2$, in which the ramping speed $\Omega$ satisfies $\Omega=0.0001$. The subgraph represents the probability distribution of evolved final state.
(f) The phase distribution of the evolved final state in Fig.~\ref{fig9}(e). (g) The fidelity between the evolved final state $|\psi_{f}^{\prime}\rangle$ and the ideal final state $|\Psi_{E=0}^{(3)\prime}\rangle$ versus the ramping speed $\Omega$ and an random long-range disorder strength $W$. (h) The evolution of the initial state $|\Psi_{E=0}^{(1)}\rangle$ with long-range disorder strength $W=0.2$, in which the ramping speed $\Omega$ satisfies $\Omega=0.0001$. The subgraph represents the probability distribution of evolved final state.
(i) The phase distribution of the evolved final state in Fig.~\ref{fig9}(h). The size of the lattice is $L=2N+1=13$. The unit is $J=1$.}\label{fig9}
\end{figure*}

However, it is worth noting that the energy gap in Fig.~\ref{fig6}(a) is nearly closed when the new long-range hopping $T_{N}$ is added into system. Due to the correspondence between the gap width and the relative robustness, a narrower energy gap implies that the gap state is extremely sensitive to the slight disorder. To show this, we plot the evolution process of the initial state $|\Psi_{E=0}^{(1)}\rangle$ versus the on-site, NN, and long-range hopping disorders in Fig.~\ref{fig7}. For the on-site disorder, we can observe that the probability distribution of evolved final state is not uniform at all $a$-type sites when the disorder strength satisfies $W=0.2$ and slow enough ramping speed satisfies $\Omega=0.0001$, as shown in Fig.~\ref{fig7}(a). The result indicates that the probability distribution of the gap state is sensitive to the on-site disorder during the whole evolution process. Furthermore, we plot the corresponding phase information of the evolved final state when $\theta=\pi$, the phase information does not follow any distribution rules, as shown in Fig.~\ref{fig7}(b). Apparently, the present gap state possesses probability distribution and phase distribution is fragile to the on-site disorder added into system. 

Similarly, we also verify the gap state of the system is sensitive to the NN and long-range disorder. As shown in Fig.~\ref{fig7}(c),  with the time-dependent Hamiltonian, the evolved final state is unable to transfer to all $a$-type sites with the same probability correspondings to the NN disorder strength $W=0.2$ and ramping speed $\Omega=0.0001$. The Fig.~\ref{fig7}(d) shows the phase information of the evolved final state, where the phase information does not keep 0 (at sites $a_{1}$ and $a_{2}$) and $\pi$ (at sites $a_{3}$, $a_{4}$,...,$a_{N}$, and $a_{N+1}$). The result shows that the state transfer between $|\Psi_{E=0}^{(1)}\rangle$ and $|\Psi_{E=0}^{(3)}\rangle$ cannot be achieved when the NN disorder added into system. Further, the Figs.~\ref{fig7}(e) and~\ref{fig7}(f) illuminate us that, the gap state of the system is sensitive to the long-range disorder. As the robustness against disorder is a characteristic feature of testing the practicality of QST schemes, phase topological router assisted by gap state, which have $N+1$ output ports, but is fragile to the on-site, NN, and long-range disorder. Thus, it is able to construct the phase non-robust topological router with $N+1$ output ports.

Despite the above scheme of the new long-range hopping $T_{N}$ introduced between the site $a_{3}$ and the site $b_{1}$ is only suitable for an ideal phase topological router [without robustness]. It is always demonstrated that generating new domain wall is an effective path to construct the phase topological router with more output ports. Therefore, following the thought of generating new domain wall, we adjust the new long-range hopping, in which the long-range hopping $T_{N}$ is added between the site $a_{4}$ and the site $b_{1}$. Similarly, when $\theta \in [\pi / 2, \pi]$, a new domain wall generated in another sites $a_{2}$, $b_{1}$ and $a_{4}$. Then, the gap state is redistributed on the site $a_{2}$ with a certain probability. Just to verify the analysis, we first draw the energy spectrum of the system with $H=\sum_{n=1}^{N}(J_{1}a_{n}^{\dag}b_{n}+J_{2}a_{n+1}^{\dag}b_{n}+\mathrm{H.c.})+\sum_{n=1}^{N-1}(T_{n}a_{1}^{\dag}b_{n+1}+\mathrm{H.c.})+T_{N}(a_{4}^{\dag}b_{1}+b_{1}^{\dag}a_{4})$ in Fig.~\ref{fig8}(a). Similar to the presence of long-range hopping $T_{N}(a_{3}^{\dag}b_{1}+b_{1}^{\dag}a_{3})$, the present energy spectrum also possesses gap state in the whole energy gap and the gap state keeps the zero energy. The energy gap between the gap state and the bulk band in Fig.~\ref{fig8}(a) exhibits an increasing behavior, which indicates that the present gap state is more immune to mild disorder due to the protection of topology. In Figs.~\ref{fig8}(b) and~\ref{fig8}(c), we plot the distribution of the gap state and the corresponding probability distribution. We can conclude that the gap state in $\theta \in[0,\pi / 2]$ is localized at site $a_{N+1}$ while it is uniformly distributed at all $a$-type sites with the same probability $1/(N+1)$ when $\theta\in[\pi / 2, \pi]$. The phase information is shown in Fig.~\ref{fig8}(d). Figure~\ref{fig8} implies that the gap state can be used as the topological channel to engineer a phase topological router with $N+1$ output ports.

Similarly, to test the robustness of phase topological router, we further study the effects of mlid disorder on QST with $N+1$ output ports in Fig.~\ref{fig9}. First, we plot the fidelity of the state transfer versus the ramping speed $\Omega$ and on-site disorder strength $W$, as shown in Fig.~\ref{fig9}(a). The numerical results reveal that, corresponding to the small enough ramping speed $\log_{10}(\Omega)< -2.7$, the mild parameter $W$ with $W<0.5$ ensures that the state transfer between $||\Psi_{E=0}^{(1)}\rangle|$ and $||\Psi_{E=0}^{(3)}\rangle|$=$|\Psi_{E=0}^{(3)\prime}\rangle$ can be realized with a high enough fidelity $F=0.99$. Here, the absolute value represents the probability density of the gap state  after ignoring the phase information. The Fig.~\ref{fig9}(b) shows a detailed state transfer process, when the on-site disorder strength is $W=0.2$ and the ramping speed corresponds to $\Omega=0.0001$.  The numerical results show that the state initially prepared at the right edge site can be transfered at $N+1$ sites uniformly. And in Fig.~\ref{fig9}(c), the phase information in $a_{3}$, $a_{5}$, ..., $a_{N}$, $a_{N+1}$ is destroyed, meaning that the present phase topological router with phase information is sensitive to on-site disorder. We also shows the effects of the NN disorder on the state transfer and phase information in Figs.~\ref{fig9}(d) to~\ref{fig9}(f). The numerical results clearly reveal that, for the slow enough ramping speed and the mild enough NN disorder strength, the probability amplitude and the phase information of the gap state are robust, which means that the mild NN disorder strength does not break the chiral symmetry. With the topological protection, the phase-robust topological router is naturally immune to the mild NN disorders. Further, we investigate the effect of long-range disorder on state transfer and phase information in Figs.~\ref{fig9}(g) to ~\ref{fig9}(i). Results are similar to those in Fig.~\ref{fig4}, the off-diagonal disorder preserves the chiral symmetry of the system and the gap state with phase information is immune to the mild long-range disorder. Thus, via designing the long-range hopping term between the site $a_{4}$ and site $b_{1}$, we can realize an optimized phase-robust topological router with $N+1$ output ports, which greatly improves the scalability of topological QST.
\begin{figure}
	\centering
     \includegraphics[width=0.80\linewidth]{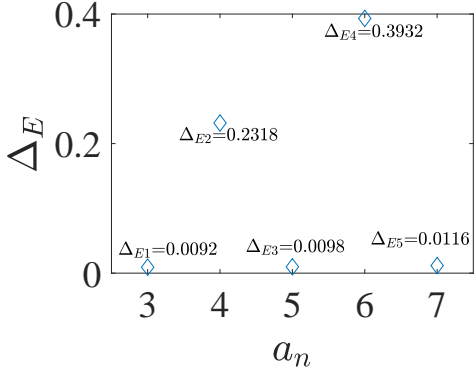}\\
     \caption{\label{f10} The minimal energy space $\Delta_{E}$ versus the location of the  new long-range hopping $T_{N}$ added on the site $b_{1}$ and the site $a_{n}(n=3,4,5,6,7)$.The size of the lattice is $L=2N+1=13$. The unit is $J=1$.}\label{fig10}
\end{figure}

The new long-range hopping $T_{N}$ connecting different sites has different minimal energy gap. In Fig.~\ref{fig10}, we further plot the minimal energy space $\Delta_{E}$ (the $\Delta_{E}$ is the maximum energy difference between the gap and the bulk.) versus the location of the new long-range hopping $T_{N}$ when the size of the extended SSH lattice is $L=2N+1=13$.  The numerical results exhibit that the width of minimal energy gap varies with the location of the $T_{N}$. For example, the $\Delta_{E1}$ for the long-range hopping $T_{N}$  added on the site $b_{1}$ and site $a_{3}$ is 0.0092. However, the $\Delta_{E2}$ attains 0.2318 when the the long-range hopping $T_{N}$ is added on the site $b_{1}$ and site $a_{4}$. More clearly, adding the long-range hopping $T_{N}$ between the site $b_{1}$ and the site $a_{N}$, $\Delta_{E4}=0.3932$ tends to be maximum. It is well known that a wide energy gap means that the gap state is more naturally immune to the mild perturbations and disorders. Therefore, when the new long-range hopping $T_{N}$ is introduced between the site $b_{1}$ and the site $a_{N}$, an optimized phase-robust topological router with $N+1$ output ports is put forward.

\section{\label{sec.4}Detection and adiabatic evolution}
\subsection{\label{sec.4A}{The detection of the zero-energy gap state}}
The superconducting circuit lattice is general and fully applicable for realizing efficient simulation for Bose system. Compared to the Fermi system, the bosonic photons can occupy one particular eigenstate at the same time, which provides the essential application for the topological features of topological states~\cite{Qi2020Engineering, Li2018Exploring}. Utilized by the lattice-based cavity input-output process~\cite{Mei2015Simulation}, we can detect the distribution of the gap state via the mean distribution of the photons. The expectation value of photons in the resonators can be expressed as [see Appendix A]
\begin{eqnarray}\label{e12}
\overrightarrow{\mathbf{R}}=-\left(\mathbf{\Delta}+\mathbf{M}-i\frac{\mathbf{K}}{2}\right)^{-1}\overrightarrow{\mathbf{\Omega}},
\end{eqnarray}
where $\overrightarrow{\mathbf{R}}=[\langle a_{1}\rangle,\langle b_{1}\rangle,\cdots,\langle a_{N}\rangle,\langle b_{N}\rangle,\langle a_{N+1}\rangle]^{T}$ represents the column vector composed by the mean value of the steady state for resonator, $\mathbf{\Delta}=\mathrm{Diag}[\Delta_{a,1},\Delta_{b,1},\cdots,\Delta_{a,N},\Delta_{b,N},\Delta_{a,N+1}]=\mathrm{Diag}[\Delta_{R},\Delta_{R},\cdots,\Delta_{R},\Delta_{R},\Delta_{R}]$ is the diagonal matrix originating from the detuning of the resonators, $\mathbf{M}$ is the coefficient matrix owning the same form as the Hamiltonian in Eq.~(\ref{e03}) [with $N$ output ports], $\mathbf{K}=\mathrm{Diag}[\kappa _{a_{1}},\kappa _{b_{1}},\cdots,\kappa _{a_{N}},\kappa _{b_{N}},\kappa _{a_{N+1}}]$ represents the diagonal matrix caused by the decay of the resonators, and $\overrightarrow{\mathbf{\Omega}}=[\Omega_{a,1},\Omega_{b,1},\cdots,\Omega_{a,N},\Omega_{b,N},\Omega_{a,N+1}]^{T}$ is the external driving.

 \begin{figure}
	\centering
	\subfigure{\includegraphics[width=0.80\linewidth]{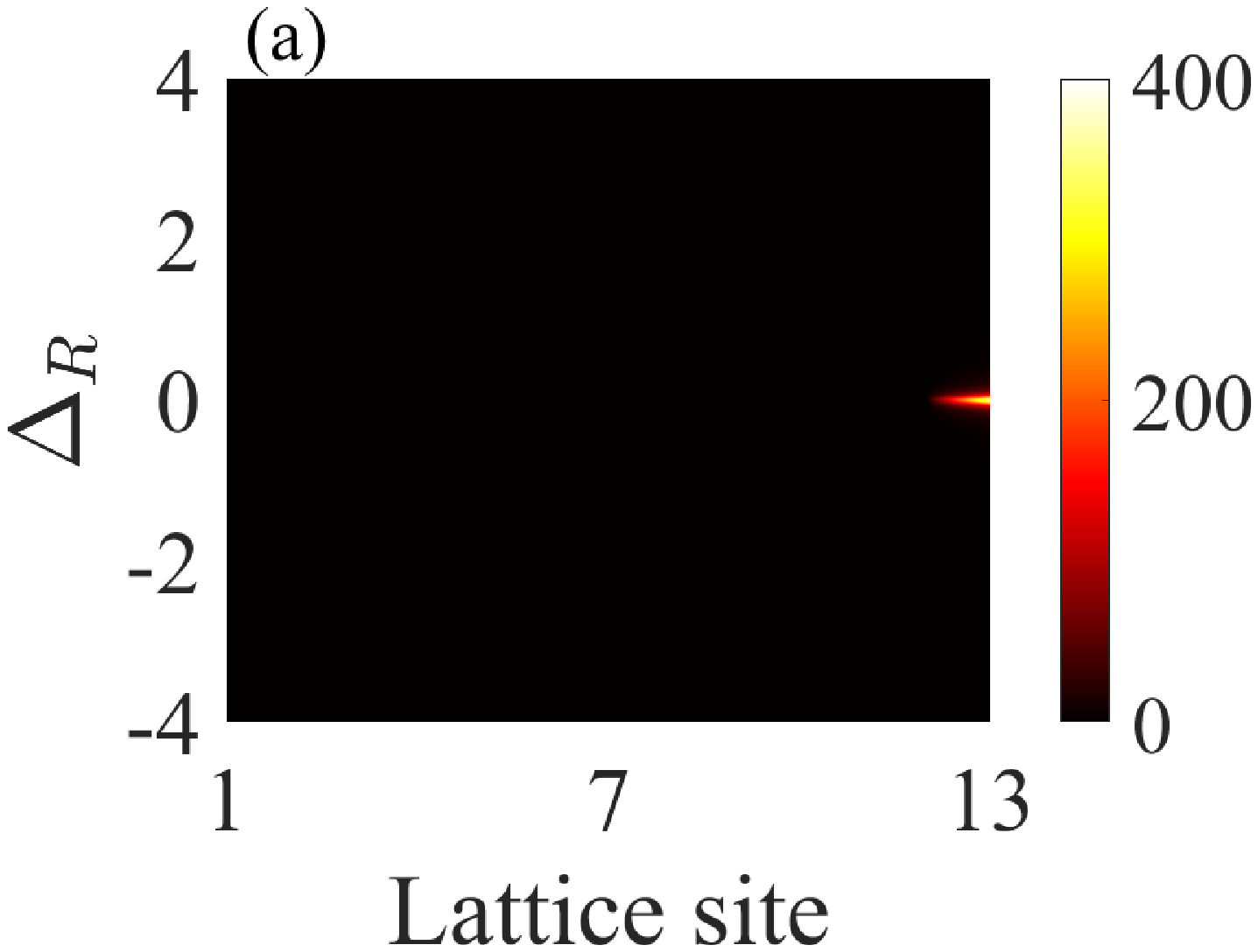}}
	\subfigure{\includegraphics[width=0.80\linewidth]{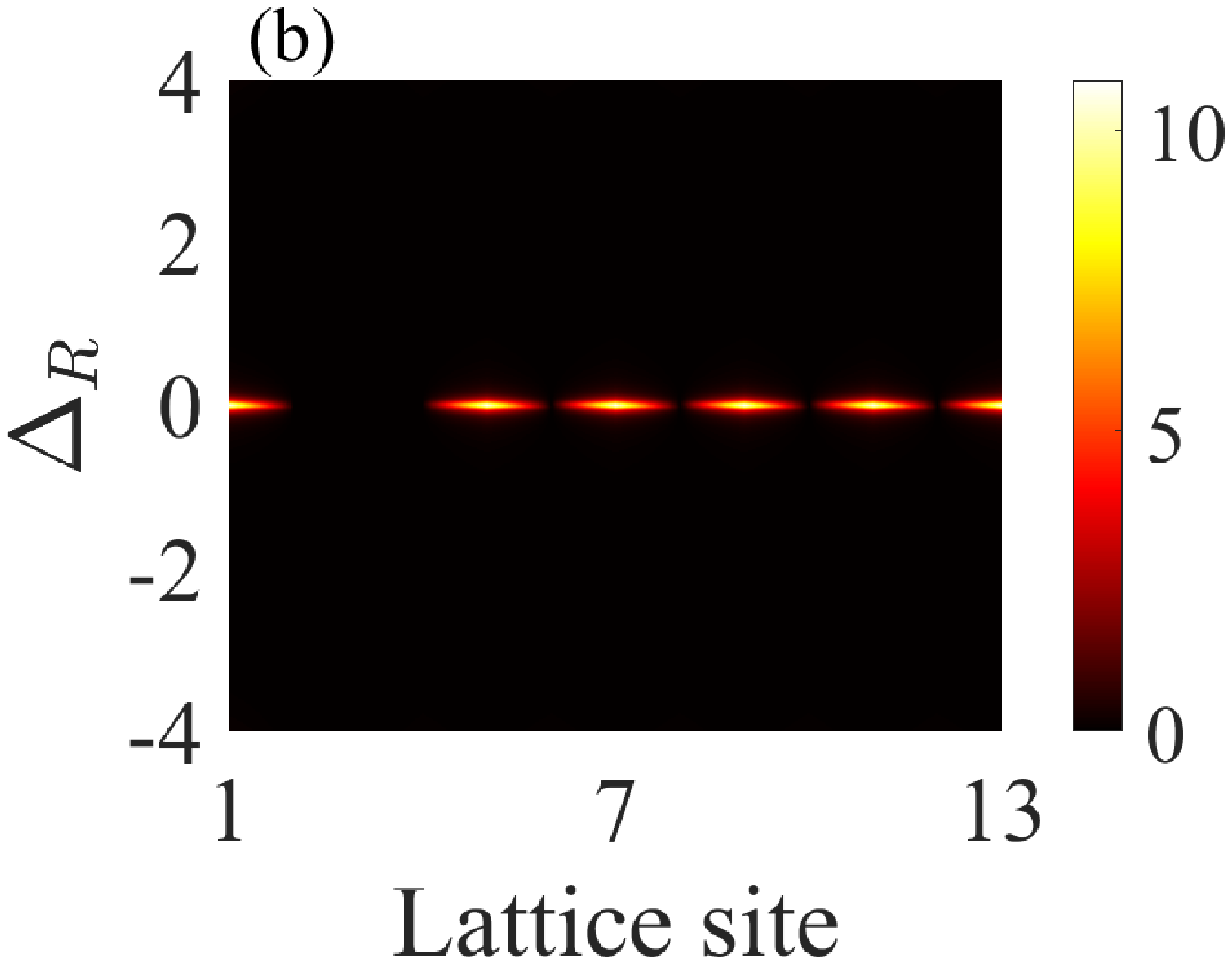}}
	\caption{The detection spectrum. (a) The external driving excites the last resonator $a_{N+1}$. The photons mainly gather into the rightmost resonator with the resonant excitation. (b) The external driving excites the first resonator $a_{1}$. The photons mainly gather into the resonators $a_{1}$, $a_{3}$, ... , $a_{N}$, and $a_{N+1}$ with the resonant excitation. The size of the lattice is $L=2N+1=13$. The decay of the resonator $\kappa_{k_{n}}(k=a,b)$ is 0.1 and the driving amplitude is $\Omega_{a,N+1}=\Omega_{a,1}=1$.}\label{fig11}
\end{figure}

To detect the gap state, one needs to firstly excite the system to occupy the gap state.  When using the external driving $\overrightarrow{\mathbf{\Omega}}=[0,0,\cdots,0,0,\Omega_{a,N+1}]^{T}$ to excite the rightmost resonator in a certain range of driving frequency, the distribution of the photons in the resonator array is plotted in the Fig.~\ref{fig11}(a). Obviously, when the scanning frequency reaches resonance with the gap state, we find the final photons of steady state mainly gather into the last resonator, which is consistent with the distribution pattern of the right edge state. Similarly, when using external driving $\overrightarrow{\mathbf{\Omega}}=[\Omega_{a,1},0,\cdots,0,0,0]^{T}$ to excite the leftmost resonator within a certain range of driving frequency, the distribution of the photons in the resonator array is plotted in the Fig.~\ref{fig11}(b). The results reveal that, when the scanning frequency reaches resonance with the gap state, the final photons of steady state now mainly gather into the resonators $a_{1}$, $a_{3}$, $a_{4}$, $\cdots$, and $a_{N+1}$ uniformly [with $N$ output ports]. In this way, the input and output signals of the phase-robust topological router can be detected via the distributions of photons. 

\begin{figure}
	\centering
	\subfigure{\includegraphics[width=0.80\linewidth]{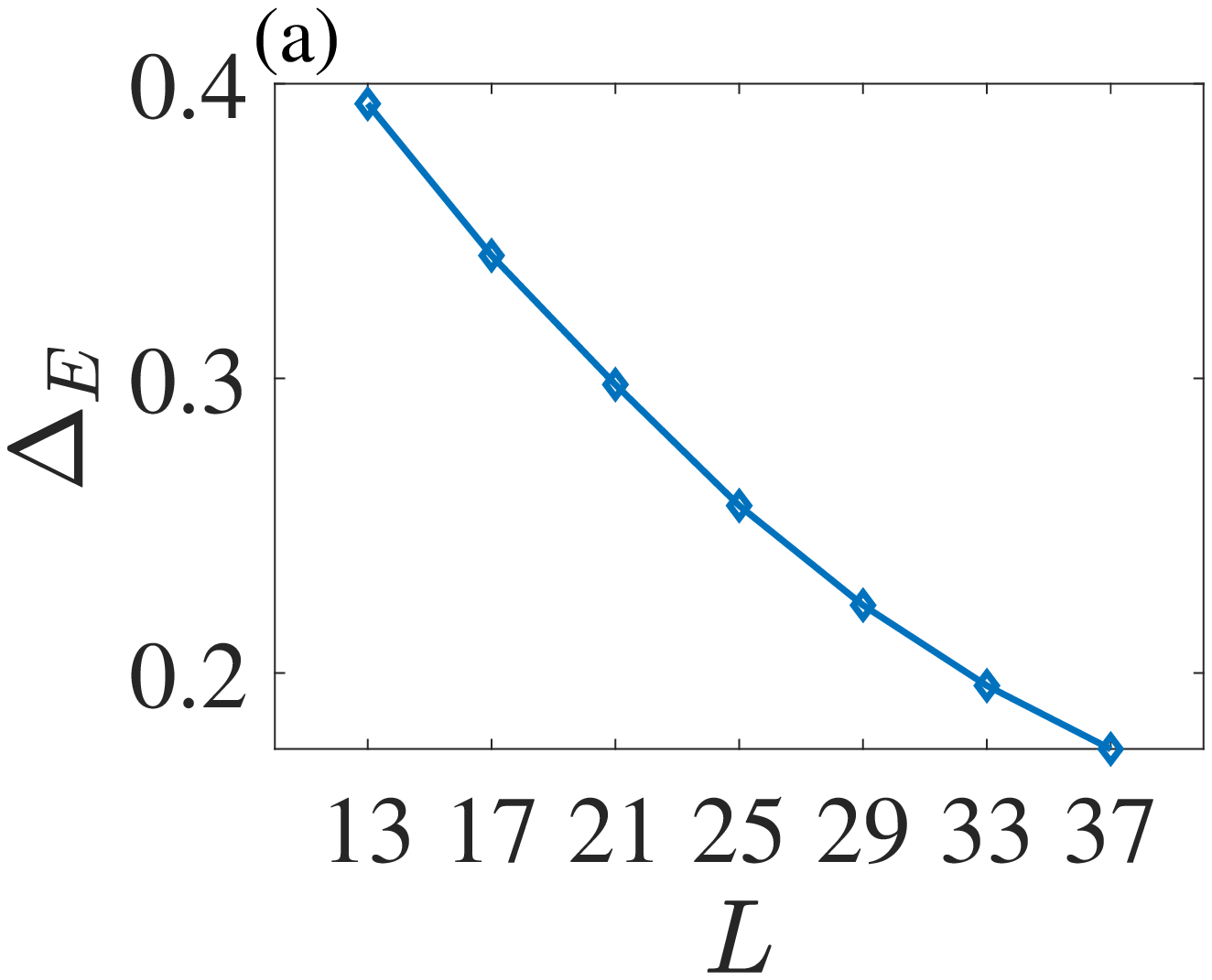}}
	\subfigure{\includegraphics[width=0.80\linewidth]{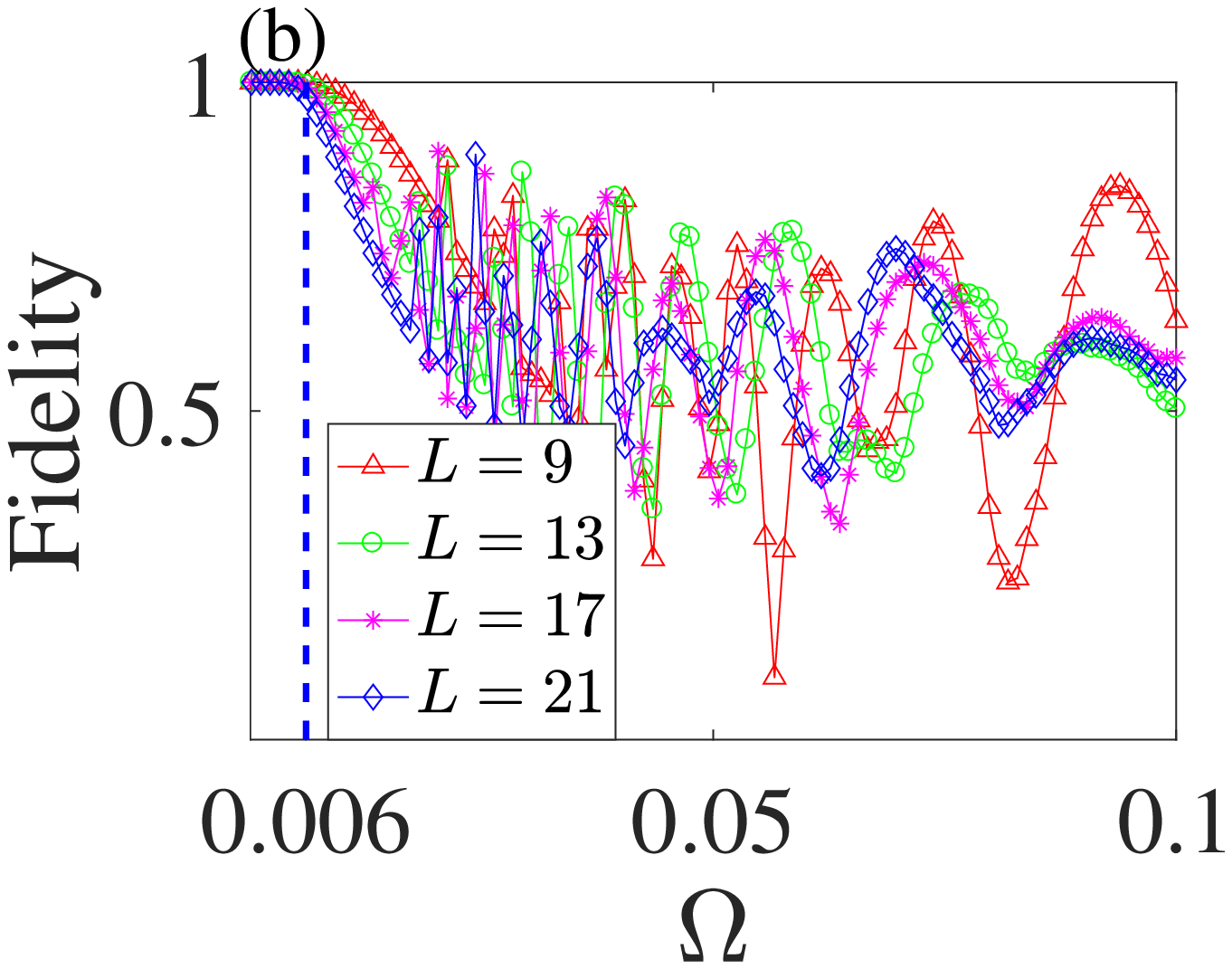}}
	\caption{\label{f11} (a) The minimal energy space $\Delta E$ versus the different sizes of the lattice $L$ when the new long-range hopping $T_{N}$ is added on the site $b_{1}$ and the site $a_{N}$. (b) The fidelities of the phase-robust topological router versus the ramping speed $\Omega$ when the new long-range hopping $T_{N}$ is added on the site $b_{1}$ and the site $a_{N}$ with the different sizes of the lattice $L$. The unit is $J=1$. }\label{fig12}
\end{figure}

\subsection{\label{sec.4B}{The adiabatic evolution of the zero-energy gap state}}
We propose to use a time-dependent Hamiltonian procedure to complete the protocol of the phase-robust topological router, where the ramping speed $\Omega$ is always to be emphasized during the evolution of the zero-energy gap state. The appropriate adiabatic  parameter $\Omega$ ensures the evolution process against the influence of the bulk state. According to the adiabatic theorem requires that $\sqrt{\Omega}<\Delta_{E}$~\cite{Mei2018Robust}, the $\Delta_{E}$ determines a safe range of the ramping speed $\Omega$. In Fig.~\ref{fig12}(a), we plot the minimal energy space $\Delta_{E}$ versus the different sizes of the lattice $L$ when the long-range hopping $T_{N}$ is added on the site $b_{1}$ and the site $a_{N}$ [mentioned in Sec.~\ref{sec.3B}]. Analysis from the trend of the minimum energy gap $\Delta_{E}$, the $\Delta_{E}$ in this paper is more larger than Ref.~\cite{Qi2021Topological} under the same size $L$ with $N+1$ output ports, perhaps the present phase topological router is more advantageous. In Fig.~\ref{fig12}(b), we also plot the fidelity of 9-, 13-, 17- and 21 sites versus the ramping speed $\Omega$ with the new long-range hopping $T_{N}$ is added on the site $b_{1}$ and the site $a_{N}$. We observe that there is a platform, labelled by $F=1$, that are fit any size for the appropriate evolution parameter $\Omega$. Here, in order to further prove the rationality of the phase-robust topological router, we take $L=21$ as an example to carry out relevant calculations. As shown in Fig.~\ref{fig12}(b), if the $ \Omega \le 0.006$, the fidelity stays almost 1, which indirectly guarantees that the evolution of the gap state follows the adiabatic evolution condition if it is carried out in the parameter range $ \Omega \le 0.006$. Then, with a typical coupling strength of $J /2 \pi=250$ MHz and the ramping speed $ \Omega= 0.006$, the operation time for phase topological router with $N+1$ output ports can be achieved in $T_{L=21}= \pi / \Omega\approx 3\mu s$, which is much shorter than the decoherence times of the superconducting qubit~\cite{Schmidt2013Circuit} and superconducting resonator~\cite{Reagor2016Quantum, Reagor2013Reaching}. 

\section{\label{sec.5}Conclusion}
In conclusion, we have proposed a model of a chiral-symmetric dimerized superconducting circuit lattice with long-range hopping. The existence of the long-range hopping induces a zero-energy gap state, which establishes the special topological channel to implement the phase topological router with $N$ or $N+1$ output ports. We demonstrate that,  based on the chiral-symmetric protection of the system, the phase topological router is robust to the mild off-diagonal disorder such as the disorder in NN and long-range hopping. Furthermore, taking advantage of the Bose statistical properties of the superconducting circuit lattice, we study that the signals of the input port and the output ports can be detected via the distributions of photons. Our scheme supplies a viable prospect with large-scale QST via the special zero-energy gap state induced by the long-range hopping in a chiral-symmetric system, and we think that the gap state owning the phase information may further supply important application for the logic gate or quantum interference.

\begin{center}$\mathbf{Acknowledgments}$\end{center}
This work was supported by the National Natural Science Foundation of China under Grants No. 12074330, No. 11775048, and No. 12047566.

\section*{APPENDIX A}
\setcounter{equation}{0}
\renewcommand\theequation{A.\arabic{equation}}
Now we give the detailed derivation from the Eq.~(\ref{e01}) to the Eq.~(\ref{e03}). The Hamiltonian of the 1D dimerized superconducting circuit lattice with the long-range hopping is shown in Eq.~(\ref{e01}). After performing a rotating frame with respect to the external driving frequency $\omega_{d,n}$ and qubit frequency $\omega_{qi,n}$, the Hamiltonian in Eq.~(\ref{e01}) can be rewritten as
\begin{eqnarray}\label{A1}
	H_{total}^{\prime}&=&\sum_{n} \left(\Delta_{a,n}a_{n}^{\dag}a_{n}+\Delta_{b,n}b_{n}^{\dag}b_{n}\right)\cr\cr
	&&+\sum_{n}\left[\frac{g_{a,n}^{2}}{\Delta_{qa,n}}\left(|e\rangle_{qa,n} \langle e| a_{n}a_{n}^{\dag}-|g\rangle_{qa,n} \langle g| a_{n}^{\dag}a_{n}\right)\right.\cr\cr
	&&\left.+\frac{g_{b,n}^{2}}{\Delta_{qb,n}}\left(|e\rangle_{qb,n} \langle e| b_{n}b_{n}^{\dag}-|g\rangle_{qb,n} \langle g| b_{n}^{\dag}b_{n}\right)\right]\cr\cr
	&&+\sum_{n}\left[\frac{g_{1,n}^{2}}{\Delta_{q1,n}}(|e\rangle_{q1,n} \langle e| a_{n}a_{n}^{\dag}-|g\rangle_{q1,n} \langle g| a_{n}^{\dag}a_{n}\right.\cr\cr
	&&\left.+|e\rangle_{q1,n}\langle e| b_{n}b_{n}^{\dag}-|g\rangle_{q1,n} \langle g| b_{n}^{\dag}b_{n})\right.\cr\cr
	&&\left.+\frac{g_{2,n}^{2}}{\Delta_{q2,n}}(|e\rangle_{q2,n} \langle e| b_{n}b_{n}^{\dag}-|g\rangle_{q2,n} \langle g| b_{n}^{\dag}b_{n}\right.\cr\cr
	&&\left.+|e\rangle_{q2,n}\langle e| a_{n+1}a_{n+1}^{\dag}-|g\rangle_{q2,n} \langle g| a_{n+1}^{\dag}a_{n+1})\right]\cr\cr
	&&+\sum_{n}\left[\frac{g_{1,n}^{2}}{\Delta_{q1,n}}(|e\rangle_{q1,n} \langle e| a_{n}^{\dag}b_{n}-|g\rangle_{q1,n} \langle g| b_{n}^{\dag}a_{n})\right.\cr\cr
	&&\left.+\frac{g_{2,n}^{2}}{\Delta_{q2,n}}(|e\rangle_{q2,n} \langle e| a_{n+1}^{\dag}b_{n}-|g\rangle_{q2,n} \langle g| b_{n}^{\dag}a_{n+1})+\mathrm{H.c.}\right]\cr\cr
	&&+\sum_{n}\left[\frac{g_{3,n}^{2}}{\Delta_{q3,n}}(|e\rangle_{q3,n} \langle e| a_{1}a_{1}^{\dag}-|g\rangle_{q3,n} \langle g| a_{1}^{\dag}a_{1}\right.\cr\cr
	&&\left.+|e\rangle_{q3,n}\langle e| b_{n}b_{n}^{\dag}-|g\rangle_{q3,n} \langle g| b_{n}^{\dag}b_{n})\right]\cr\cr
	&&+\sum_{n}\left[\frac{g_{3,n}^{2}}{\Delta_{q3,n}} (|e\rangle_{q3,n} \langle e| a_{1}^{\dag}b_{n}-|g\rangle_{q3,n} \langle g| b_{n}^{\dag}a_{1})+\mathrm{H.c.}\right]\cr\cr
	&&+\sum_{n}\left(\Omega_{a,n}a_{n}^{\dag}+\Omega_{b,n}b_{n}^{\dag}+\mathrm{H.c.}\right).
\end{eqnarray}

In the dispersive regime, when all of the qubits are prepared in their ground states, then we can get the Eq.~(\ref{e02}). After engineering the coupling strength between resonator and the nearest neighbor qubit as
 $-\frac{g_{1,n}^{2}}{\Delta_{q1,n}}=J_{1}$, $-\frac{g_{2,n}^{2}}{\Delta_{q2,n}}=-\frac{g_{3,n}^{2}}{\Delta_{q3,n}}=J_{2}$, the Hamiltonian becomes
\begin{eqnarray}\label{A2}
	H_{eff}^{1}&=&\left (\Delta_{a,1}-\frac{g_{a,1}^{2}}{\Delta_{qa,1}}+J_{1}+NJ_{2}\right)a_{1}^{\dag}a_{1}\cr\cr
	&&+\left(\Delta_{b,1}-\frac{g_{b,1}^{2}}{\Delta_{qb,1}}+J_{1}+J_{2}\right)b_{1}^{\dag}b_{1}\cr\cr
	&&+\sum_{n}\left[\left(\Delta_{a,n}-\frac{g_{a,n}^{2}}{\Delta_{qa,n}}+J_{1}+J_{2}\right)a_{n}^{\dag}a_{n}\right.\cr\cr
	&&\left.+\left(\Delta_{b,n}-\frac{g_{b,n}^{2}}{\Delta_{qb,n}}+J_{1}+2J_{2}\right)b_{n}^{\dag}b_{n}\right]\cr\cr
	&&+\sum_{n}\left(J_{1}b_{n}^{\dag}a_{n}+J_{2}a_{n+1}^{\dag}b_{n}+\mathrm{H.c.}\right)\cr\cr
	&&+\sum_{n}J_{2}\left(b_{n}^{\dag}a_{1}+a_{1}^{\dag}b_{n}\right)\cr\cr
	&&+\sum_{n}\left(\Omega_{a,n}a_{n}^{\dag}+\Omega_{b,n}b_{n}^{\dag}+\mathrm{H.c.}\right).
\end{eqnarray}
And, when the coupling strength between resonator and the embedded qubit further satisfies $\frac{g_{a,1}^{2}}{\Delta_{qa,1}}=J_{1}+NJ_{2}$, $\frac{g_{b,1}^{2}}{\Delta_{qb,1}}=\frac{g_{a,n}^{2}}{\Delta_{qa,n}}=J_{1}+J_{2}$, and $\frac{g_{b,n}^{2}}{\Delta_{qb,n}}=J_{1}+2J_{2}$, the above Hamiltonian can be written as
\begin{eqnarray}\label{A3}
	H_{eff}&=&\sum_{n}\left(\Delta_{a,n}a_{n}^{\dag}a_{n}+\Delta_{b,n}b_{n}^{\dag}b_{n}+\mathrm{H.c.}\right)\cr\cr
	&&+\sum_{n}\left(J_{1}b_{n}^{\dag}a_{n}+J_{2}a_{n+1}^{\dag}b_{n}+\mathrm{H.c.}\right)\cr\cr
	&&+\sum_{n}J_{2}\left(b_{n}^{\dag}a_{1}+a_{1}^{\dag}b_{n}\right)\cr\cr
	&&+\sum_{n}\left(\Omega_{a,n}a_{n}^{\dag}+\Omega_{b,n}b_{n}^{\dag}+\mathrm{H.c.}\right).
\end{eqnarray}
Note that, the effective dynamics of the superconducting circuit lattice can be described by the quantum Langevin equations [ignoring the input noise], i.e., $\dot {\rho_{n} } = i[ H_{eff}, \rho_{n}] -\frac{\kappa _{\rho_{n}} }{2}\rho_{n}$. Here, $\rho_{n}$ ($\rho=a,b$) represents the annihilation operator of resonator, $\kappa _{\rho_{n}}$ is the decay rate of the resonator. Thus, we can obtain a set of dynamic equations of operators
\begin{eqnarray}\label{A4}
	\dot{a_{n}} &=&-i\left(\Delta_{a,n}-i\frac{\kappa _{{a_{n}}} }{2}\right)a_{n}-iJ_{2}b_{n-1}-iJ_{1} b_{n} -i\Omega _{a,n},\cr\cr
	\dot{b_{n}} &=&-i\left(\Delta_{b,n}-i\frac{\kappa _{{b_{n}}} }{2}\right)b_{n}-iJ_{2}a_{1}-iJ_{1} a_{n}-iJ_{2}a_{n+1}-i\Omega _{b,n}\cr\cr
	&&(n=2,3,\cdots,N),
\end{eqnarray}
with the boundary condition $\dot{a_{1}}=-i(\Delta_{a,1}-i\frac{\kappa _{{a_{1}}} }{2})a_{1}-iJ_{1} b_{1}-iJ_{2} \left ( b_{2}+b_{3}+\cdots+b_{N} \right ) -i\Omega _{a,1}$,  $\dot{b_{1}}=-i(\Delta_{b,1}-i\frac{\kappa _{{b_{1}}} }{2})b_{1}-iJ_{1} a_{1}-iJ_{2}a_{2}-i\Omega _{b,1}$, and $\dot{a}_{N+1} =-i(\Delta_{a,N+1}-i\frac{\kappa _{{a_{N+1}}} }{2})a_{N+1}-iJ_{2}b_{N} -i\Omega _{a,N+1}$.
Under the condition of the strong driving amplitudes, we can implement the standard linearization process via rewriting the  operators as the summation of the mean value and quantum fluctuations, i.e., $\rho_{n}=\langle\rho_{n}\rangle+\delta\rho_{n}$. After that, the quantum fluctuations of the operators satisfy
\begin{eqnarray}\label{A5}
	\dot{\delta a_{n}} &=&-i\left(\Delta_{a,n}-i\frac{\kappa _{{a_{n}}} }{2}\right)\delta a_{n}-iJ_{2}\delta b_{n-1}-iJ_{1}\delta b_{n},\cr\cr
	\dot{\delta b_{n}} &=&-i\left(\Delta_{b,n}-i\frac{\kappa _{{b_{n}}} }{2}\right)\delta b_{n}-iJ_{2}\delta a_{1}-iJ_{1} \delta a_{n}-iJ_{2}\delta a_{n+1}\cr\cr
	&&(n=2,3,\cdots,N),
\end{eqnarray}
accompanied with the boundary condition $\dot{\delta a_{1}}=-i(\Delta_{a,1}-i\frac{\kappa _{{a_{1}}} }{2})\delta a_{1}-iJ_{1}\delta b_{1}-iJ_{2} \left (\delta b_{2}+\delta b_{3}+\cdots+\delta b_{N} \right )$,  $\dot{\delta b_{1}}=-i(\Delta_{b,1}-i\frac{\kappa _{{b_{1}}} }{2})\delta b_{1}-iJ_{1}\delta a_{1}-iJ_{2}\delta a_{2}$, and $\delta\dot{ a}_{N+1}=-i(\Delta_{a,N+1}-i\frac{\kappa _{{a_{N+1}}} }{2})\delta a_{N+1}-iJ_{2}\delta b_{N}$. Obviously, if removing the notation ''$\delta$'' and deriving the Hamiltonian inversely, the linearized Hamiltonian describing the intrinsic interactions of the quantum operators can be given by
\begin{eqnarray}\label{A6}
	H&=&\sum_{n} [\Delta_{a,n}a_{n}^{\dag}a_{n}+\Delta_{b,n}b_{n}^{\dag}b_{n}\cr\cr
	&&+ (J_{1}b_{n}^{\dag}a_{n}+J_{2}a_{n+1}^{\dag}b_{n}+\mathrm{H.c.}]\cr\cr
	&&+\sum_{n}J_{2}\left(b_{n}^{\dag}a_{1}+a_{1}^{\dag}b_{n}\right).
\end{eqnarray}
If we further set the detuning of the resonators $\Delta_{a,n}=\Delta_{b,n}=\Delta_{R}$ as the zero-point of the energy, the above Hamiltonian has the identical form as the Hamiltonian in Eq.~(\ref{e03}).

Besides the quantum fluctuations of the operators, the mean value of the operators also satisfy a set of differential equations
\begin{eqnarray}\label{A7}
	\dot{\langle a_{n}\rangle} &=&-i\left(\Delta_{a,n}-i\frac{\kappa _{{a_{n}}} }{2}\right)\langle a_{n}\rangle-iJ_{2}\langle b_{n-1}\rangle-iJ_{1}\langle b_{n}\rangle-i\Omega_{a,n},\cr\cr
	\dot{\langle b_{n}\rangle} &=&-i\left(\Delta_{b,n}-i\frac{\kappa _{{b_{n}}} }{2}\right)\langle b_{n}\rangle-iJ_{2} \langle a_{1}\rangle-iJ_{1} \langle a_{n}\rangle\cr\cr
	&&-iJ_{2}\langle a_{n+1}\rangle-i\Omega_{b,n}~~~(n=2,3,\cdots,N),
\end{eqnarray}
with the boundary condition  $\dot{\langle a_{1}\rangle}=-i\left(\Delta_{a,1}-i\frac{\kappa _{{a_{1}}} }{2}\right)\langle a_{1}\rangle-iJ_{1}\langle b_{1}\rangle-iJ_{2} \left (\langle b_{2}\rangle+\langle b_{3}\rangle+\cdots+\langle b_{N}\rangle\right )-i\Omega _{a,1}$, $\dot{\langle b_{1}\rangle}=-i\left(\Delta_{b,1}-i\frac{\kappa _{{b_{1}}} }{2}\right)\langle b_{1}\rangle-iJ_{1}\langle a_{1}\rangle-iJ_{2}\langle a_{2}\rangle-i\Omega _{b,1}$, and $\dot{ \langle a_{N+1}\rangle}=-i \left(\Delta_{a,N+1}-i\frac{\kappa _{{a_{N+1}}}}{2}\right)\langle a_{N+1}\rangle-iJ_{2}\langle b_{N}\rangle-i\Omega _{a,N+1}$. For the superconducting resonator possessing the external driving and decay simultaneously, the injected photons and the leaked photons may reach balance, making the system enter the steady state, namely, $\dot{\langle\rho\rangle}=0$. Thus, under the assumption of steady state, we have
\begin{eqnarray}\label{A8}
	\overrightarrow{\mathbf{R}}=-\left(\mathbf{\Delta}+\mathbf{M}-i\frac{\mathbf{K}}{2}\right) ^{-1}\overrightarrow{\mathbf{\Omega}},
\end{eqnarray}
where $\overrightarrow{\mathbf{R}}=[\langle a_{1}\rangle,\langle b_{1}\rangle,\cdots,\langle a_{N}\rangle,\langle b_{N}\rangle,\langle a_{N+1}\rangle]^{T}$ represents the column vector composed by the mean value of the steady state for resonator, $\mathbf{\Delta}=\mathrm{Diag}[\Delta_{a,1},\Delta_{b,1},\cdots,\Delta_{a,N},\Delta_{b,N},\Delta_{a,N+1}]=\mathrm{Diag}[\Delta_{R},\Delta_{R},\cdots,\Delta_{R},\Delta_{R},\Delta_{R}]$ is the diagonal matrix originating from the detuning of the resonators, $\mathbf{M}$ is the coefficient matrix owning the same form as the Hamiltonian in Eq.~(\ref{e03}), $\mathbf{K}=\mathrm{Diag}[\kappa _{a_{1}},\kappa _{b_{1}},\cdots,\kappa _{a_{N}},\kappa _{b_{N}},\kappa _{a_{N+1}}]$ represents the diagonal matrix caused by the decay of the resonators, and $\overrightarrow{\mathbf{\Omega}}=[\Omega_{a,1},\Omega_{b,1},\cdots,\Omega_{a,N},\Omega_{b,N},\Omega_{a,N+1}]^{T}$ is the external driving.

\bibliographystyle{apsjnl}

\end{document}